\documentclass[aps,twocolumn,showpacs,nofootinbib]{revtex4-2}
\usepackage{graphicx}
\usepackage{textcomp}
\usepackage{psfrag}
\usepackage{epstopdf}
\usepackage{amssymb}
\usepackage{amsmath}
\usepackage{xcolor}
\usepackage{float}
\usepackage{lineno}
\usepackage{makecell}
\usepackage{setspace}
\usepackage{lipsum}%
\usepackage{inputenc}
\usepackage{soul,collcell,datatool}
\usepackage[colorlinks=true,linkcolor=blue,citecolor=blue,urlcolor=blue]{hyperref}
\usepackage[normalem]{ulem}

\begin{document}
\setstretch{1.0}
\renewcommand\linenumberfont{\normalfont\scriptsize\sffamily\color{red}}

\title{Structural, electronic, vibrational, optical, piezoelectric, thermal and thermoelectric properties of BCZT from first-principles calculations} 

\author{Debidutta Pradhan and Jagadish Kumar$^{}$}
\email{Corresponding author	jagadish.physics@utkaluniversity.ac.in}
\affiliation{Center of Excellence in High Energy and Condensed Matter Physics, Department of Physics, Utkal University, Bhubaneswar 751004, India 
}
\begin{abstract}

Perovskite material such as BCZT (Ba$_{0.875}$Ca$_{0.125}$(Zr$_{0.125}$Ti$_{0.875}$)O$_{3}$) is well known for its high value of piezoceramic properties and Curie temperature which has potential applications in sensors, actuators, optoelectronic and thermoelectric devices. Based on its composition and physical parameters such as pressure and temperature, experimentally, BCZT shows different crystal structures (rhombohedral, tetragonal and orthorhombic) with multiferroic properties. Here, we have designed these materials by comparing experimental stoichiometry and evaluated their stability by calculating the tolerance factor, formation energy, and cohesive energy. 
The structural, electronic and vibrational properties of BCZT are explored using generalized gradient approximation (GGA) within the framework of density functional theory. We have also shown variation of piezoelectric and optical properties through multiple phases using time dependent density functional theory. The electronic band gap, optical response in the visible light range, as well as piezoelectric, electrical, thermal, and thermoelectric properties, demonstrate excellent characteristics, making this material a promising lead-free ferroelectric candidate for various energy harvesting applications. Boltzmann transport theory is used for the calculation of Seebeck coefficient, electron thermal conductivity, and electrical conductivity to estimate the power factor and figure of merit which represents the efficiency of the material. The high values at the Fermi level suggest that these materials are well-suited for future device applications.
\end{abstract}
\maketitle
\section{Introduction} 
Piezoelectric materials are widely used in the area of sensors, actuators, resonators, capacitors, medical imaging systems and energy harvesting devices \cite{Martin72,Uchino10,HLi14,Zaszczynska20,Habib22}. This includes quartz, Ba titanate (BTO), lead zirconate, lead metaniobate (PbNb$_{2}$O$_{6}$), lead zirconate titanate(PZT) etc. PZT is one of the most accepted stable piezoelectrics with wide applicability in multiple electro-mechanical devices because of its high permittivity, piezoelectric coefficient (d$_{33} \thicksim $300-500 pC$N^{-1}$) and greater sensitivity even at a high operating temperature \cite{Gubinyi08,Zou23,Huang23}. This piezoceramic has low manufacturing cost, high bending ability and better physical strength. Commercially available PZT also possesses high Curie temperature ($\thicksim$ 386$^{o}$C) making it irreplaceable, however, in $2002$ lead was restricted because of its hazardous nature \cite{Eitel01,EUdir02,EUdir11,Bellaiche00}. So, intensive research on lead free ferroelectric materials with piezoelectric properties was necessary and one of the alternatives is to investigate suitable perovskite materials \cite{Acosta17}. Alkaline Niobate was the first promising alternative as a lead free perovskite but has less efficiency with low operating temperature \cite{Saito04}. This work improved significantly with alkaline bismuth and titanium composites, which demonstrated higher efficiency \cite{Takenaka91,Kanie11,kobayashi04}. Further, Ba Titanate (BTO) was investigated which has significantly high dielectric properties than a piezoelectric ceramic ($d_{33} \thicksim 200$ pCN$^{-1}$) \cite{Heartling99,Zhu14,Ma12}. This lead free ferroelectric perovskite is found to be more effective than PZT, by suitable doping and forming the composites. The elemental composition, atomic positions and structural symmetry can influence the piezo-electric coefficient as well as Curie temperature. Greater efficiency can be achieved by combining different bi-valent and tetra-valent dopants at A and B sites of the perovskite ($ABO_3$) respectively \cite{Zhu14,Huan13}. Possible dopants at the Ba site include Ca$^{2+}$, Sr$^{2+}$ and La$^{2+}$ while, at the Ti site, the possible dopants are Zr$^{4+}$, Sn$^{4+}$ and Hf$^{4+}$ \cite{Zhu14,Kimura13,Yang12,Chen14}. Liu et al. found that doping of BTO with Ca and Zr simultaneously forming Ba(Zr$_{0.2}$Ti$_{0.8})$O$_{3}$-x(Ba$_{0.7}$Ca$_{0.3}$)TiO$_{3}$ or (BZT-xBCT) can result in high d$_{33}$ value ($\thicksim$ 620 pC$N^{-1}$) \cite{Liu09}. 
Substituting Ca$^{2+}$ in place of Ba$^{2+}$ in BTO is found to induce better dielectric permittivity and optical properties than other dopants. This also reduces Curie temperature and increases relaxor behaviour by decreasing grain size resulting high polarization \cite{Pullar09,Panigrahi10}. Barium calcium titanate (BCT) shows faster and efficient transfer of energy among optical waves at different frequencies \cite{Xu18}. On the other hand, Zr is a suitable substituent to Ti, as it is chemically more stable with larger ionic radii to expand the lattice size and simultaneously change the ferroelectric behaviour significantly \cite{Moura08}. Hybridization with higher d-orbitals due to Zr, can cause high spontaneous polarization at optimum conditions. Optimized tuning of mutual contributions of BZT (Barium zirconate titanate) and BCT to the perovskite crystal can result in a material with enhanced properties \citep{Danila18}. Again, the tri-critical point for BCZT is around room temperature, which is lower than PZT, BTO and other standard piezoceramics \cite{Aksel10,Ghayour16,Gao17}.
It is found that the maximum dielectric and piezoelectric effects are observed at multiphase boundaries than that of single phases. In case of BCZT, the piezoelectric constant $d_{33}$ is high near the morphotropic phase boundary (MPB) between rhombohedral and tetragonal phases \cite{Liu09,Ahart08}. Stabilizing this boundary over a wide range of temperatures and pressures can be useful for efficient applications in transducers. BCZT shows four crystal phases such as rhombohedral, cubic, tetragonal and orthorhombic. Electronic properties of the material may vary in the presence of multiple phases. Doping of Ca in place of Ba reduces the polymeric phase transition temperature while increasing spontaneous polarization and nonlocal optical response. Reliability on these properties depends on stability of the phases at specific conditions like temperature, pressure and molecular composition. Distortion of octahedral voids in the perovskite structure along with orbital hybridization in the compound can induce polarization, enhancing optoelectronic properties. 
Perovskite structures with metallic character can achieve high thermoelectric properties for applications like active cooling, energy harvesting and thermal imaging \cite{Okuda01,Lee09,Jong20,Peng18}. 
 Thermoelectric properties depend more on lattice structure than that of electronic configuration, making it more effective to calculate in real space. Replacing Zr with Ti increases the stability of the compound and also induces higher thermodynamic potential \cite{Ma12}. BCT has higher polarization than BZT and the band gap is lower but with proper adjustment of these combinations better results can be achieved \cite{Rusevich19,Levin13,Zhao11,Suzuki12}. The composite $0.5$BZT-$0.5$BCT switches its symmetry with some remnant polarization at room temperature \cite{Bao10,Dey21} and large piezoelectric response due to tricritical point between rhombohedral, cubic and tetragonal phases \cite{Liu09}. Another explanation for the enhanced d$_{33}$ value is supported by structural distortion in the oxygen octahedra that increases strain in the system, resulting in reverse domain wall motion \cite{Gao14,Tutuncu14}. Here, we have presented the first-principles calculation results of 0.5-BCT-0.5-BZT (BCZT) by taking its orthorhombic, tetragonal and rhombohedral structures. Our study investigates variation of the electronic, vibrational, dielectric, optical, piezoelectric and thermoelectric properties through different phases, to understand the possible applications of the material in different fields.

\section{Computational Method}
First principles calculations are performed within the framework of density functional theory, implemented in open source software, QuantumESPRESSO and SIESTA package to get the information about materials properties \cite{Giannozzi09,Garcia20}. 
Perovskite structures are designed using VESTA software from barium titanate (BTO) of tetragonal(T), rhombohedral(R) and orthorhombic crystal structure, obtained from crystallographic open database COD$-9014668$, COD$-9015236$ and COD$-9014645$ respectively \cite{Momma11,COD}. A suitable supercell has been taken and substitution was made to design the BCZT composites based on the experimental stoichiometry \cite{Liu09}. It contains two cuboidal structures of Ba and Ti with overlapping diagonals where one corner atom of Ti resides at the center of Ba-cuboidal structure and vice versa. 
The structures are optimized by using Broyden, Fletcher, Goldfarb, Shanno (BFGS) algorithm implemented in QuantumESPRESSO package \cite{Broyden70,Fletcher70,Goldfarb70,Shanno70}. The generalized gradient approximation (GGA) with Perdew-Burke-Ernzerhof (PBE) functional was considered for exchange and correlation energies \cite{Vanderbilt90,PBE96}. Ultra-soft Pseudo potentials are used for all the atoms where the valence electrons are for Ba-($5s$, $6s$, $5p$), Ca-($3s$, $4s$, $3p$, $4p$), Zr-($4s$, $5s$, $4p$, $5p$, $4d$), Ti-($3s$, $4s$, $3p$, $3d$), O-($2s$, $2p$). A suitable K-point mesh was generated using the Monkhorst-Pack scheme and Methfessel-Paxton scheme with broadening of 0.003 Ry is used for the smearing of occupation number to obtain more accurate results in the calculation \cite{MP89,Monkhorst76}. Out of all possible structures, we found from energy minimization calculations, Ba$_{6}$Ca$_{2}$Zr$_{2}$Ti$_{6}$O$_{24}$ (BCZT6) and Ba$_{6}$Ca$_{2}$Zr$_{1}$Ti$_{7}$O$_{24}$ (BCZT7) are more stable. So we have calculated the electronic, vibrational, optical, piezoelectric, thermal and thermoelectric properties of BCZT6 and BCZT7 for rhombohedral (R3m), tetragonal (P4mm) and orthorhombic (Amm2) structures. BCZT6 and BCZT7 composites are abbreviated as $a6$ and $a7$ for orthorhombic, $p6$ and $p7$ for tetragonal, $r6$ and $r7$ for rhombohedral structures. Electronic and phonon band structure, density of states are calculated by using QuantumESPRESSO. For optical properties, we used the Spanish Initiative for Electronic Simulations with Thousands of Atoms (SIESTA) simulation package, as it allows the inclusion of multiple orbitals of choice in the calculations. \cite{Garcia20}. It uses a pseudo-atomic-orbital basis set including first and second zeta potential. Atomic orbitals of all core and valence electrons are considered along with immediate orbitals next to valence electrons. GGA(PBE) functional is used with mesh cut-off $300$Ry with k-grid Monkhorst matrix of $5\times 5 \times 5$. Using conjugate gradient approximation for optical calculation, a $5\times 5 \times 5$ optical mesh is used along the [0 0 1] and 1400 optical bands with 0.5eV optical broadening is used. The calculated result is confined in an energy range of $0.0$ eV to $20$ eV. We have also calculated piezoelectric constant and Born effective charge (BEC) using the Vienna Ab Initio Simulation Package (VASP) with a mesh grid of $4\times 4 \times 4$ centered at $\Gamma$ point\cite{Kresse93,Kresse96}. Further, thermal, thermoelectric and transport properties are characterized using PHONOPY and Wannier90 codes \cite{Togo15,Kresse95,Pizzi14,Pizzi20}.
\begin{figure}
\includegraphics[height=4.0cm,width=4.55cm]{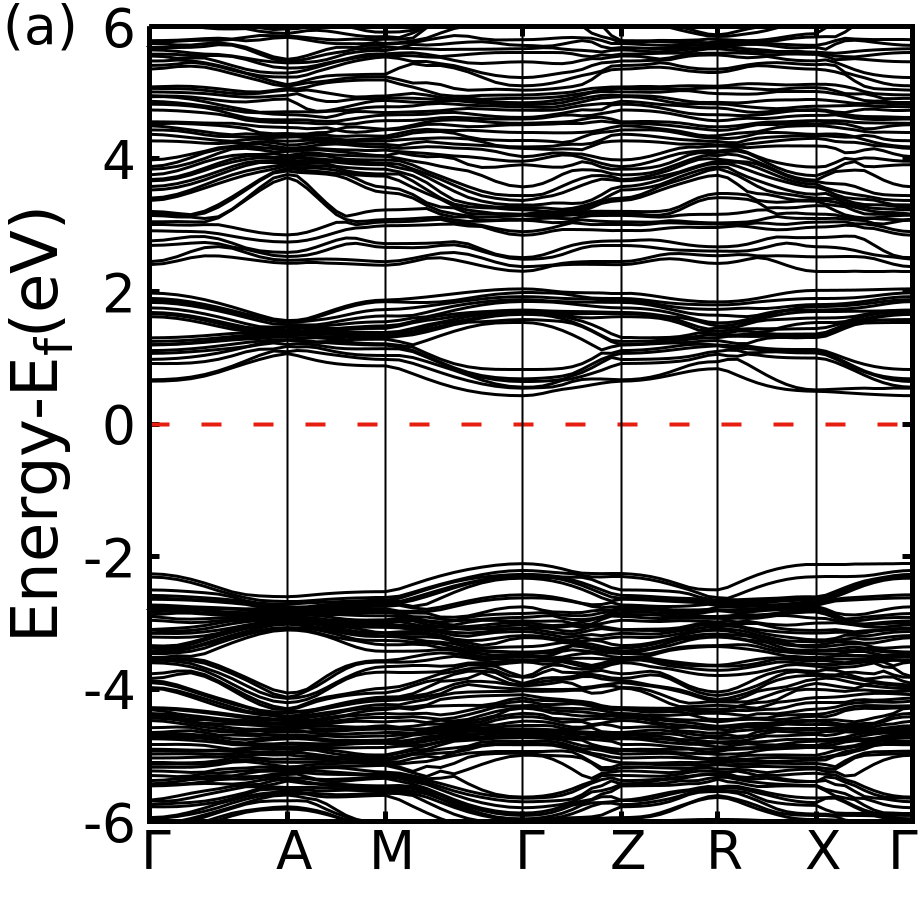}
\includegraphics[height=3.9cm,width=2.6cm]{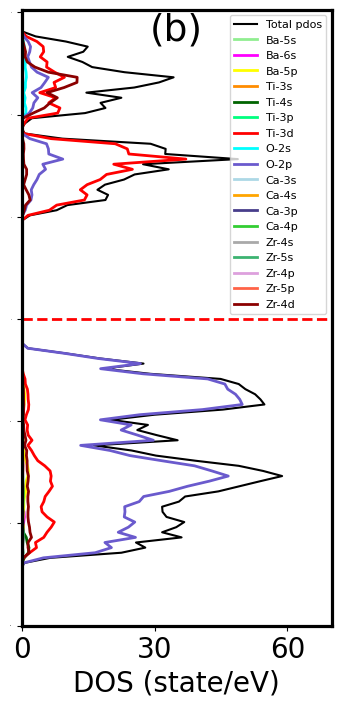}
\includegraphics[height=4.0cm,width=4.55cm]{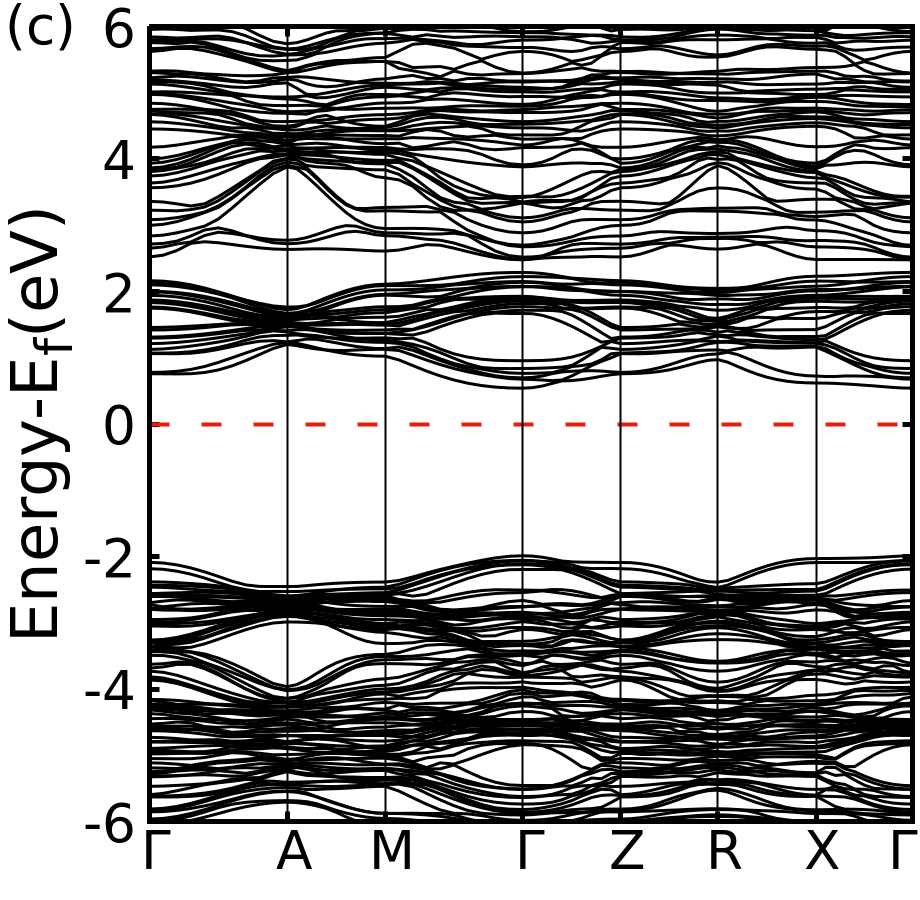}
\includegraphics[height=3.9cm,width=2.6cm]{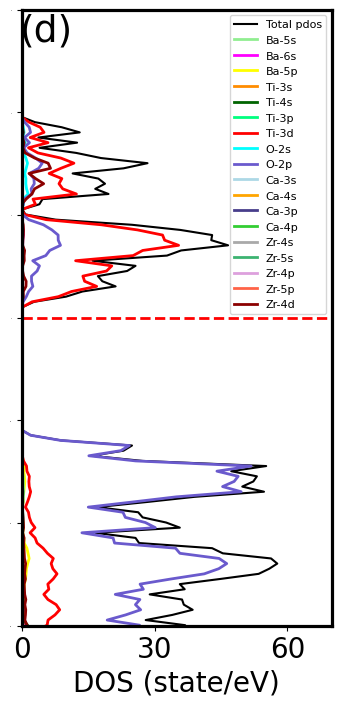}
\includegraphics[height=4.0cm,width=4.55cm]{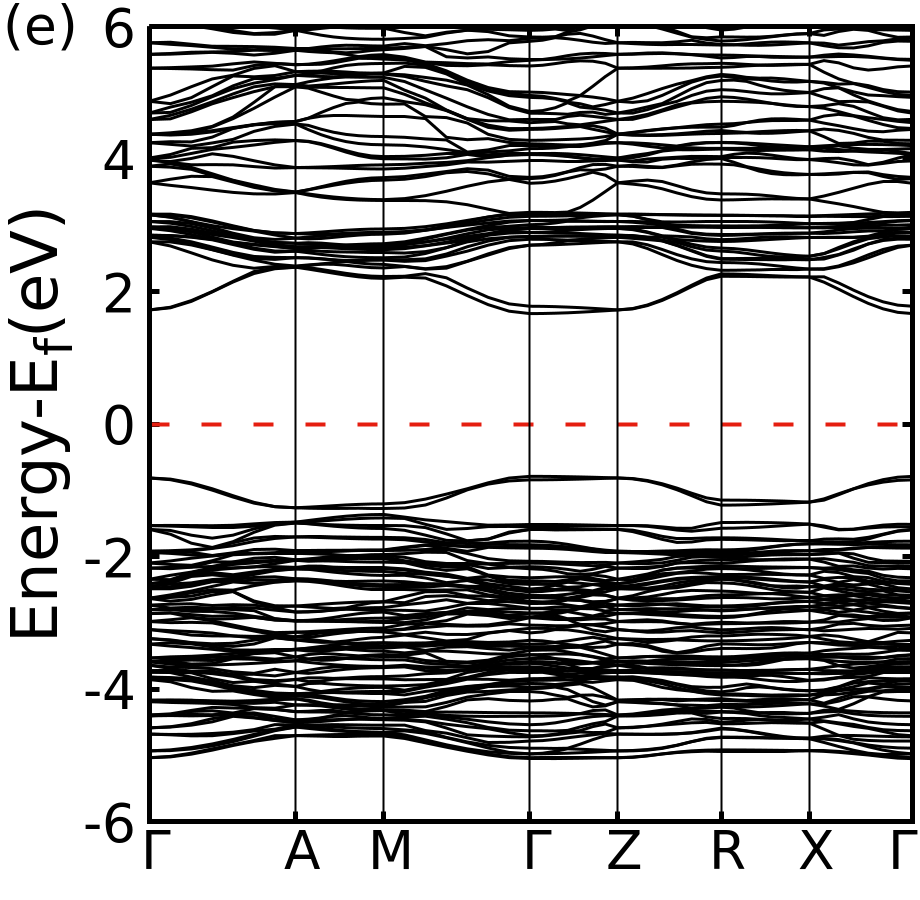}
\includegraphics[height=3.9cm,width=2.6cm]{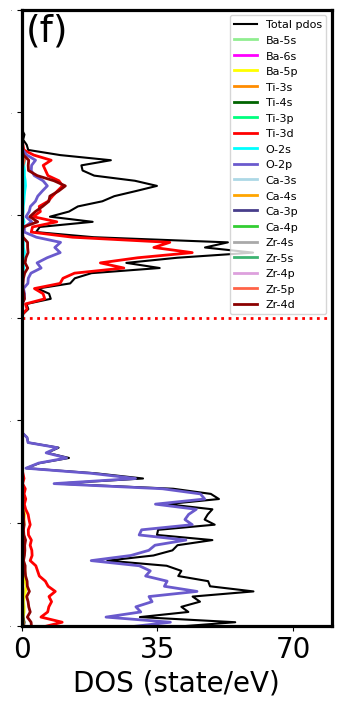}
\includegraphics[height=4.0cm,width=4.55cm]{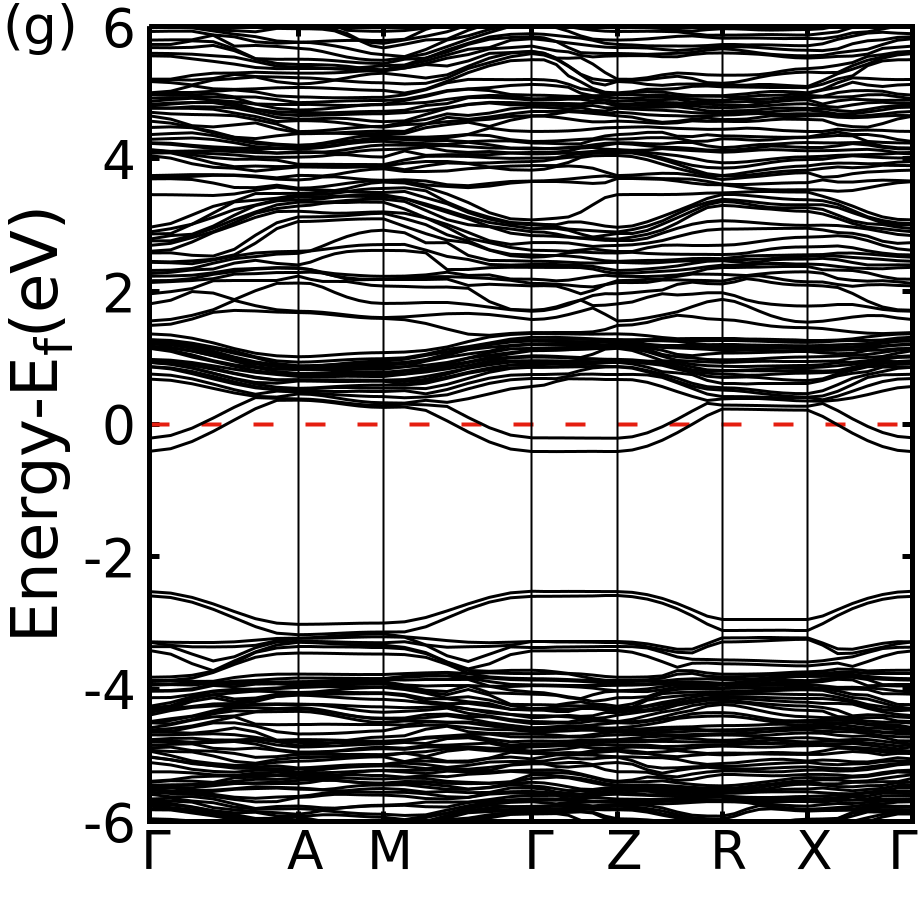}
\includegraphics[height=3.9cm,width=2.6cm]{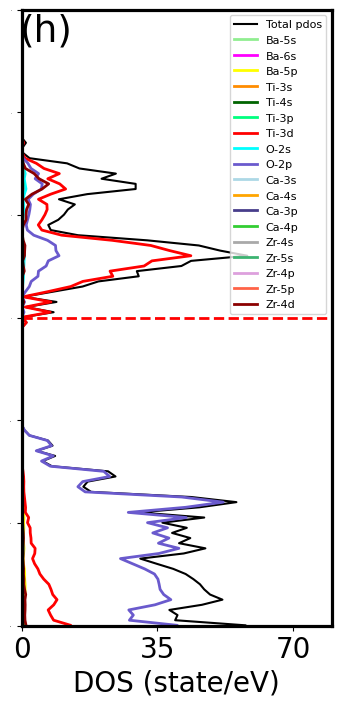}
\includegraphics[height=4.0cm,width=4.55cm]{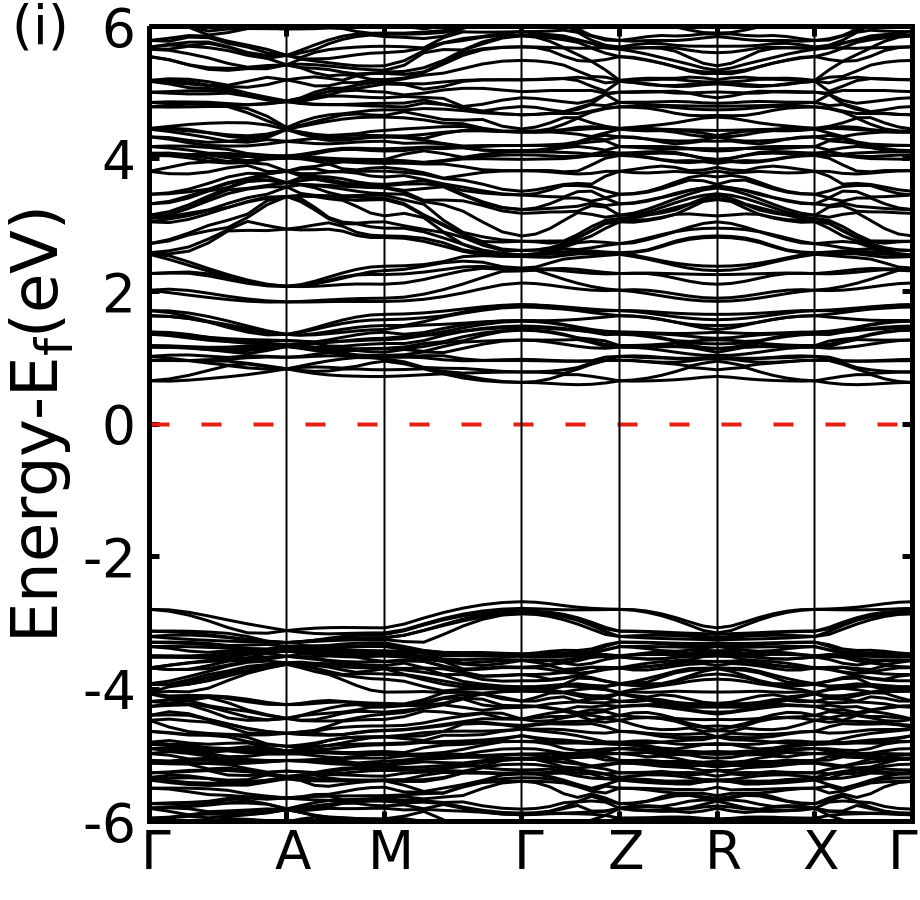}
\includegraphics[height=3.9cm,width=2.6cm]{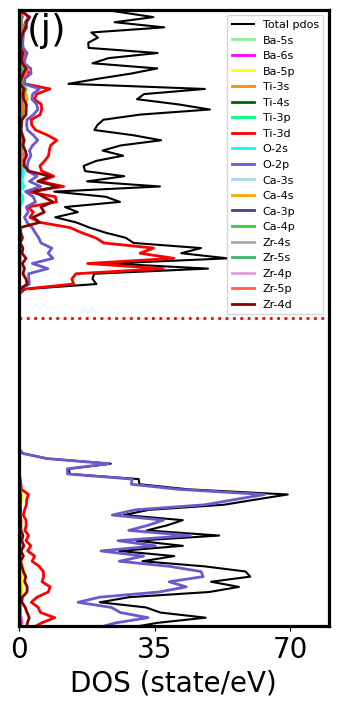}
\includegraphics[height=4.0cm,width=4.55cm]{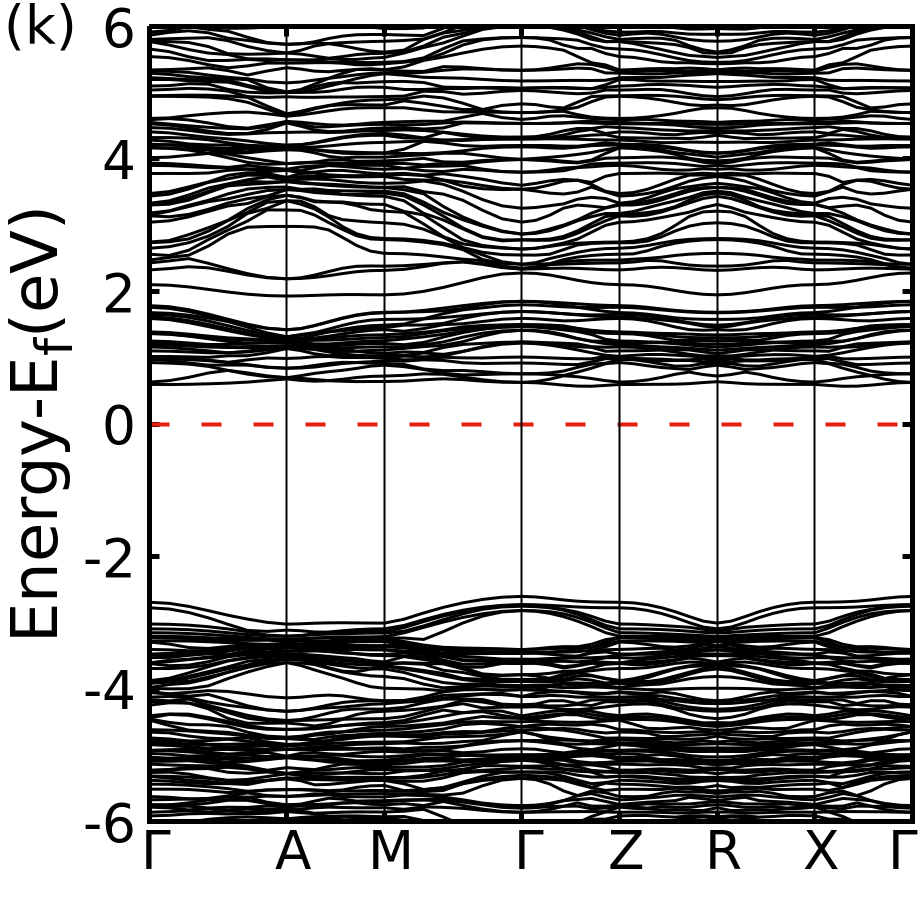}
\includegraphics[height=3.9cm,width=2.6cm]{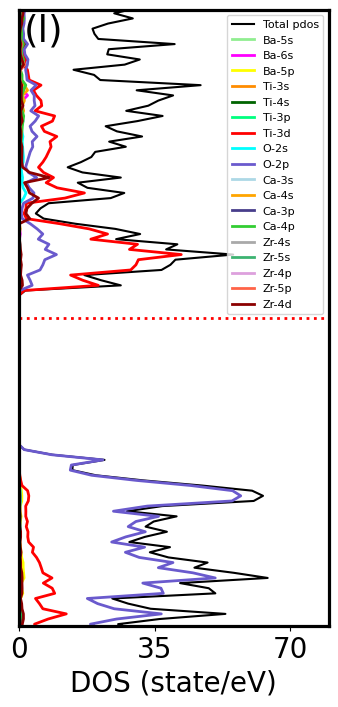}
\caption{Electronic band structures and corresponding density of states for (a,b) $a6$, (c,d) $a7$, (e,f) $p6$, (g,h) $p7$, (i,j) $r6$ and (k,l) $r7$}
\label{band}
\end{figure}
\section{Results and Discussion}
We present the detailed analysis of stability, electronic, optical, piezoelectric, thermal, thermoelectric and transport properties of the selected BCZT composites for the possible potential applications of the material. The above properties are discussed in detail in the following subsections.
\subsection{Tolerance Factor, Formation and Cohesive Energy}
The stability of a material is inherently challenging due to its compositional complexity and structural modifications, which influence its thermodynamic and kinetic properties. Measure of Goldschmidt tolerance factor indicates the stability of a perovskite structure (ABO$_{3}$) and that is efficient for nearly 70-72$\%$ of perovskites. It is defined as,
\begin{equation}
t=\dfrac{r_{A}+r_{O}}{\sqrt{2}(r_{B}+r_{O})},
\end{equation}
where $r_{A}$, $r_{B}$ and $r_{O}$ are the average ionic radii of atoms based on their coordination number \cite{Goldschmidt26,Shannon76}. In our case the calculated value radii $r_{A}=1.5425$, $r_{B_{6}}=0.6337$ and $r_{B_{7}}=0.620$, where $r_{B_{6}}$ and $r_{B_{7}}$ are the average ionic radii for Zr and Ti atoms in BCZT6 and BCZT7 structures respectively. The values of $t$ are $1.023$ and $1.030$ for BCZT6 and BCZT7 which are in the stable range of $0.825 < t < 1.059$. Bartel et al., reported another formula by including oxidation state with ionic radii given by,
\begin{equation}
t^{\prime}=\dfrac{r_{O}}{r_{B}}-n_{A}\Big(n_{A}-\dfrac{r_{A}/r_{B}}{ln(r_{A}/r_{B})}\Big)
\end{equation}
where $n_{A}=2$ is the oxidation state of atoms in site A \cite{Bartel19}. If $t^{\prime}$ is less than 4.18 then the structure is considered to be stable. Here, the value of $t^{\prime}$= 3.682 and 3.719 for the BCZT6 and BCZT7 respectively, confirms the structural stability.

Further, the composite is analyzed by calculating the formation and cohesive energies which are defined as, 
\begin{eqnarray}
\nonumber
E_{formation} &=& E^{system}_{total}-\sum_{constituent}n E_{bulk} \hspace{0.5cm} \text{and}, \\
E_{cohesive} &=& E^{system}_{total}-\sum_{constituent}n E_{atomic}.
\end{eqnarray}
where, $E_{bulk}$ is the energy of the bulk individual system per atom and $E_{atomic}$ is the energy of the isolated atom. The calculated values shown in Table \ref{formation} indicate the stability of BCZT composites.
\begin{table}
\begin{center}
\caption{\label{formation} Formation and cohesive energy of different BCZT composite}
\begin{tabular}{|c|c|c|c|c|c|c|}
\hline 
BCZT & $a6$ & $a7$ & $p6$ & $p7$ & $r6$ & $r7$ \\ 
\hline 
\thead{$E_{formation}$ \\ (eV)/atom} & -3.031 & -3.010 & -2.990 &-2.9730 & -3.030 & -3.018 \\
\hline 
\thead{$E_{cohesive}$\\ (eV)/atom} & 29.627 & 29.635 & 29.669 & 29.672 & 29.628 & 29.627 \\ 
\hline 
\end{tabular} 
\end{center}
\end{table}
\begin{figure}
\includegraphics[height=3.9cm,width=4.5cm, angle =0]{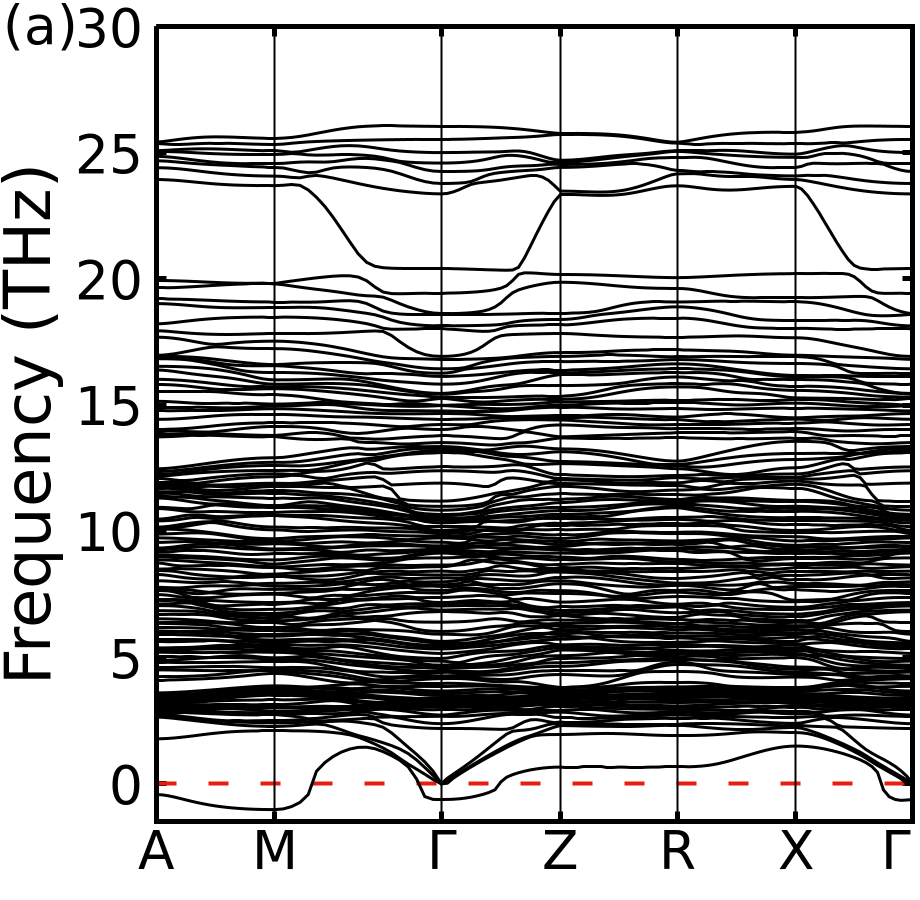}
\includegraphics[height=3.85cm,width=2.5cm, angle =0]{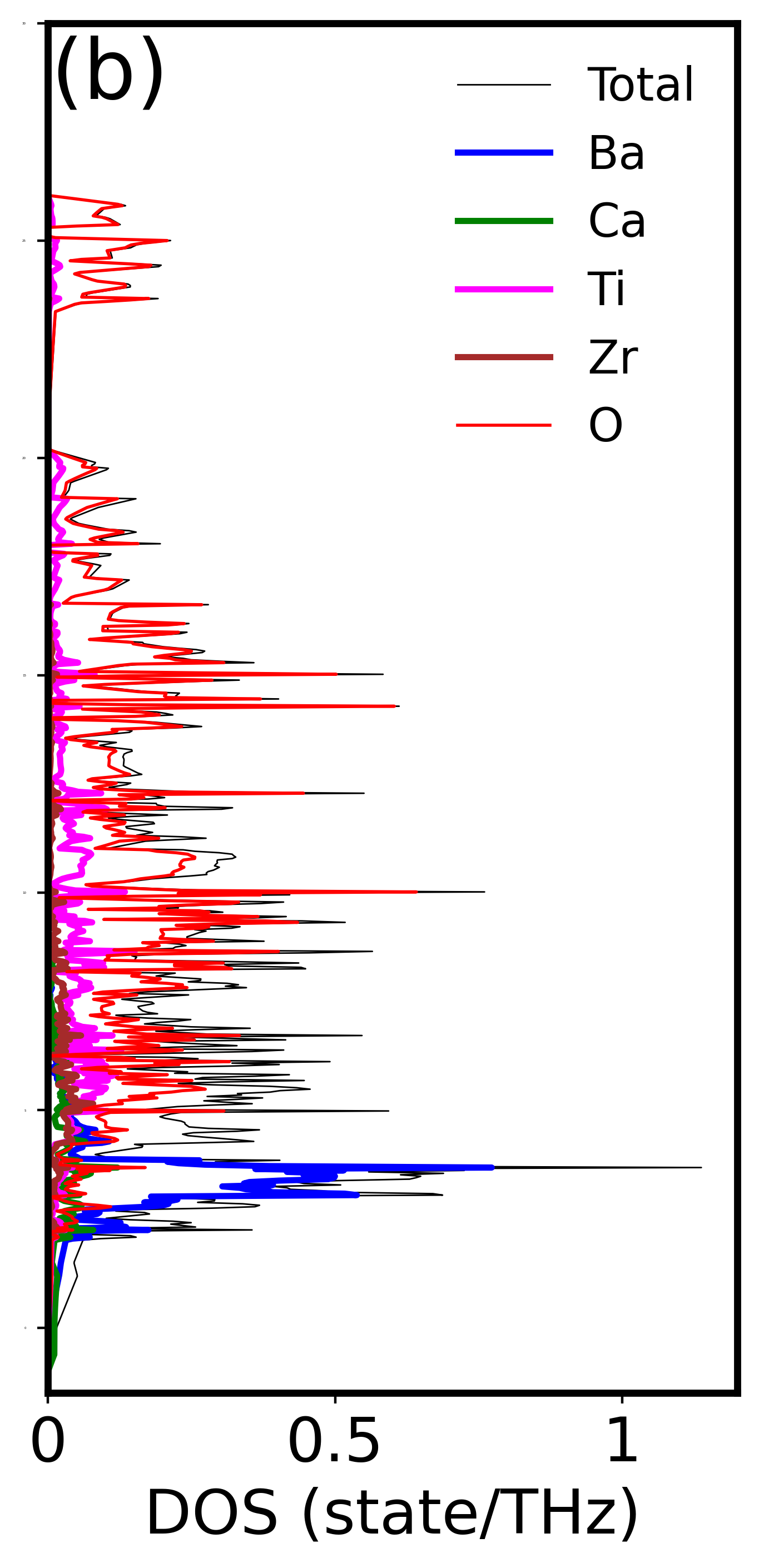}
\includegraphics[height=3.9cm,width=4.5cm, angle =0]{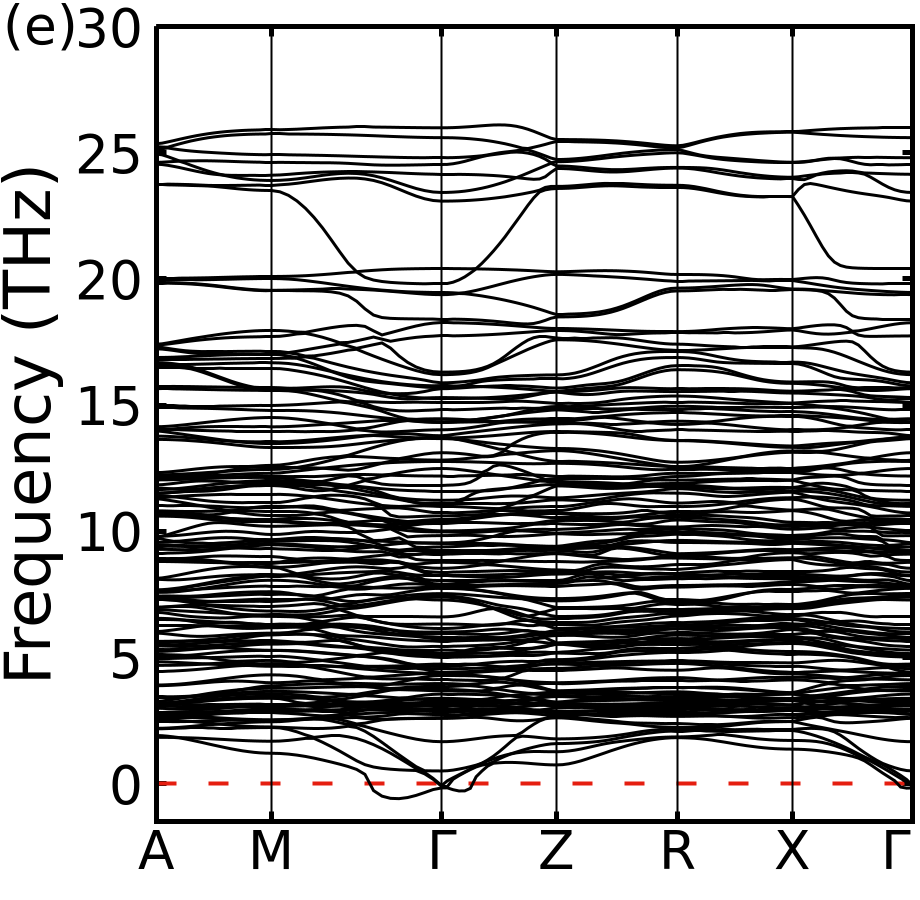}
\includegraphics[height=3.85cm,width=2.5cm, angle =0]{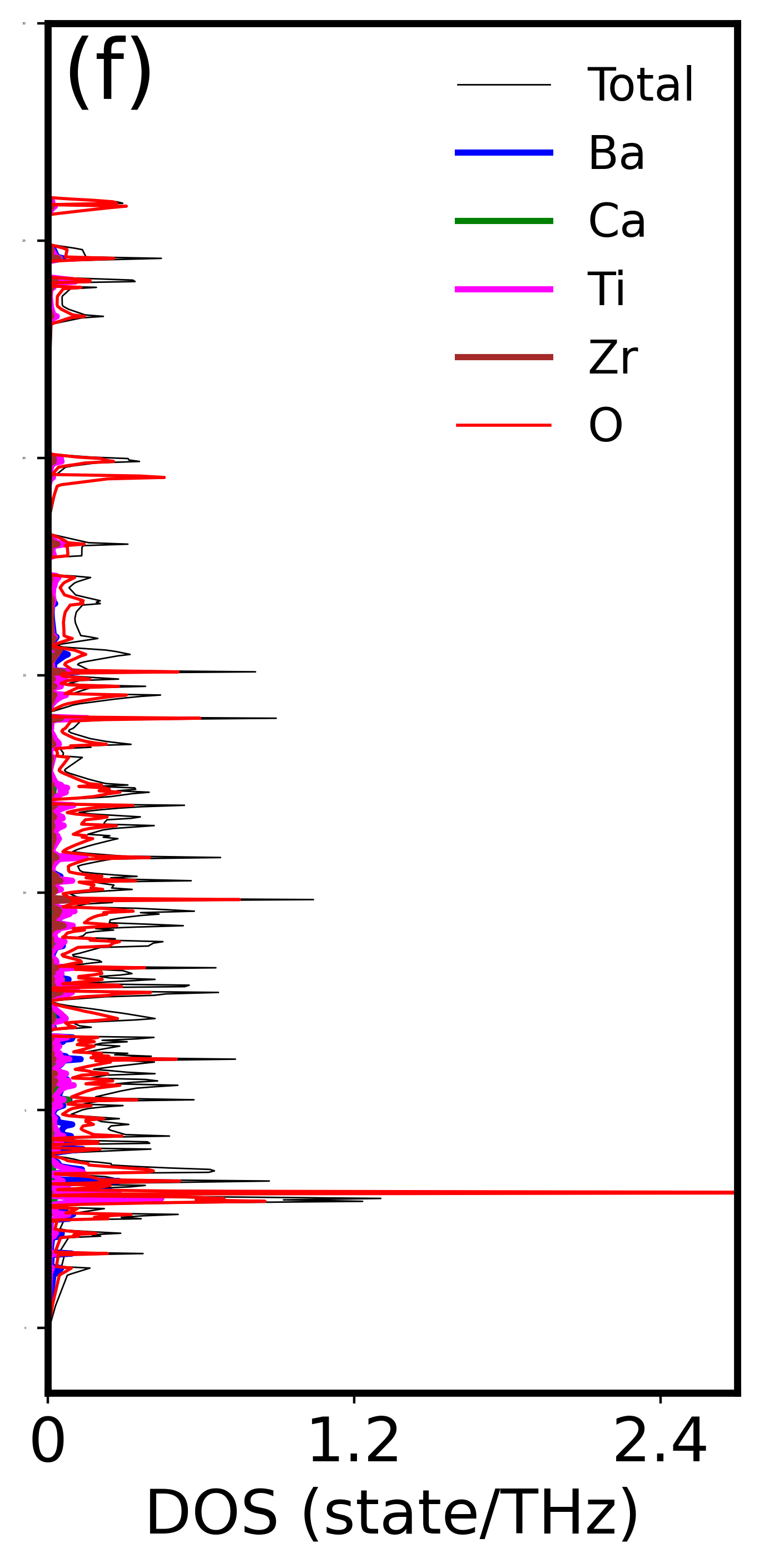}
\includegraphics[height=3.9cm,width=4.5cm, angle =0]{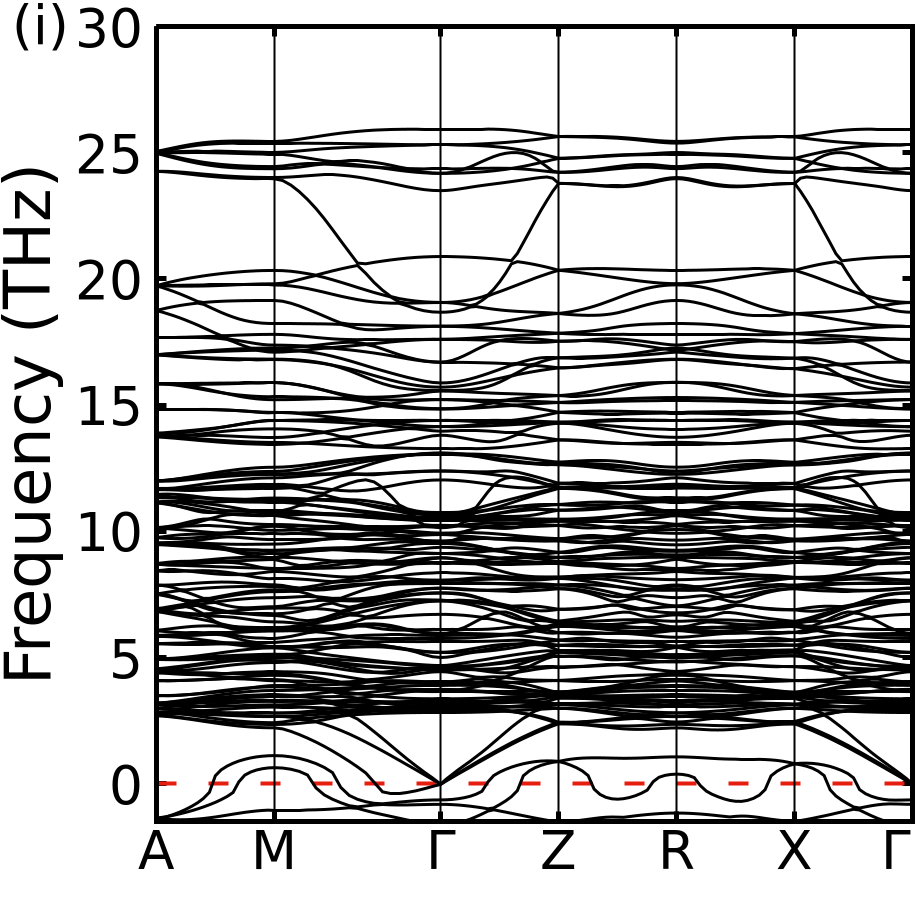}
\includegraphics[height=3.85cm,width=2.5cm, angle =0]{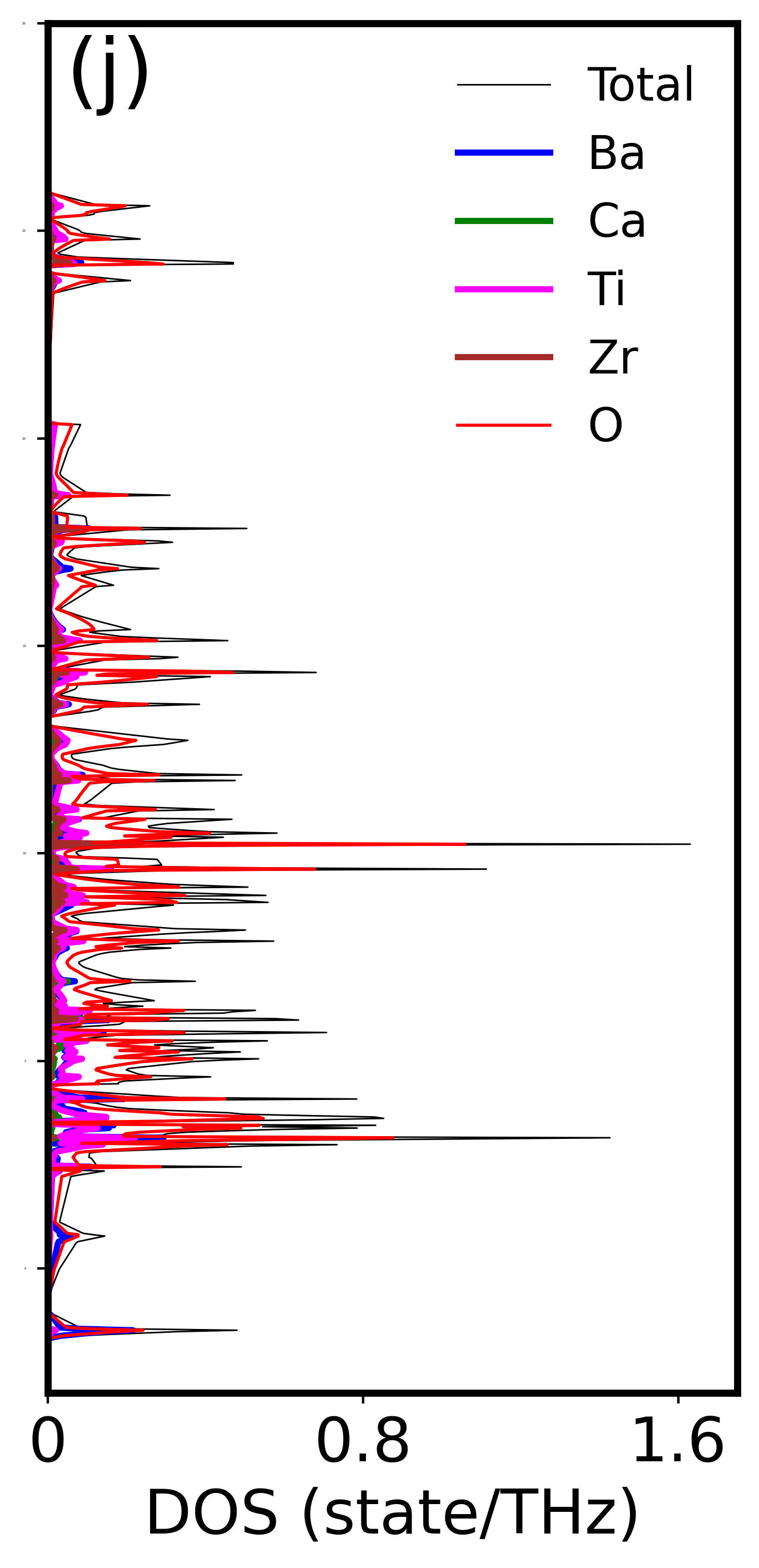}
\includegraphics[height=3.9cm,width=4.5cm, angle =0]{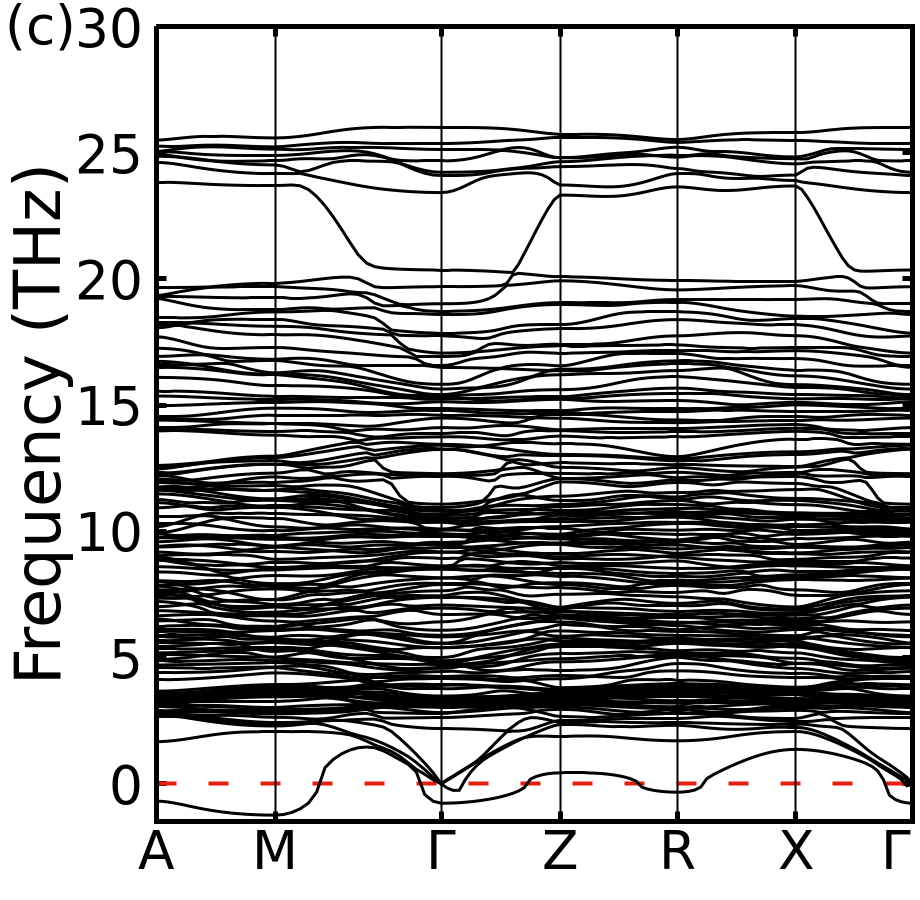}
\includegraphics[height=3.85cm,width=2.5cm, angle =0]{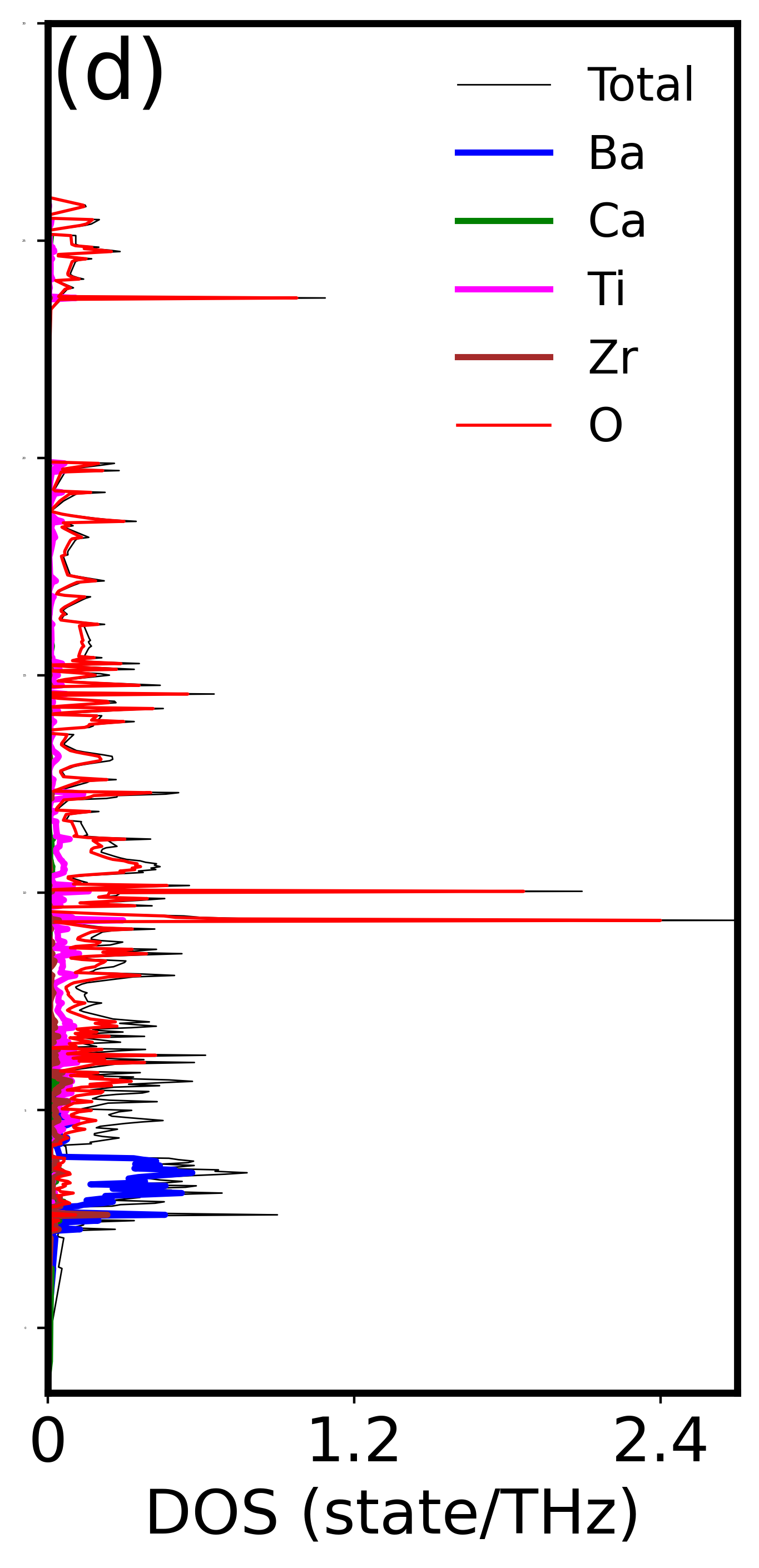}
\includegraphics[height=3.9cm,width=4.5cm, angle =0]{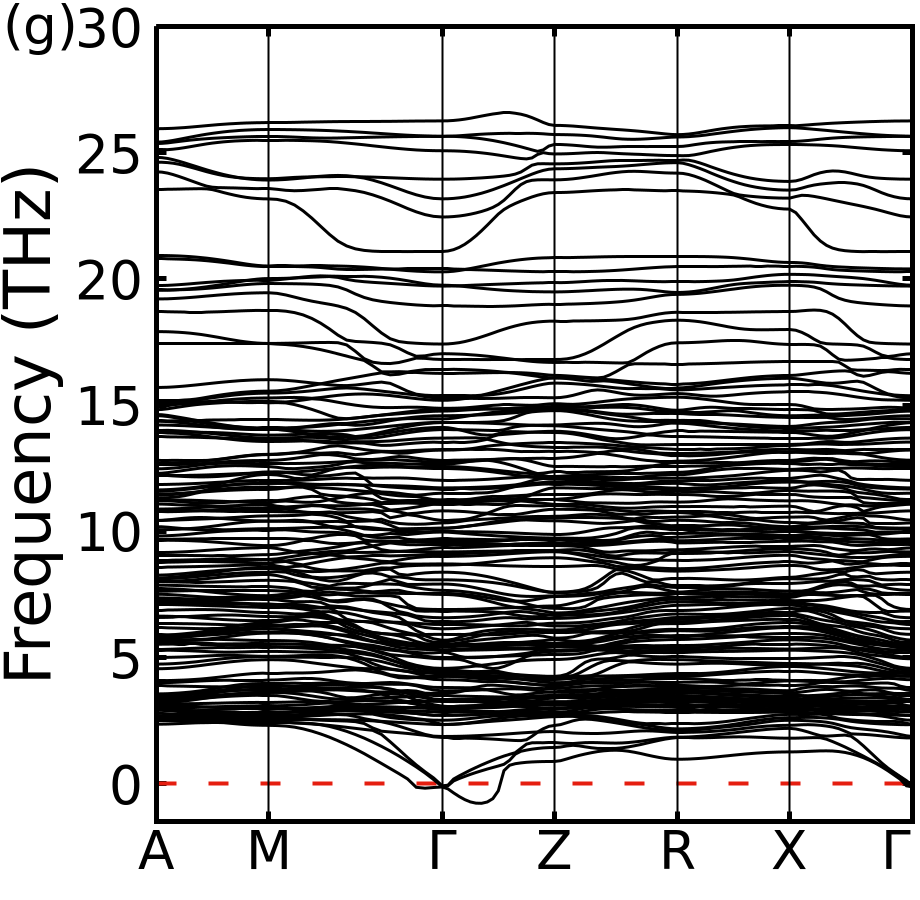}
\includegraphics[height=3.85cm,width=2.5cm, angle =0]{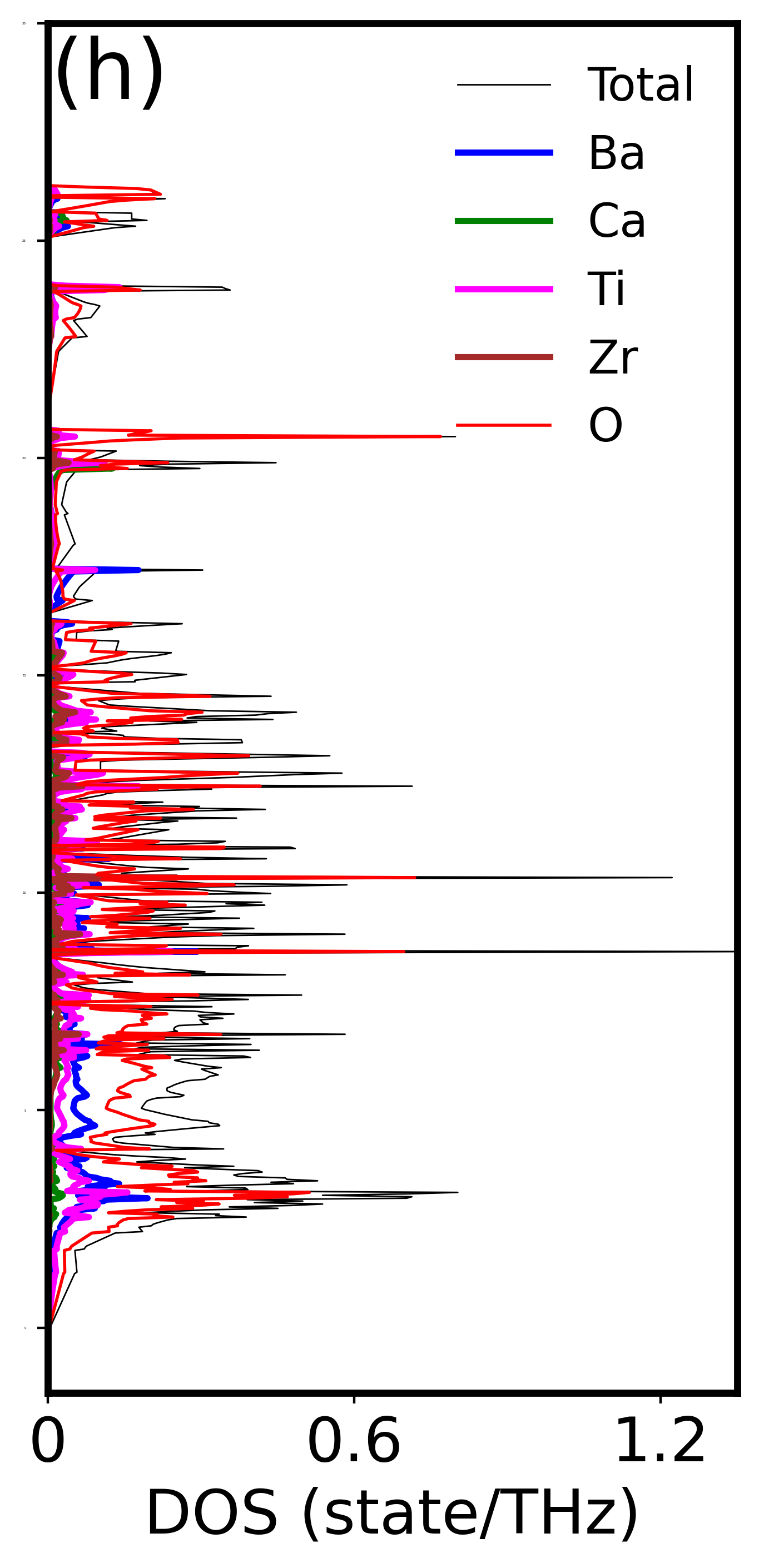}
\includegraphics[height=3.9cm,width=4.5cm, angle =0]{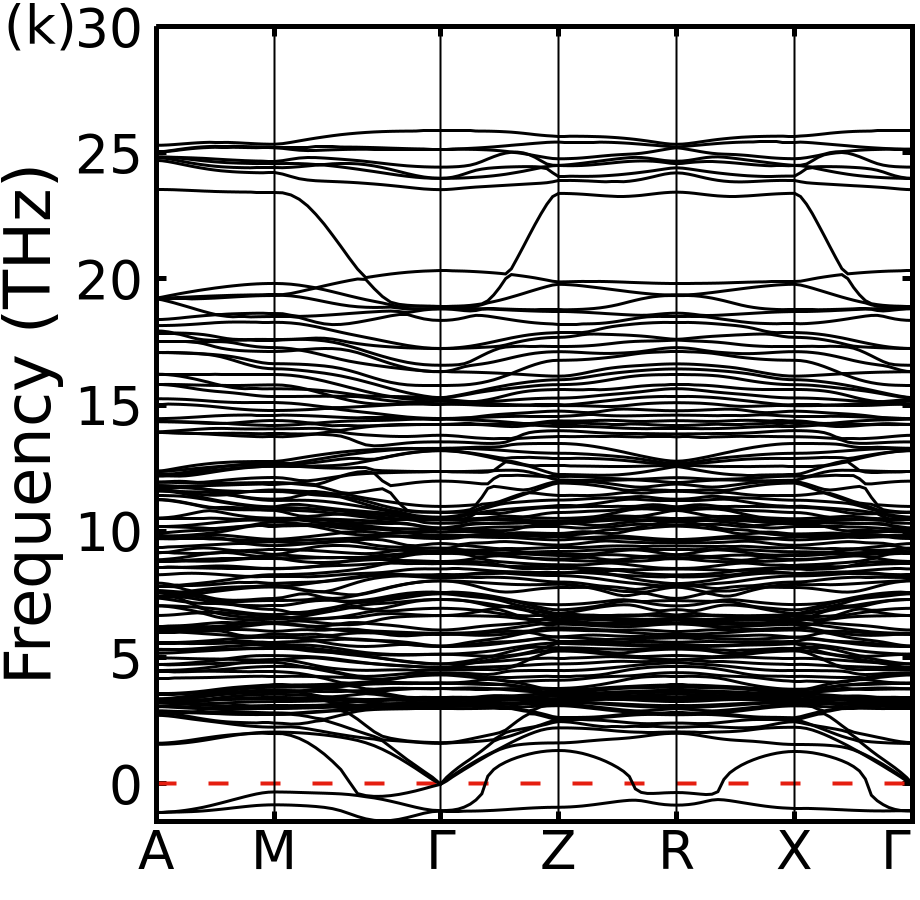}
\includegraphics[height=3.85cm,width=2.5cm, angle =0]{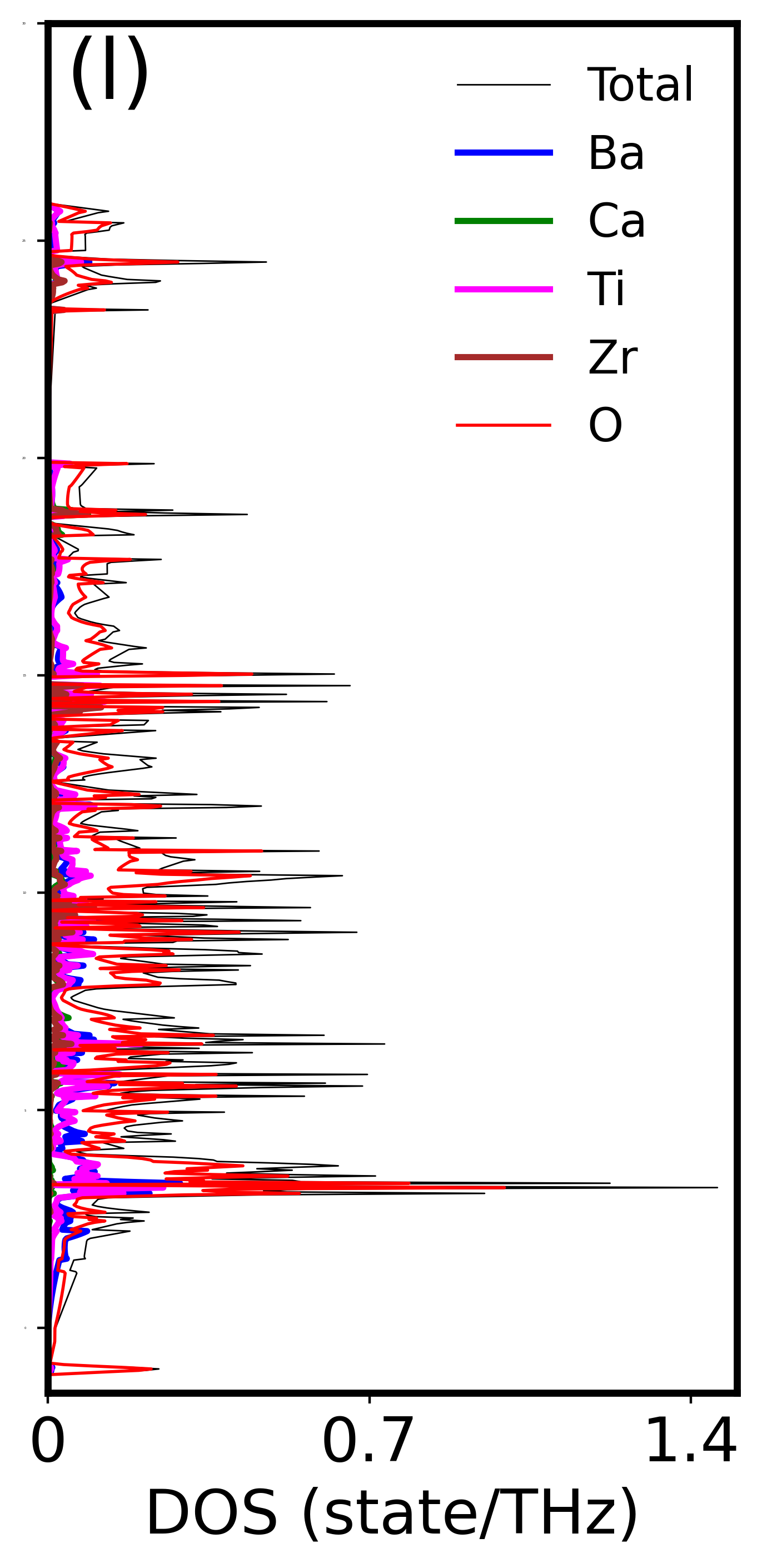}
\caption{Phonon dispersion and density of states of (a,b) $a6$, (c,d) $a7$, (e,f) $p6$, (g,h) $p7$, (i,j) $r6$ and (k,l) $r7$}
\label{phband}
\end{figure}
\begin{table}[!b]
\begin{center}
\caption{\label{bandgap} Electronic band gap of BCZT}
\begin{tabular}{|c|c|c|c|c|c|}
\hline 
Amm2 & $E_{g}(eV)$ & P4mm & $E_{g}(eV)$ & R3m & $E_{g}(eV)$ \\ 
\hline 
 $a6$ & 2.90658 & $p6$ & 2.5358 & $r6$ & 3.4484 \\
\hline 
 $a7$ & 2.84657 & $p7$ & 2.117 & $r7$ & 3.2799 \\ 
\hline 
\end{tabular} 
\end{center}
\end{table}
\subsection{Band structures and density of states}
The band structure (BS) and partial density of state (PDoS) are calculated for all compositions (Ba$_{6}$Ca$_{2}$Zr$_{2}$Ti$_{6}$O$_{24}$ and Ba$_{6}$Ca$_{2}$Zr$_{1}$Ti$_{7}$O$_{24}$) of different possible space groups. The band structures are shown in Fig. \ref{band} along with the high symmetric path $\Gamma$-A-M-$\Gamma$-Z-R-X-$\Gamma$ and the band gap values are shown in the Table \ref{bandgap}. As the number of Ti atom increases, the band gap decreases, which is expected because of increase in Ti$-3d$ orbital contribution. In all cases the conduction band minima is mostly dominated by Ti$-3d$ and O$-2p$ electrons whereas the valance band maxima is with O$-2p$ electrons near the Fermi level. This indicates the hybridization between valance band O-2p electrons and conduction band Ti-3d electrons. The band gap in rhombohedral structure is high but it reduces in orthorhombic and further decrease in tetragonal symmetry (see Table \ref{bandgap}). The corresponding density of states are also shown in Fig. \ref{band}. Further, the effective mass of charge carriers $m^{*} = {\hbar^{2}}/\Big({\dfrac{\partial^{2}E}{\partial k^{2}}}\Big)$ is calculated by fitting with a suitable second degree nonlinear polynomial at symmetric point (see Table \ref{effective}). Its value depends on the dispersive nature of the energy band both in conduction as well as in the valence band and having low value of effective mass increases the mobility of photo-excited charge carriers \cite{Atal19}.
\begin{table}[!b]
\begin{center}
\caption{\label{effective} Effective mass ($m^{*}$ in $\times 10^{-30}$) of charge carrier in conduction and valence band of BCZT. }
\begin{tabular}{|c|c|c|c|c|c|c|}
\hline 
BCZT & $a6$ & $a7$ & $p6$ & $p7$ & $r6$ & $r7$\\ 
\hline 
$m^*$ (CB) & 0.0391 & 0.0387 & 0.0389 & 0.0358 & 0.0235 & -0.5648 \\ 
\hline 
$m^*$ (VB)& -0.0385 & -0.0389 & -0.0644 & -0.0560 & -0.5657 & -0.5960 \\ 
\hline 
\end{tabular} 
\end{center}
\end{table}
The dynamical behavior and thermal properties of solids can be studied from the phonon dispersion relation. The dispersion curve along the high symmetry lines in the Brillouin zone and PDOS are shown in Fig. \ref{phband}. The imaginary acoustic mode is due to the structural instability that can lead to phase transitions \cite{Danila18}. There are $40$ atoms in the composite resulting in $120$ branches ($117$ optical and $3$ acoustic) and it has been seen that there is overlap between mid-optical and acoustic branches. The PDOS represents the contribution of different atoms to optical and acoustic branches. The acoustic mode phonons are due to heavier atoms like Ba and optic mode is from other atoms. 

\begin{figure}[!h]
\includegraphics[height=3.5cm,width=4.0cm]{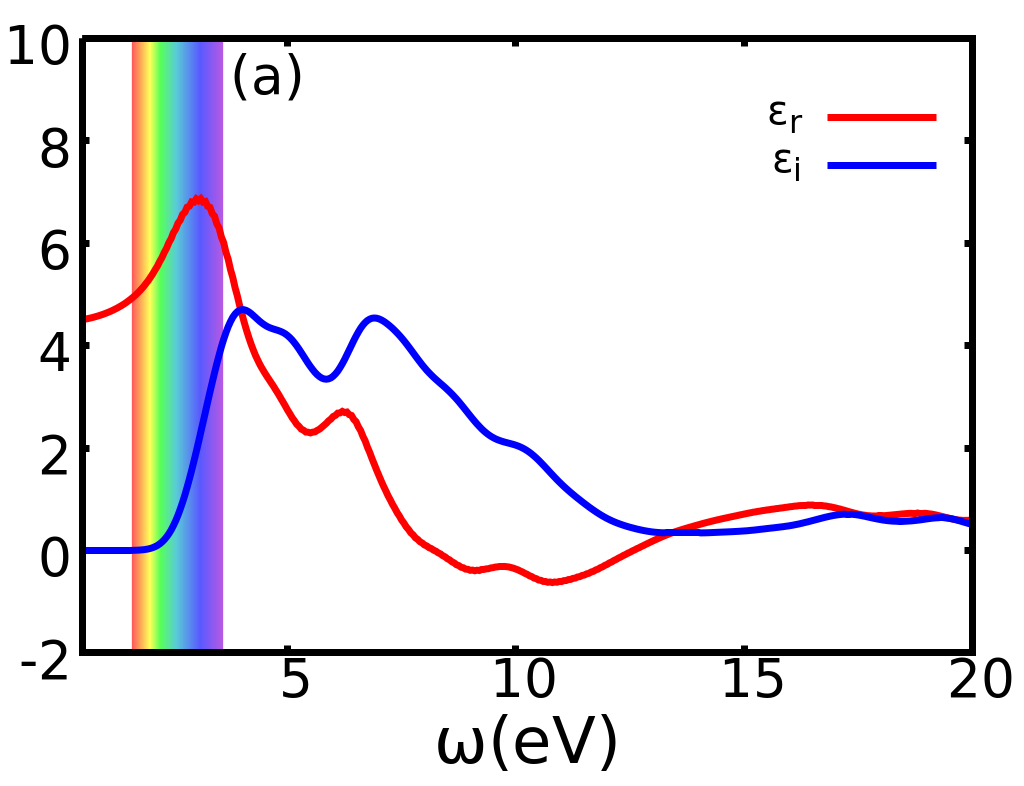}
\includegraphics[height=3.5cm,width=4.0cm]{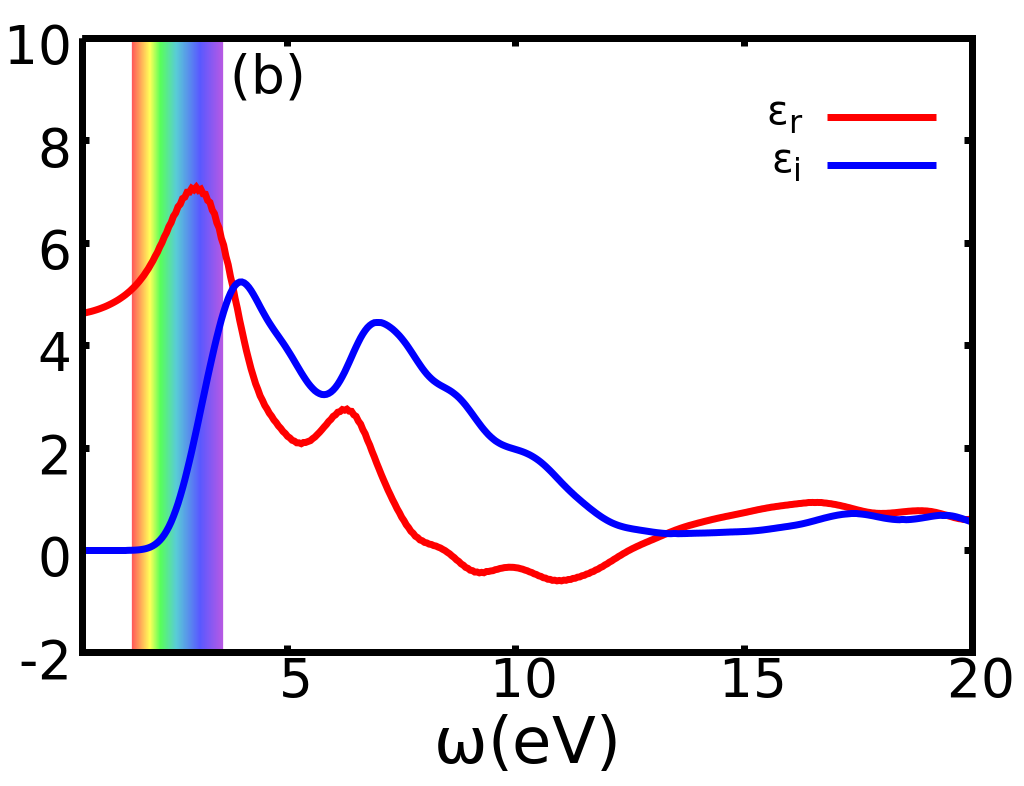}
\includegraphics[height=3.5cm,width=4.0cm]{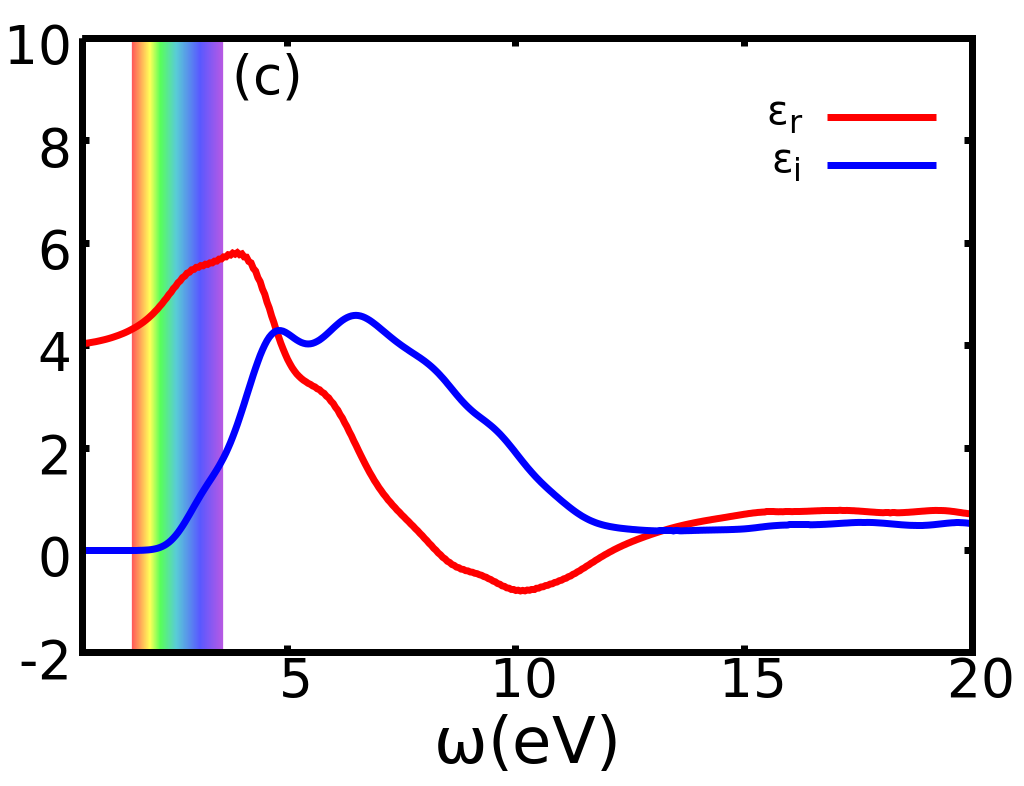}
\includegraphics[height=3.5cm,width=4.0cm]{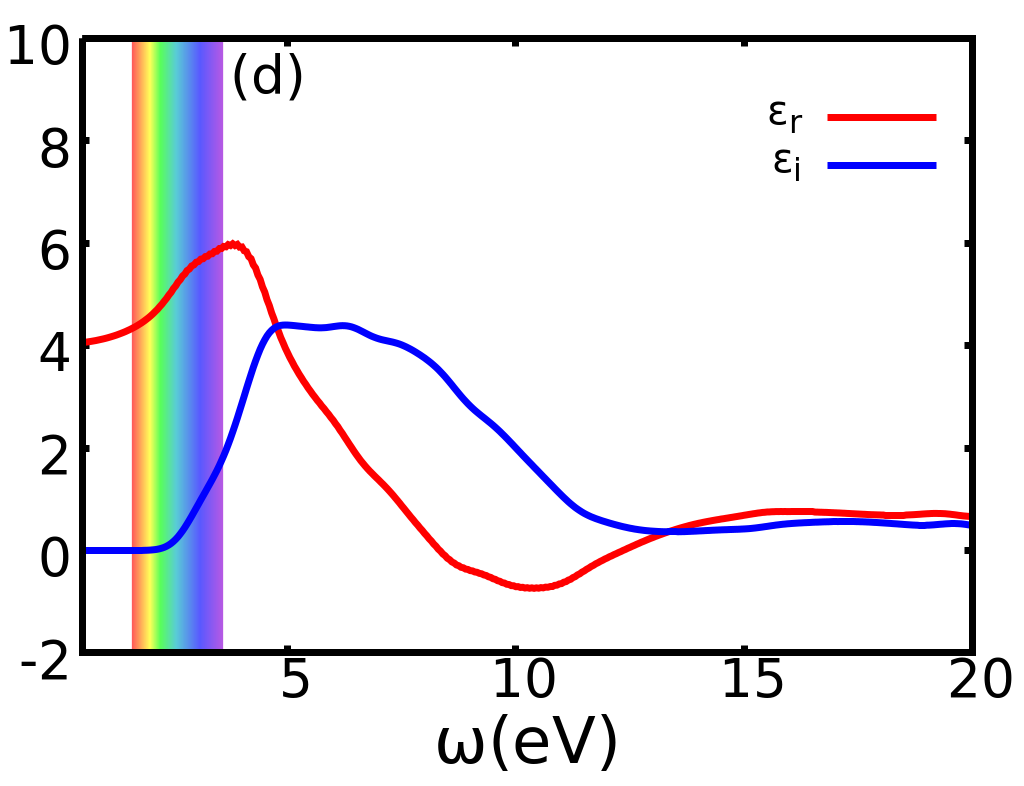}
\includegraphics[height=3.5cm,width=4.0cm]{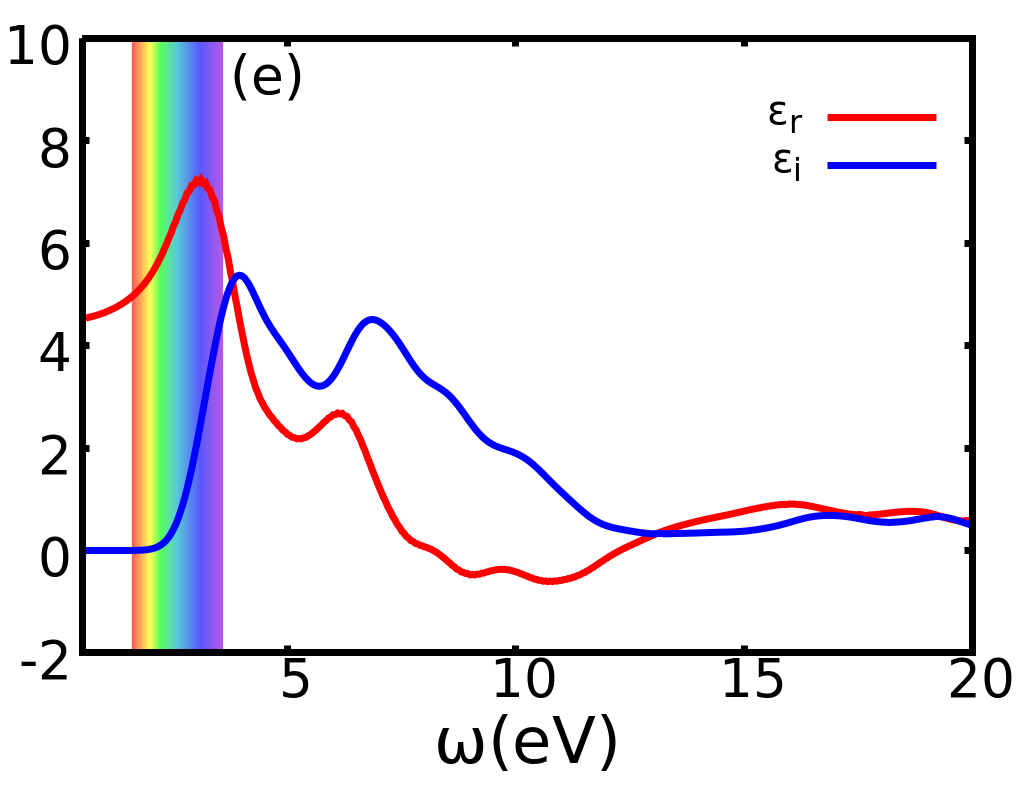}
\includegraphics[height=3.5cm,width=4.0cm]{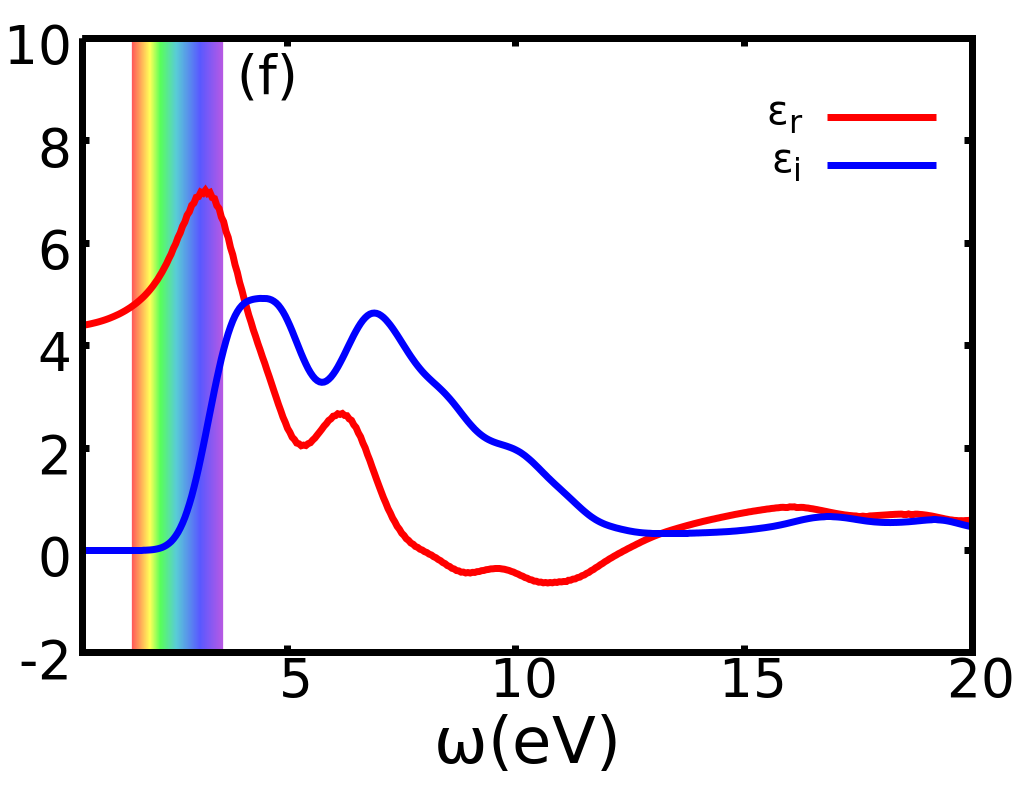}
\caption{Real and imaginary the dielectric functions for (a,b) $a6$, $a7$, (c,d) $p6$, $p7$, (e,f) $r6$ and $r7$.}
\label{optic}
\end{figure}

\subsection{Optical Properties} 
The optical properties of the materials are calculated using SIESTA software package in the range of $0-20$ eV. Dielectric function is computed in a finite system using adiabatic approximation in the time dependent density functional theory \cite{Ordejon96,Artacho08,Artacho08}. The imaginary part of dielectric constant is defined as,
\begin{equation}
\varepsilon_{i}(\omega)=\dfrac{4\pi^{2}e^{2}}{V}\sum_{i,n}f_{i}(1-f_{n})|\langle n|p|i\rangle|^{2}\delta(E_{n}-E_{i}-\hbar\omega)
\end{equation}
$f_{i}$ and $f_{n}$ are the initial and final states $|i\rangle$ and $|n\rangle$ of the Fermi function, with corresponding energies $E_{i}$ and $E_{n}$ and $p$ is the momentum operator. The real part can be computed using Kramers-Kroning relation,
\begin{equation}
\varepsilon_{r}(\omega)=1+\dfrac{2}{\pi}P\int_{0}^{\infty}\dfrac{\omega^{\prime}\varepsilon_{i}(\omega^{\prime})}{\omega^{\prime 2}-\omega^{2}}d\omega^{\prime}
\end{equation}
Real part explains the electronic polarizability and anomalous dispersion, whereas the imaginary part is related to dissipation of energy into the medium. The peaks of the real part are in the UV and visible region while the static dielectric constant ranges between $4.01-4.60$ (highest is for $a7$ and lowest for $p6$) (Table \ref{dielectric}). The maxima lies in the visible region(except for tetragonal structure), while other peaks appear in near and far UV regions[Fig. \ref{optic} (a-f)]. In rhombohedral structure we see multiple peaks and the highest peak lies in the visible region. Two peaks are observed in the imaginary part of the dielectric function for all structures except $p7$ where the second peak vanishes. The index of refraction is also complex function $\textbf{n} = \eta_{opt} + i\kappa_{opt}$ which is a combination of real part ($\eta_{opt}$) as refractive index and imaginary part ($\kappa_{opt}$) as extinction coefficient. These quantities are defined as,
\begin{gather}
\eta_{opt}(\omega) =\sqrt{\dfrac{|\varepsilon(\omega)|+Re(\varepsilon(\omega))}{2}},\notag\\
\kappa_{opt}(\omega) =\sqrt{\dfrac{|\varepsilon(\omega)|-Re(\varepsilon(\omega))}{2}}.
\end{gather}
BCZT has an average refractive index($\eta_{opt}$) of 2.1 and a maximum value of 2.7 (for $a7$) in the visible region, indicating high light absorption capacity with minimal loss.
The absorption coefficient can be derived from the extinction coefficient by using relation, $\alpha = \dfrac{2\omega\kappa_{opt}}{c}$, where, $c$ is the velocity of light. Absorption coefficient signifies the frequency range of light in which the material can absorb and it depends on the energy of incident photons. It increases gradually and the threshold is measured in the range of $2.0-2.25$ eV followed by some peaks that indicate the possibility of direct electronic transitions from valence to conduction band (see Table \ref{dielectric} for details). Materials with higher absorbance in visible range can be useful for optical and optoelectronic applications. In rhombohedral structure, slope of the coefficient in the visible region is 1.5 times higher than tetragonal. In orthorhombic and rhombohedral crystals more peaks appeared in near UV region than in tetragonal crystals. This confirms the existence of higher number of transition states in amm2 and r3m structures \cite{Schwale16}. The details of the absorption coefficient are given in Table \ref{dielectric}. The optical properties confirm that these materials absorb visible light, which can enhance the efficiency of solar cells and optoelectronic devices.
\begin{figure}[!h]
\includegraphics[height=3.5cm,width=4.0cm]{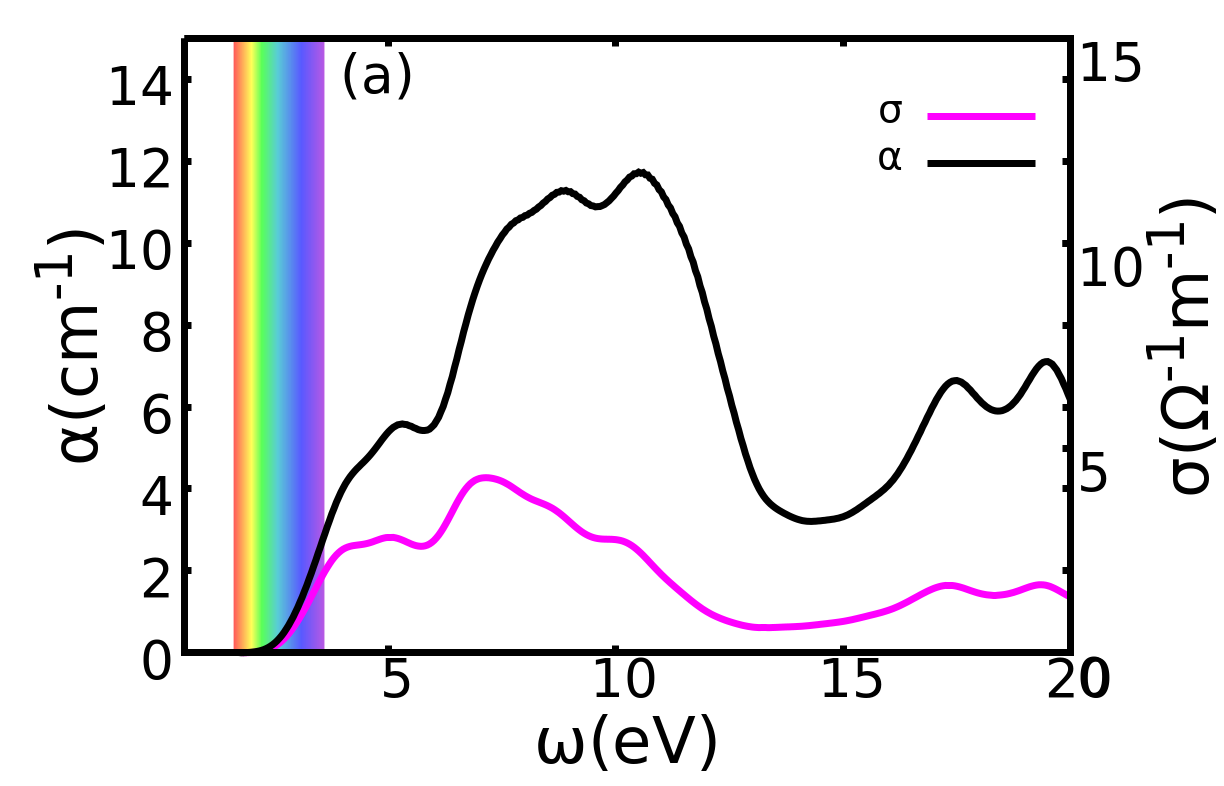}
\includegraphics[height=3.5cm,width=4.0cm]{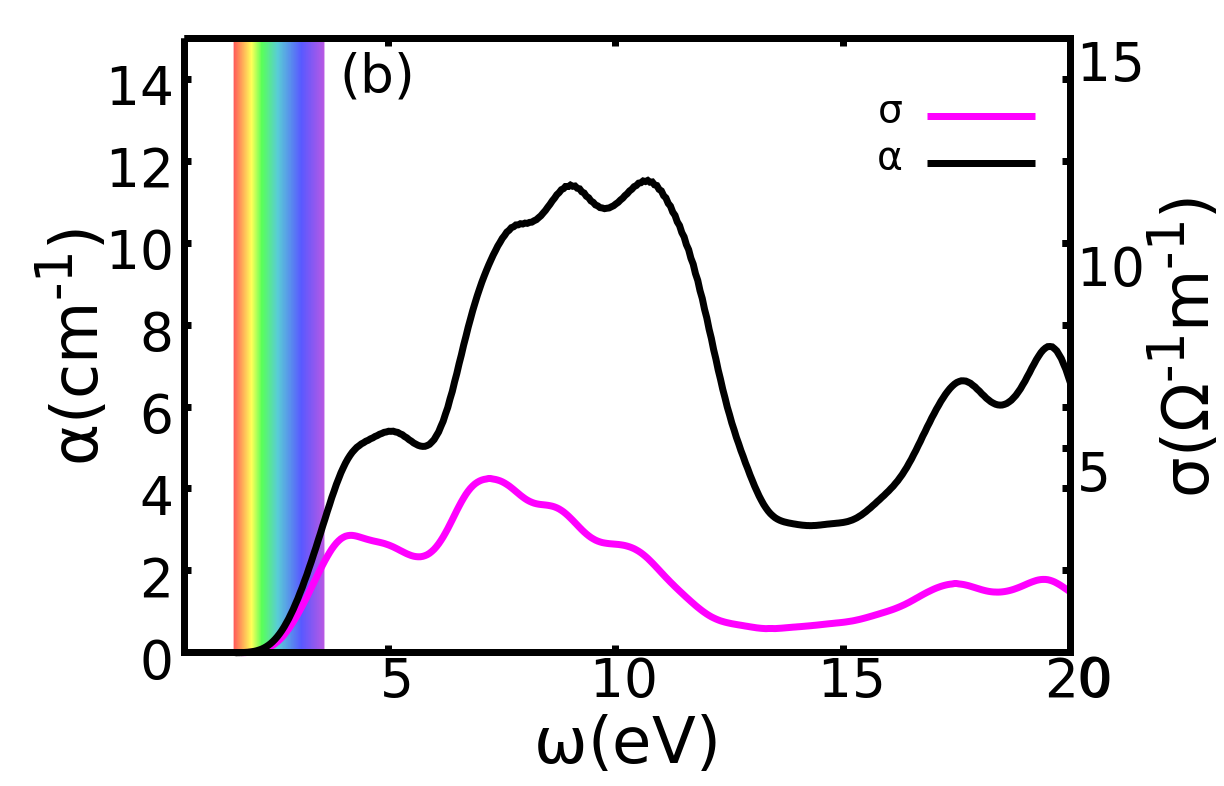}
\includegraphics[height=3.5cm,width=4.0cm]{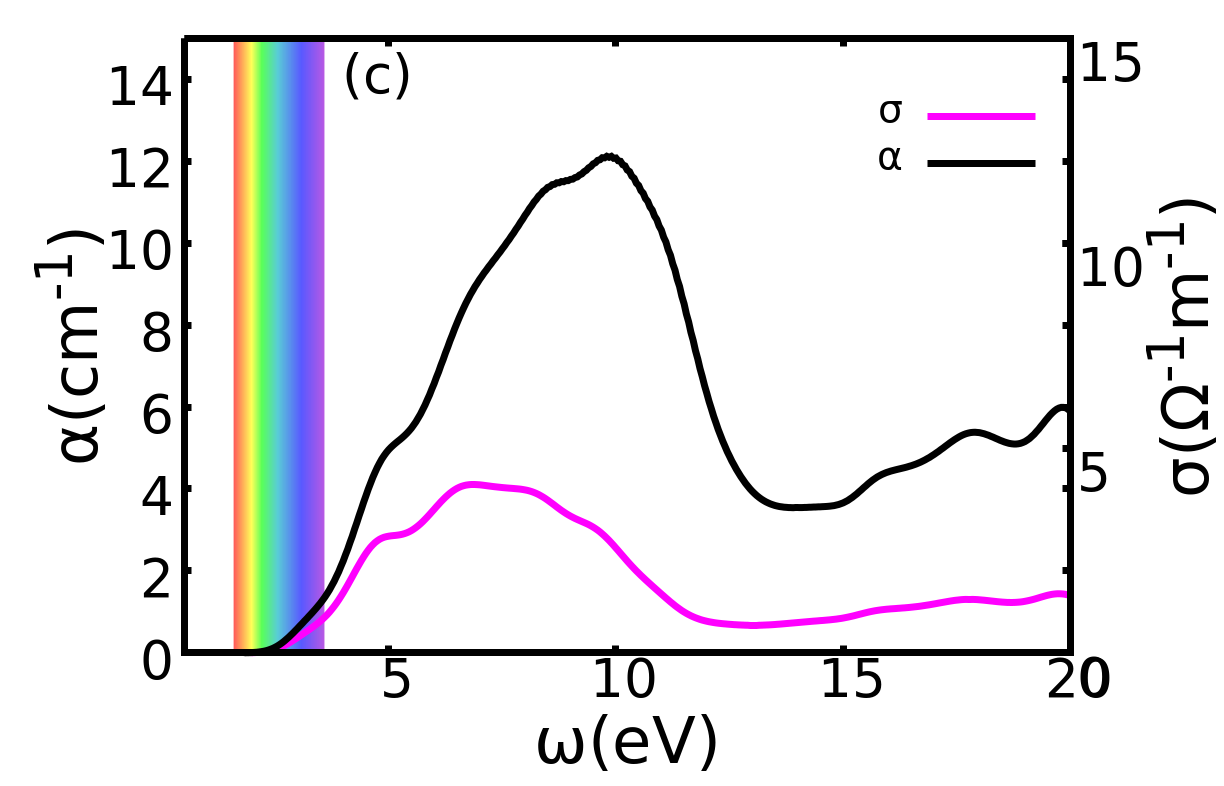}
\includegraphics[height=3.5cm,width=4.0cm]{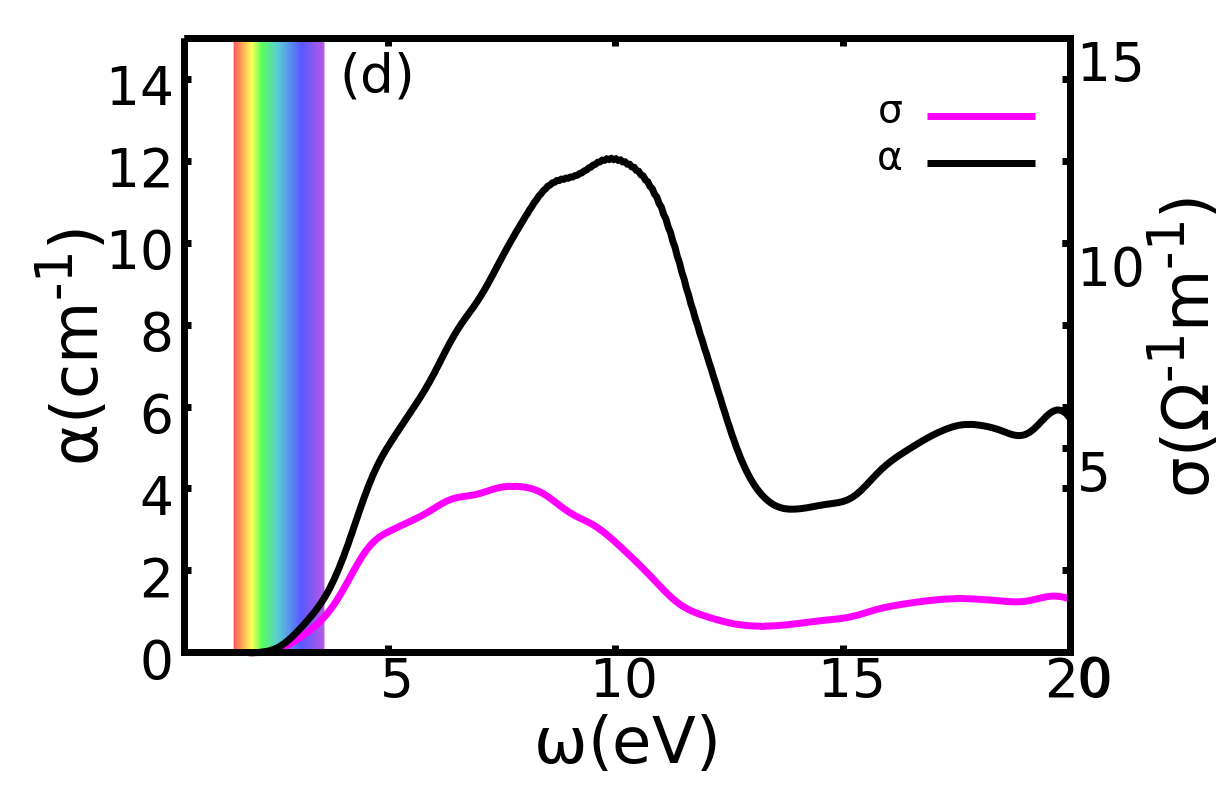}
\includegraphics[height=3.5cm,width=4.0cm]{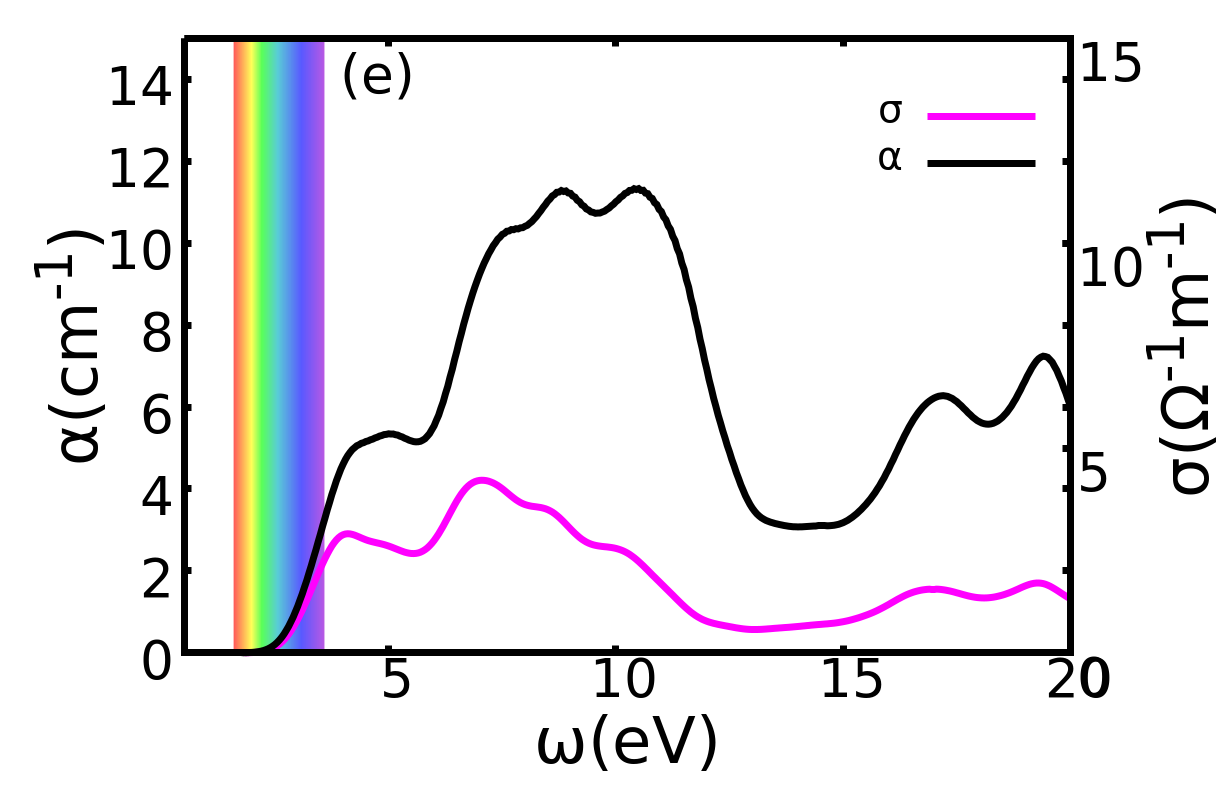}
\includegraphics[height=3.5cm,width=4.0cm]{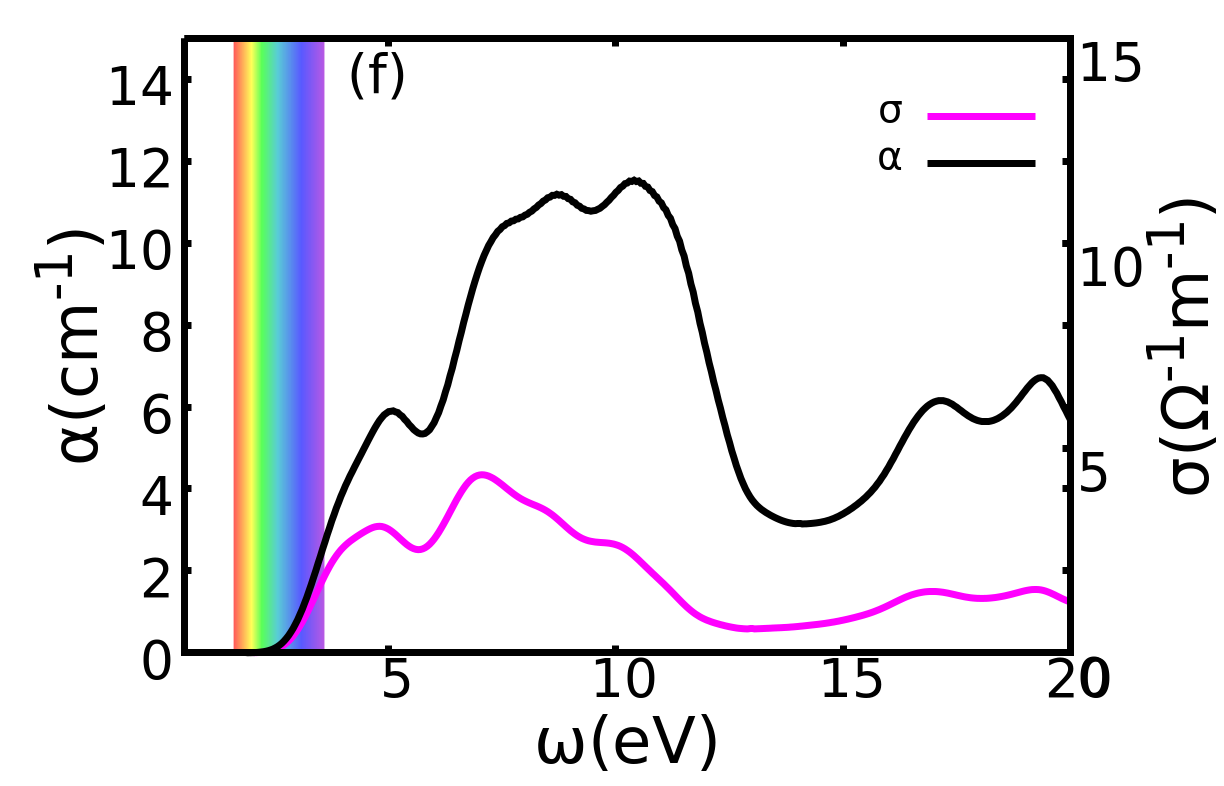}
\caption{Absorption coefficient and optical conductivity for (a,b) $a6$ and $a7$, (c,d) $p6$ and $p7$, (e,f) $r6$ and $r7$.}
\label{optic1}
\end{figure}
\begin{table}[!t]
\begin{center}
\caption{\label{dielectric} Real part of dielectric function $\varepsilon_{r}(\omega)$, absorption coefficient ($\alpha$), index of refraction ($\eta_{opt}$) and Reflectance (R) of BCZT}
\begin{tabular}{|c|c|c|c|c|c|c|}
\hline
BCZT & $a6$ & $a7$ & $p6$ & $p7$ & $r6$ & $r7$\\ 
\hline 
\thead{$\varepsilon_{r}(\omega)$ at edge ($\omega=0)$ \\ peak position in VR (eV)} & \thead{4.47 \\ 3.10} & \thead{4.60 \\ 3.00} & \thead{4.01 \\ 3.02} & \thead{4.03 \\ 2.86} & \thead{4.49 \\ 3.10} & \thead{4.36 \\ 3.20}\\ 
\hline 
\thead{ $\alpha$ at edge ($\omega=0)$ \\ peak position in VR (eV)} & \thead{2.10 \\ 4.36} & \thead{2.00 \\ 4.36} & \thead{2.20 \\ 5.06} & \thead{2.25 \\ 9.89} & \thead{2.20 \\ 4.322} & \thead{ 2.15 \\ 5.09} \\ 
\hline 
\thead{ $\eta_{opt}$ at edge ($\omega=0)$ \\ peak position in VR (eV)} & \thead{2.114 \\ 3.12} & \thead{2.14 \\ 3.10} & \thead{2.002 \\ 3.06} & \thead{2.00 \\ 2.96} & \thead{2.118 \\ 3.19} & \thead{2.08 \\ 3.30} \\ 
\hline 
 \thead{ R at edge ($\omega=0)$ \\ peak position in VR (eV)} & \thead{0.130 \\ 3.80} & \thead{0.132 \\ 3.83} & \thead{0.111 \\ 4.634} & \thead{0.112 \\ 4.625} & \thead{0.128 \\ 3.80} & \thead{0.124 \\ 3.90} \\ 
\hline 
\end{tabular}
\end{center}
\end{table}
\begin{figure}[!h]
\includegraphics[height=3.5cm,width=4.0cm]{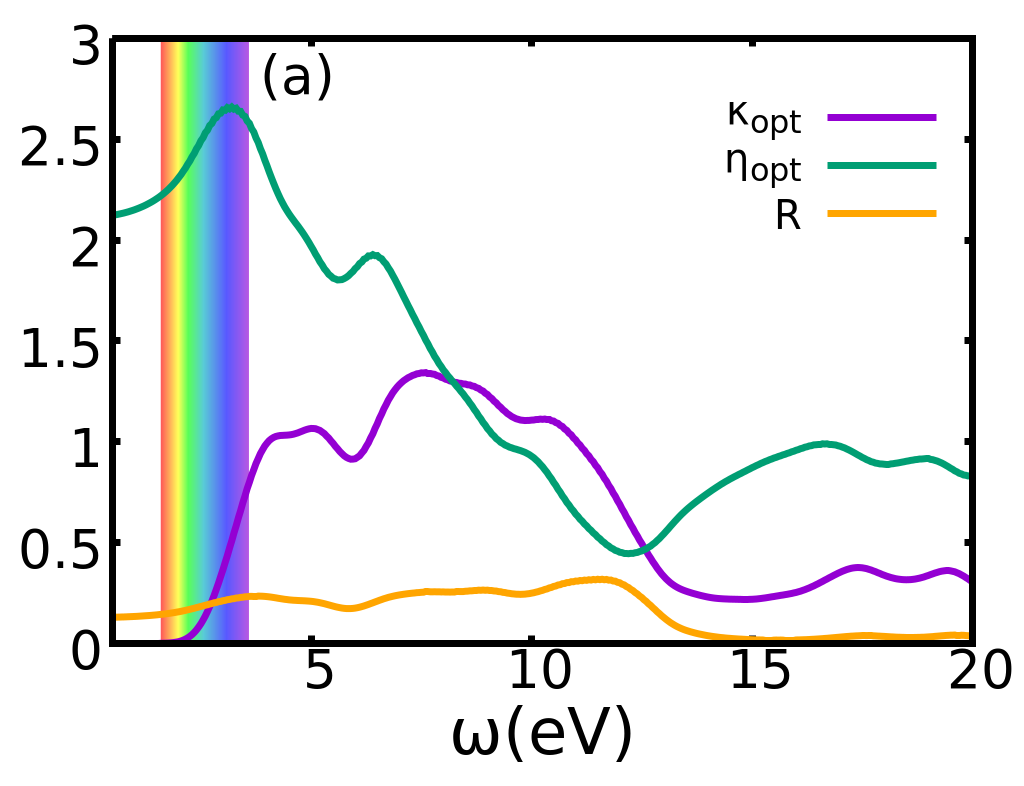}
\includegraphics[height=3.5cm,width=4.0cm]{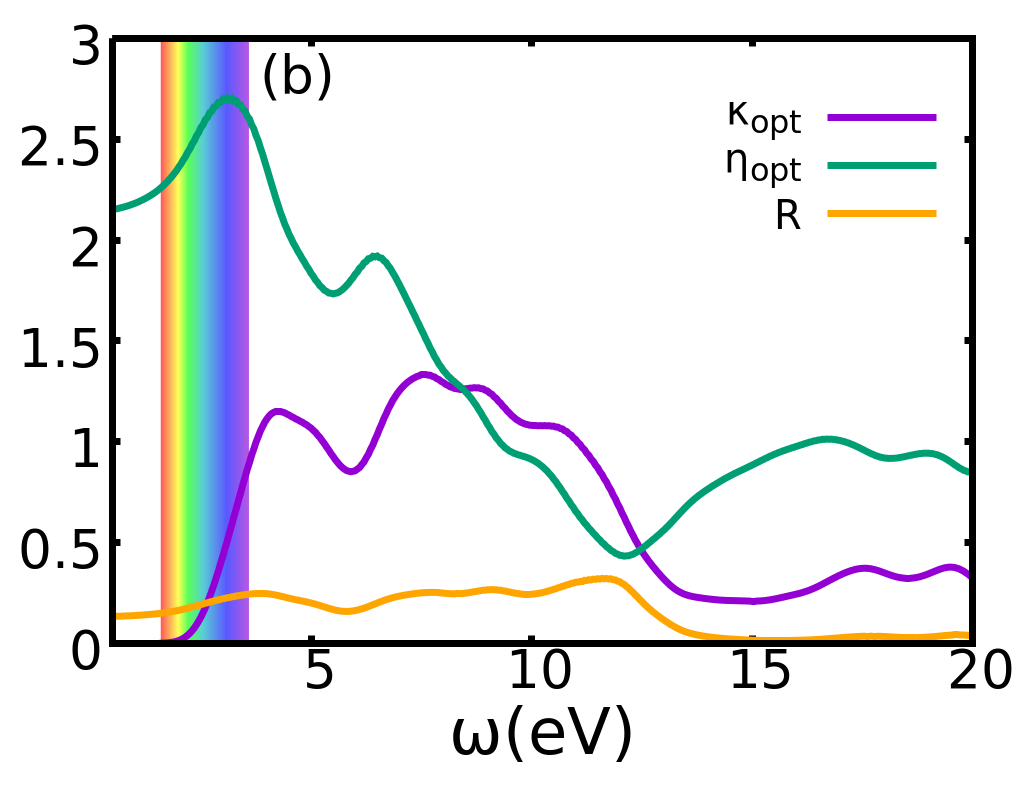}
\includegraphics[height=3.5cm,width=4.0cm]{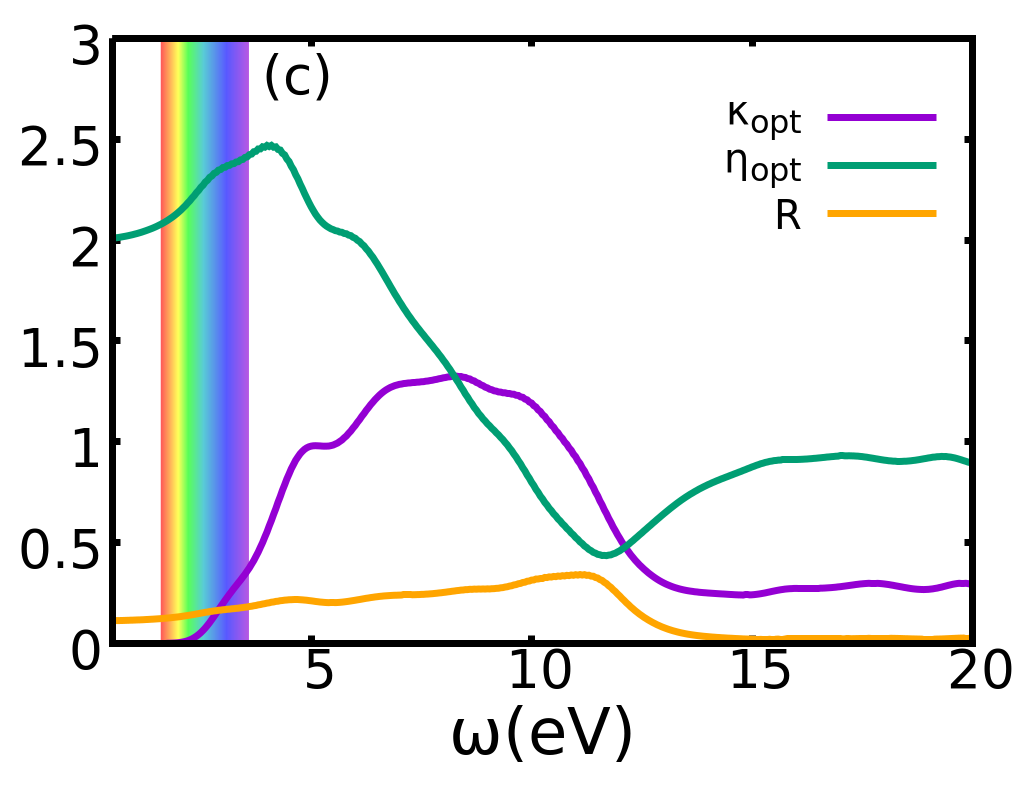}
\includegraphics[height=3.5cm,width=4.0cm]{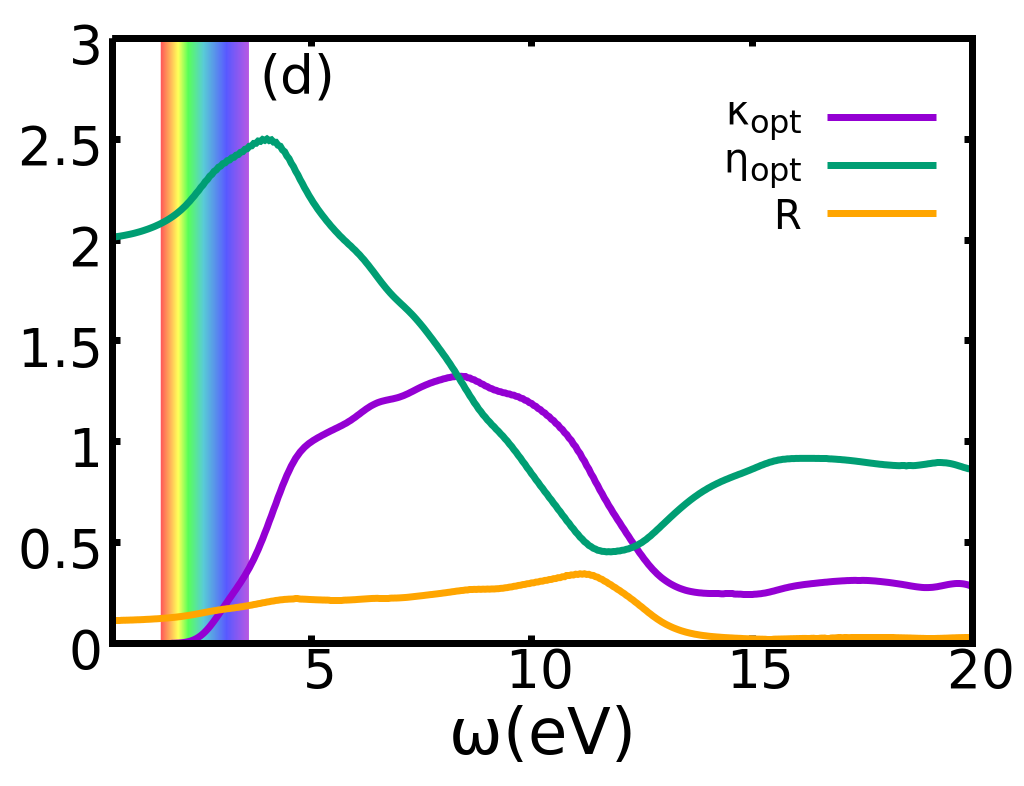}
\includegraphics[height=3.5cm,width=4.0cm]{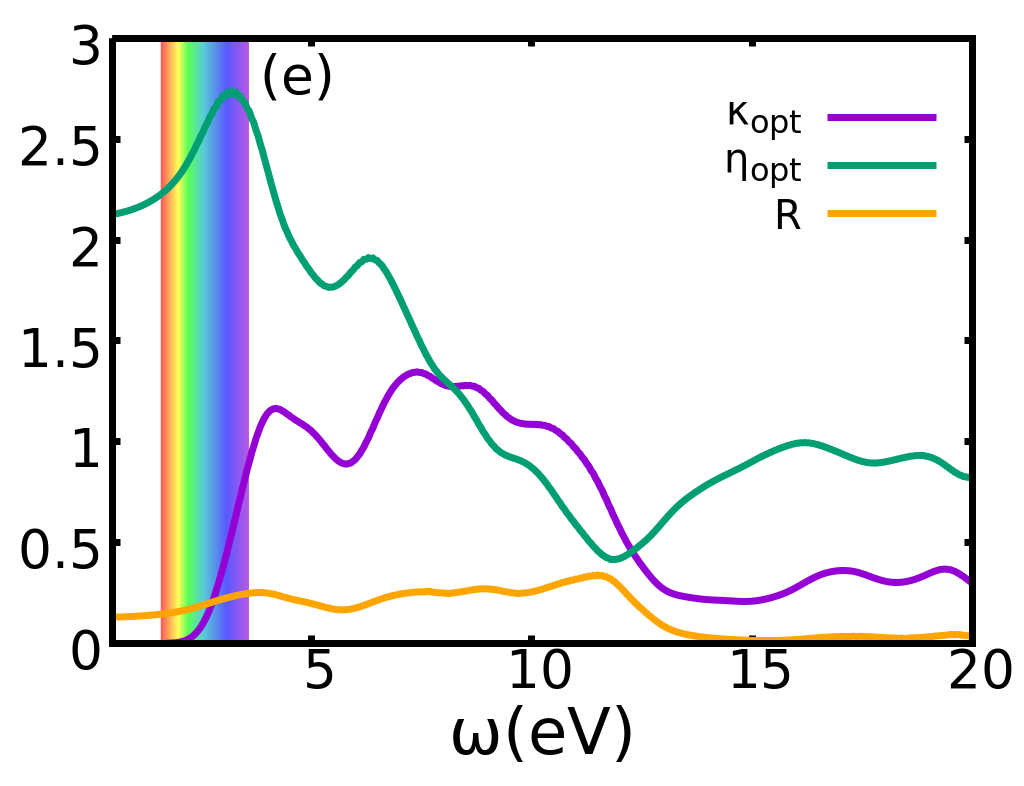}
\includegraphics[height=3.5cm,width=4.0cm]{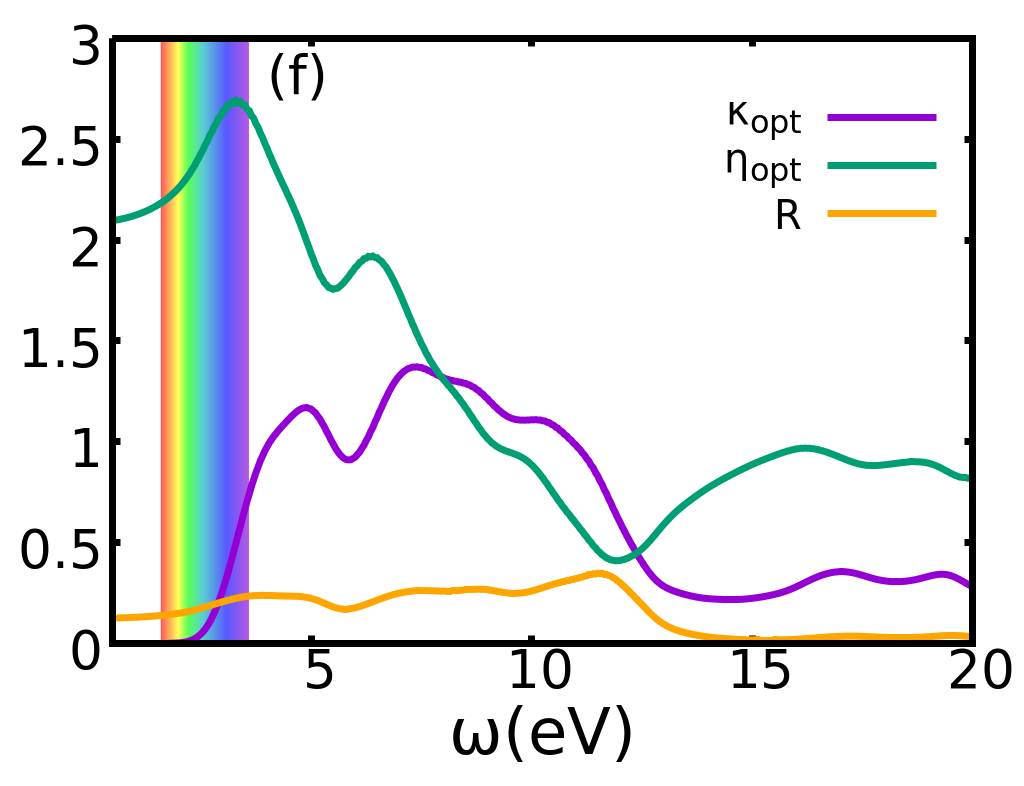}
\caption{ Index of refraction and loss functions for (a,b) $a6$ and $a7$, (c,d) $p6$ and $p7$, (e,f) $r6$ and $r7$.}
\label{optic2}
\end{figure}
The optical conductivity , $\sigma_{opt}(\omega)=\dfrac{\varepsilon_{o} c \eta_{opt} \alpha}{\omega}$ is zero in the energy range of 0-2.0 eV as it falls within the forbidden region. However, it increases beyond this range, with peaks observed between 2.0 and 14.0 eV, corresponding to the photoconductivity region. $\eta_{opt}$ is high in R3m and Amm2 structures as compared to P4mm structure(see Table \ref{dielectric}). The peaks in the $\kappa_{opt}$ gives information about direct electronic transitions between the valence and conduction bands. Again the nature of the refractive index is quite similar to the real part of the dielectric constant.
 Reflectance is another optical property, that can be calculated using the formula,
\begin{gather}
R = \dfrac{(\eta_{opt} - 1)^{2} + \kappa_{opt} ^{2}}{(\eta_{opt} + 1)^{2} + \kappa_{opt} ^{2}}.
\end{gather}
The calculated value of reflectances are provided in the Table \ref{dielectric}. From overall analysis we found that rhombohedral and orthorhombic structures are better optical materials than tetragonal structures. Absorbance, low reflectance in the visible region is a better choice for optoelectronic devices.
\begin{table}
\caption{\label{piezo} Piezoelectric constant ($\vert e_{ij}\vert _{max}$) in $C/m^{2}$}
\begin{center}
\begin{tabular}{|c|c|c|c|c|c|c|c|c|}
\hline 
BCZT & $a6$ & $a7$ & $p6$ & $p7$ & $r6$ & $r7$ & PZT & a6-mod \\
\hline 
PBEsol & 4.767 & 7.545 & 2.636 & 2.727 & 4.725 & 4.615 & 3.649 & 4.621 \\ 
\hline 
PBE & 5.287 & 24.430 & 2.793 & 2.918 & 5.713 & 5.657 & 3.502 & 32.413 \\ 
\hline 
\end{tabular} 
\end{center}
\end{table}
\subsection{Piezoelectric Properties}
\begin{figure}[!h]
\includegraphics[height=3.5cm,width=8.0cm]{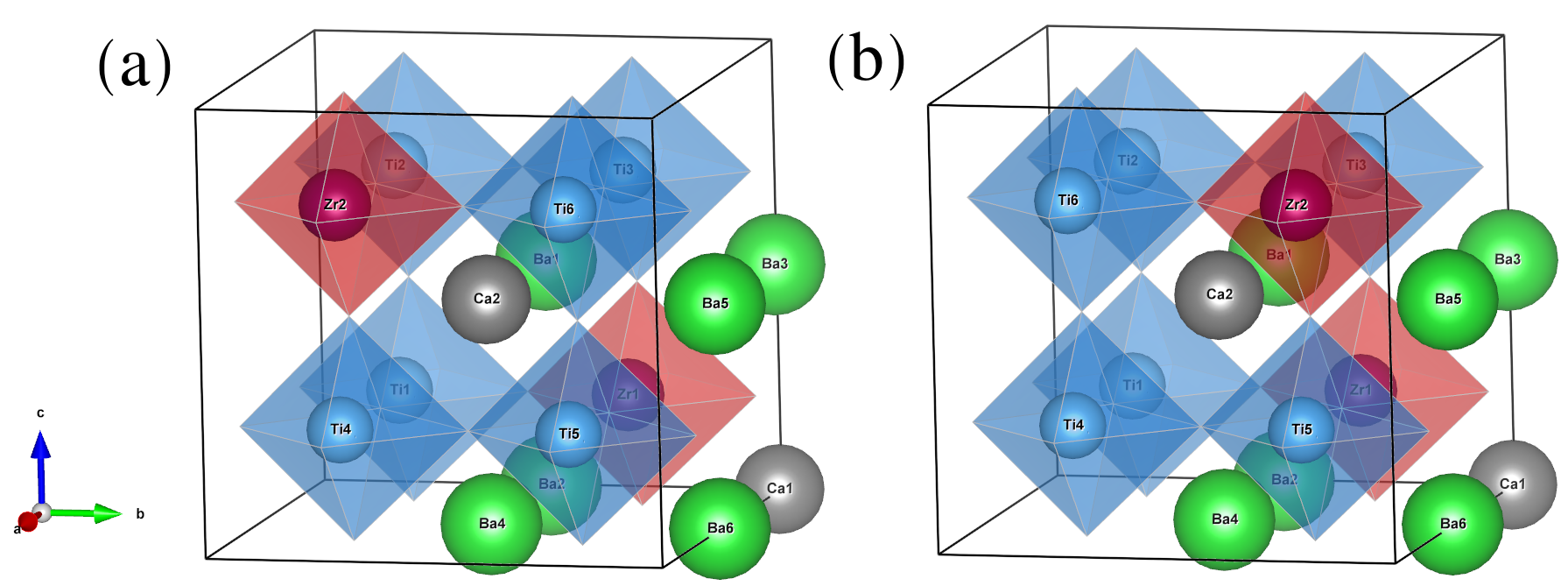}
\caption{\label{moda6} Geometrical structure of (a) $a6$ and (b) modified $a6$ for piezoelectric coefficient.}
\end{figure}

Recent developments on application of piezoelectric materials in various compact smart devices is in high demand for technological advancements \cite{Akdogan05,Bansevicius11}. The Piezoelectric coefficient tensor is defined as $\epsilon_{ij} = \left(\dfrac{\partial \mathbf{D}_{i}}{\partial \epsilon_{j}}\right)_{\mathbf{E}} = -\left(\dfrac{\partial \sigma_{j}}{\partial \mathbf{E}_{i}}\right)_{\epsilon} $, where $\mathbf{D}$, $\epsilon$, $\mathbf{E}$ and $\sigma$ represent electric displacement field, strain tensor, electric field and stress tensor respectively \cite{Maarten15}. The indices represent the direction of the vector component, where i = 1, 2, 3 and j= 11, 22, 33,(12 or 21),(13 or 31), (23 or 31). 
We calculated the intrinsic piezoelectric constant using the Projected Augmented Wave (PAW) method with the Generalized Gradient Approximation (GGA) for the exchange-correlation functional, employing both PBE and PBEsol, as implemented in the Vienna Ab initio Simulation Package (VASP). GGA-PBE was used to compare the results with previously reported data, while the improved GGA-PBEsol method was also employed for higher accuracy \cite{Wu05,Alyoruk21,Shi24}. A plane wave cutoff of 400 eV with dense k-points, first-order Methfessel-Paxton smearing with a width of 0.2 eV, and the linear-tetrahedron method with Bloechl corrections in reciprocal space projection operators were used for high-precision calculations. Tensor components are calculated for both clamped and relaxed ion configuration in a strain-free condition. $\vert e_{ij}\vert _{max}$ represents the maximum piezoelectric constant in all directions, and a value greater than $ > 3 C/m^{2}$ is considered highly piezoelectric. For some of the widely used piezoceramics such as PbTiO$_{3}$, BaNiO$_{3}$, RbTaO$_{3}$ and SrHfO$_{3}$, even the most efficient commercial material(PZT), piezoelectric constant ranges between 6-12 $C/m^{2}$ \cite{Maarten15}. In modified $a6$ structure, position of one diagonal Zr is exchanged with non-diagonal Ti atom to observe the shift in polarisation. BCZT shows higher $\vert e_{ij}\vert_{max}$ than PZT (3.648 $C/m^{2}$) for all structure except tetragonal (see Table \ref{piezo}). The details of piezoelectric tensor for different crystal calculated with PBESol and PBE are given in the Table S I \& II. 
Structural deformation leads to increased polarizability, and in BCZT7, this results in a higher piezoelectric property compared to BCZT6 \cite{Prem21}. In the case of $r6$ and $r7$ structures, it is observed that the diagonal elements exhibit minimal variation, because of its highly symmetric structure. Tetragonal structures have low atomic density with minimal distortion leading to lower piezo response, a corresponding behavior also observed in optical properties. $a6$ and $a7$ structures are the most appropriate piezoceramic as they have comparatively high atomic density with better stability. $a7$ shows the highest piezo response of 7.454 $C/m^{2}$, among considered structures. Piezo response using PBE is higher than PBEsol for most cases. Comparing the values for GGA-PBE with previous calculations \cite{Maarten15}, $a7$ structure has the highest piezoelectric effect of 24.430 $C/m^{2}$, among all the compounds. The modified $a6$ structure exhibits an extremely high piezoelectric response of 32.413 $C/m^{2}$, as exchanging the position of the transition element induces a significantly high static polarization.
 Polarization increases as the inter-atomic distance between the two Zr atoms decreases, increasing stress in that plane causing greater BEC value for both Zr atoms in modified structure compared to the previous (see Fig. \ref{moda6}).

\subsection{Born Effective Charge}
Born effective charges (BECs) quantifies the polarization due to the atomic displacements which disrupts the charge distribution. It is defined as, $Z_{\alpha\beta}^{*}=\Omega\dfrac{\partial P_{\alpha}}{\partial u_{\beta}}$ where, $P_{\alpha} $ represents polarization in $\alpha$ direction per unit volume, $u_{\beta}$ is the displacement along $\beta$ direction and $\Omega$ is the total volume \cite{Ravindran06}. In this work, the BEC values of the composite are determined from the highest diagonal component of the BEC tensor of the Zr atom. 
 Ba and Ca exhibit BEC values close to the nominal charge, indicating minimal induced dipole moments, while Zr shows the highest BEC $(> 6 e)$ in all structures due to strong $d$-orbital hybridization and shielding effect. Zr and Ti have partially filled d orbitals, which strongly hybridize with surrounding oxygen, enhancing the charge transfer and increasing the effective charge. This causes a stronger distortion effect in the surrounding oxygen octahedra. With shielding effect, valence electrons exhibit high covalent contribution inducing greater polarizability.

The BEC value of Ti is higher than Ba and Ca but lower than Zr, as Zr possesses high $d$ orbitals. Oxygen has a higher BEC value $(>5 e)$ between Zr and Ti compared to the two Ti atoms, whereas it is lower near Ba and Ca $( <2 e)$. Oxygen in the body diagonal has higher BECs than other positions, confirming the occurrence of spontaneous polarization because of Zr and nearby Ti and O atoms. Large diagonal components indicate strong coupling between atomic displacement and polarization which is crucial for ferroelectric and piezoelectric properties \cite{Hong24,Gonze97}. Based on the crystal structure it is found that all the off-diagonal values of rhombohedral structures are more significant $(>0.40)$ than others, indicating higher anisotropy and better piezoelectric properties in all directions \cite{Macheda24}. BECs values are correlated to the piezoelectric response tensor linearly. For a7, the piezoelectric coefficient is 7.454 $C/m^{2}$ with a BEC of 6.45 e using PBEsol, whereas with PBE, it is 24.430 e $C/m^{2}$ with a BEC of 6.47 e. In case of modified a6, a high piezoelectric coefficient of 32.4124 $C/m^{2}$ corresponds to a BEC of 6.78 e, whereas in $a6$ structure it is 5.287$C/m^{2}$ with BEC of 6.433e (see Table S III). All other BEC values correspond to the highest diagonal components of the Zr atom in their respective structures [Table S IV]. 
\begin{figure}[!h]
\includegraphics[height=6.6cm,width=8.0cm]{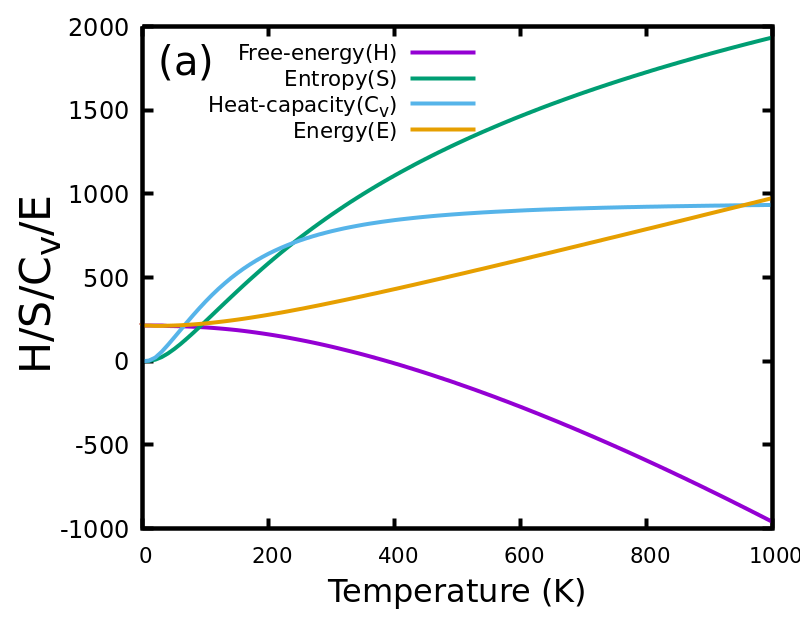}
\includegraphics[scale=0.15]{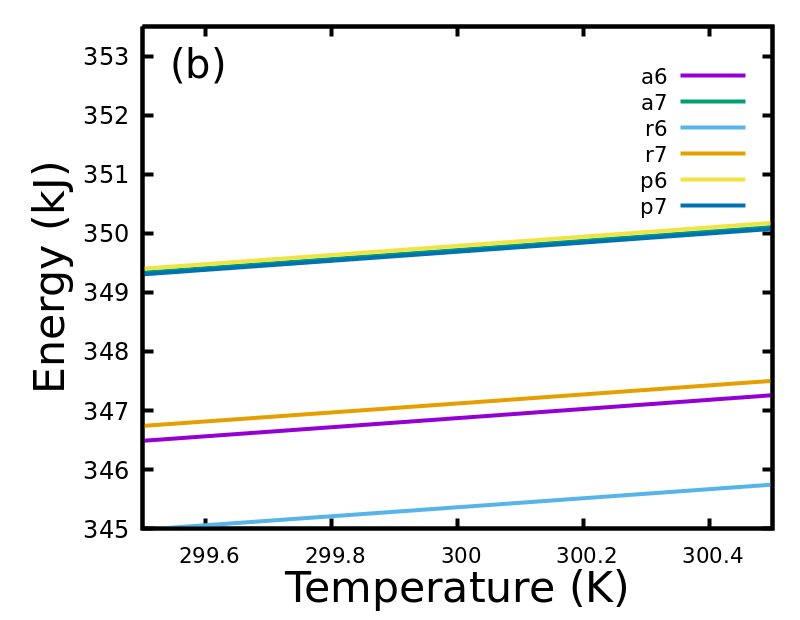}
\includegraphics[scale=0.15]{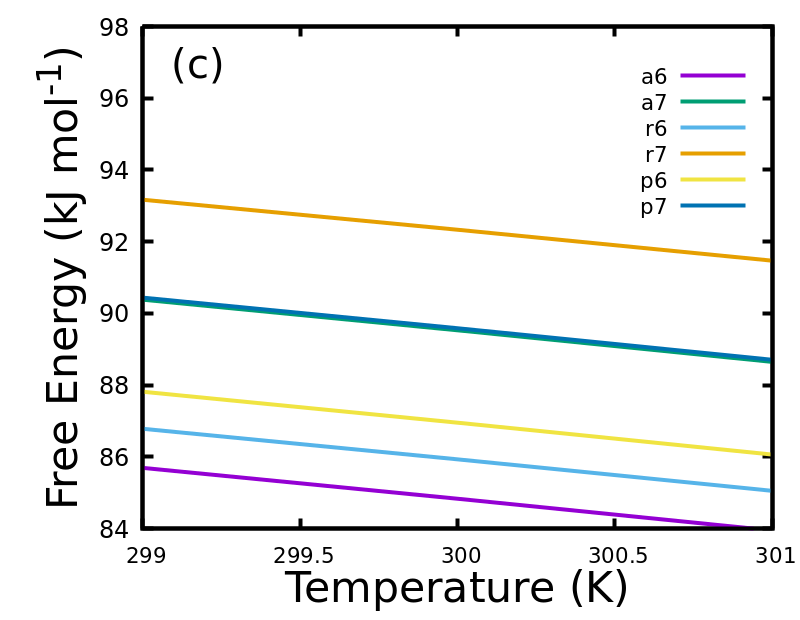}
\includegraphics[scale=0.15]{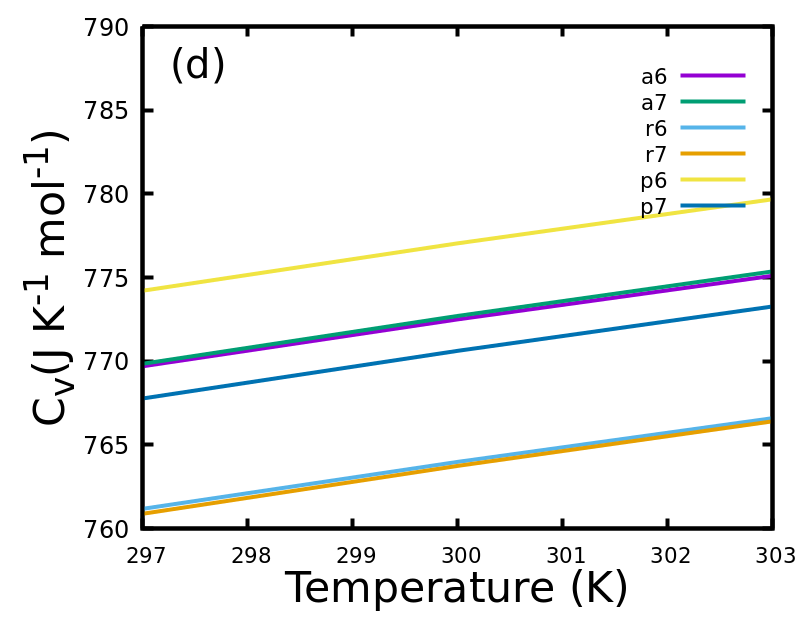}
\includegraphics[scale=0.15]{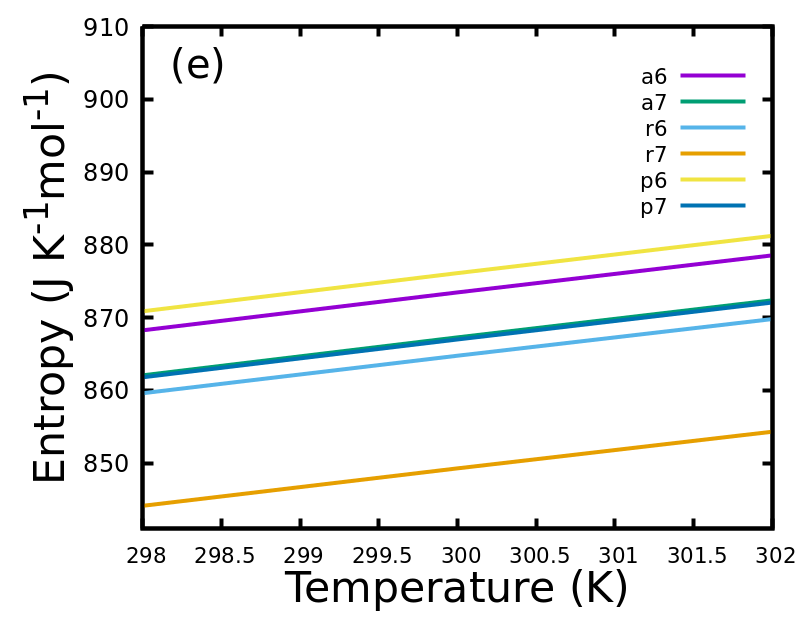}
\caption{\label{phthermal} (a) Calculated phonon thermal properties of BCZT with quasi-harmonic approximation. (b,c,d,e) Magnified plot of Helmholtz free energy, heat capacity and entropy at constant volume and total energy around 300K.}
\end{figure}

\subsection{Thermal properties}
Phonon thermal properties of BCZT are calculated using PHONOPY software where the supercell approach is incorporated along with the finite displacement method \cite{Togo15,Kresse95}. Displacement of one atom in a crystal can produce dissipative forces, along all directions, applied to all the atoms in the crystal. Displacing an atom from its equilibrium position can undergo various small displacements. So, potential energy can be expanded with respect to these small displacements in the form of Taylor series. This set of displacements is also linked with respective forces, which can give a series of frequencies associated with phonons. Density Functional Perturbation Theory(DFPT) method is another promising method for higher orders potentials $(>2)$. PHONOPY gathers force constants from both supercell and DFPT methods, for high accuracy within the limit of quasi-harmonic approximation\cite{Togo23}. 
Derived harmonic phonon energy from canonical distribution of statistical mechanics is given by,
\begin{equation}
E = \sum_{\mathbf{q}\nu}\hbar\omega_{\mathbf{q}\nu}\left[\frac{1}{2} +
\frac{1}{\exp(\hbar\omega_{\mathbf{q}\nu}/k_\mathrm{B} T)-1}\right]
\end{equation}
At absolute zero $p7$ has the highest energy followed by $a7$, $r7$ and $p6$ structure whereas, $r6$ and $a6$ remain low for the entire temperature range. The specific heat capacity at constant volume($C_{V}$) can be derived as,
\begin{eqnarray}
C_V &=& \left(\frac{\partial E}{\partial T} \right)_V \nonumber \\
 &=& \sum_{\mathbf{q}\nu} k_\mathrm{B}
 \left(\frac{\hbar\omega_{\mathbf{q}\nu}}{k_\mathrm{B} T} \right)^2
 \frac{\exp(\hbar\omega_{\mathbf{q}\nu}/k_\mathrm{B}
 T)}{[\exp(\hbar\omega_{\mathbf{q}\nu}/k_\mathrm{B} T)-1]^2}
\end{eqnarray}
Distribution of the particles along with its degeneracy can be calculated using the partition function which is given by,
\begin{equation}
Z = \exp(-\varphi/k_\mathrm{B} T)\prod_{\mathbf{q}\nu}
 \frac{\exp(-\hbar\omega_{\mathbf{q}\nu}/2k_\mathrm{B}
 T)}{1-\exp(-\hbar\omega_{\mathbf{q}\nu}/k_\mathrm{B} T)}
\end{equation}
where $\varphi$ represents the static energy (or ground state energy). Helmholtz free energy is formulated as,
\begin{eqnarray}
F &= -k_\mathrm{B} T \ln Z 
= \varphi + \frac{1}{2} \sum_{\mathbf{q}\nu}
\hbar\omega_{\mathbf{q}\nu} \nonumber \\ &+ k_\mathrm{B} T \sum_{\mathbf{q}\nu} \ln
\bigl[1 -\exp(-\hbar\omega_{\mathbf{q}\nu}/k_\mathrm{B} T) \bigr]
\end{eqnarray}
Again, the entropy of the system is
\begin{eqnarray}
S &= -\Big(\frac{\partial F}{\partial T}\Big) = \frac{1}{2T}
\sum_{\mathbf{q}\nu} \hbar\omega_{\mathbf{q}\nu}
\coth(\hbar\omega_{\mathbf{q}\nu}/2k_{\mathrm{B}}T) \nonumber \\ &-k_{\mathrm{B}}
\sum_{\mathbf{q}\nu}
\ln\left[2\sinh(\hbar\omega_{\mathbf{q}\nu}/2k_{\mathrm{B}}T)\right]
\end{eqnarray}

The calculated phonon thermal properties as total energy, Helmholtz free energy, specific heat capacity and entropy are shown in Fig. \ref{phthermal}(a) and a magnified version around room temperature, near 300K are shown in Fig. \ref{phthermal}(b,c,d,e). With increase in temperature, internal energy increases and free energy decreases gradually. Specific heat tends to attain saturation around 270K. At low temperature it satisfies Debye T$^{3}$ law and at high temperatures C$_{v}$ tends to 3NK$_{b}$ as in Dulong-Petit law. 
Free energy approaches zero at 388K and lesser the value, greater the stability and we found that r$6$ has the lowest total energy followed by a$6$, a$7$, p$6$, p$7$ and r$7$.

\subsection{Thermoelectric and electronic transport properties}
Thermal energy is also a good source of electricity and can be harnessed with suitable thermoelectric materials. Here, we have calculated the thermoelectric and transport properties by using the BoltzWann code implemented in Wannier90 \cite{Pizzi14,Pizzi20}. The computation was performed on a homogeneous finite system using constant relaxation time, obtained by Wannier interpolation. Maximally Localized Wannier Function (MLWF) basis set and finite difference method are incorporated in Boltzman transport equation (BTE), in which thermoelectric coefficients are calculated \cite{Marzari12,Li14}. Seebeck coefficient(S), electrical conductivity $(\boldsymbol{\sigma})$, electronic contribution of thermal conductivity $(\boldsymbol{\kappa})$, BoltzWann-DOS and transport distribution function (TDF) are evaluated for all compositions. From these results, figure of merit (ZT) and power factor(PF) are derived and analysed for possible commercial applications. Efficiency of a thermoelectric material is estimated using figure of merit value, which is defined as $ZT=\dfrac{\sigma S^{2}T}{\textbf{K}}$, where \textbf{K} represents total thermal conductivity. 
\textbf{K} is a combination of electronic$(\boldsymbol{\kappa})$ and lattice $(\boldsymbol{\kappa}_l)$ contributions of thermal conductivity. A constant relaxation time($\tau_{n\mathbf{q}}$) is assumed, for one electron considering $n$ bands, at the wave vector $\mathbf{q}$ \cite{Mishima06}. 
It can also be calculated experimentally from conductivity experiments, at a fixed temperature and carrier density. 
Electrical conductivity ($\boldsymbol{\sigma}$) can be obtained as a function of chemical potential($\mu$) and temperature, by the following relation,		
\begin{equation}
[\boldsymbol{\sigma}] _{ij} (\mu,T) = e^2 \int_{-\infty}^{+\infty} dE \left(-\frac{\partial f(E, \mu, T)}{\partial E}\right)\sum _{ij}(E)
\end{equation}
where, $e$ is the electronic charge, $i$ and $j$ are Cartesian indices, and $f(E, \mu, T)$ is the Fermi-Dirac distribution function. 
\begin{figure}
\includegraphics[scale=0.15]{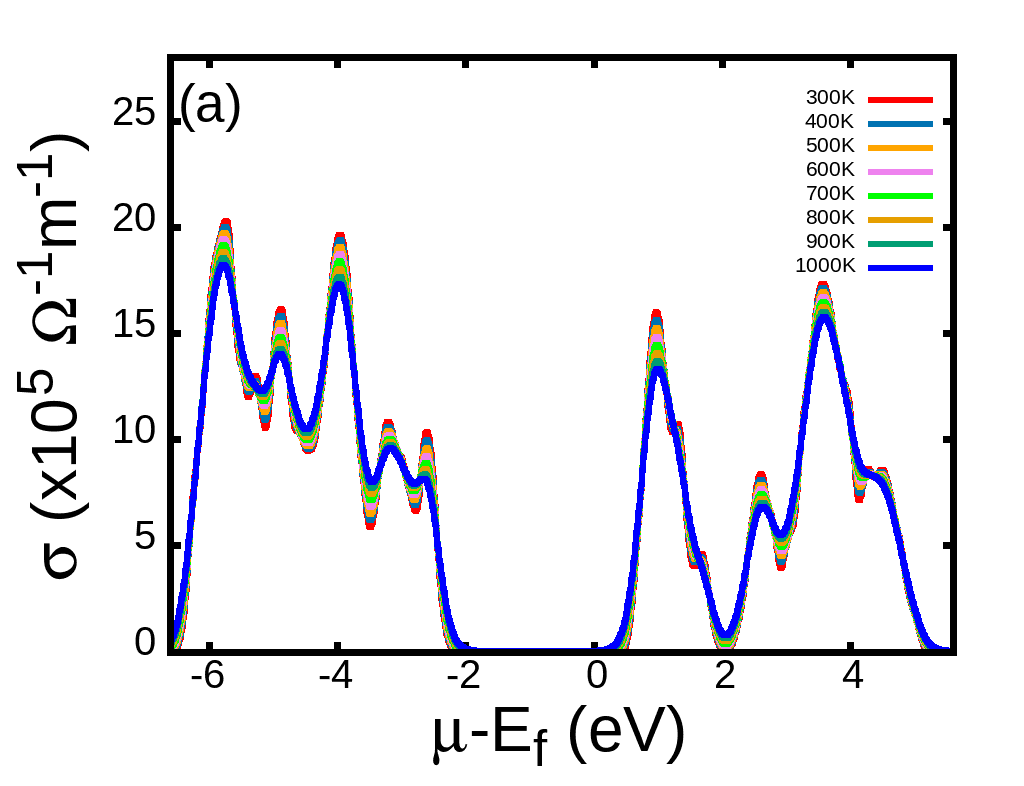}
\includegraphics[scale=0.15]{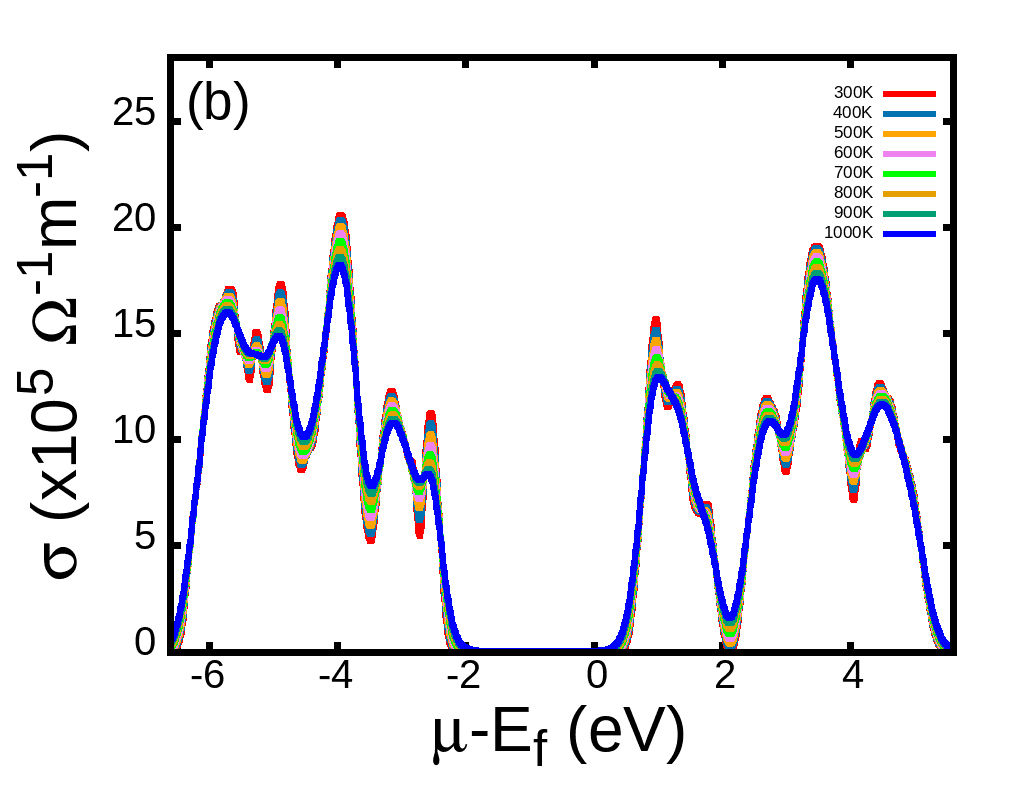}
\includegraphics[scale=0.15]{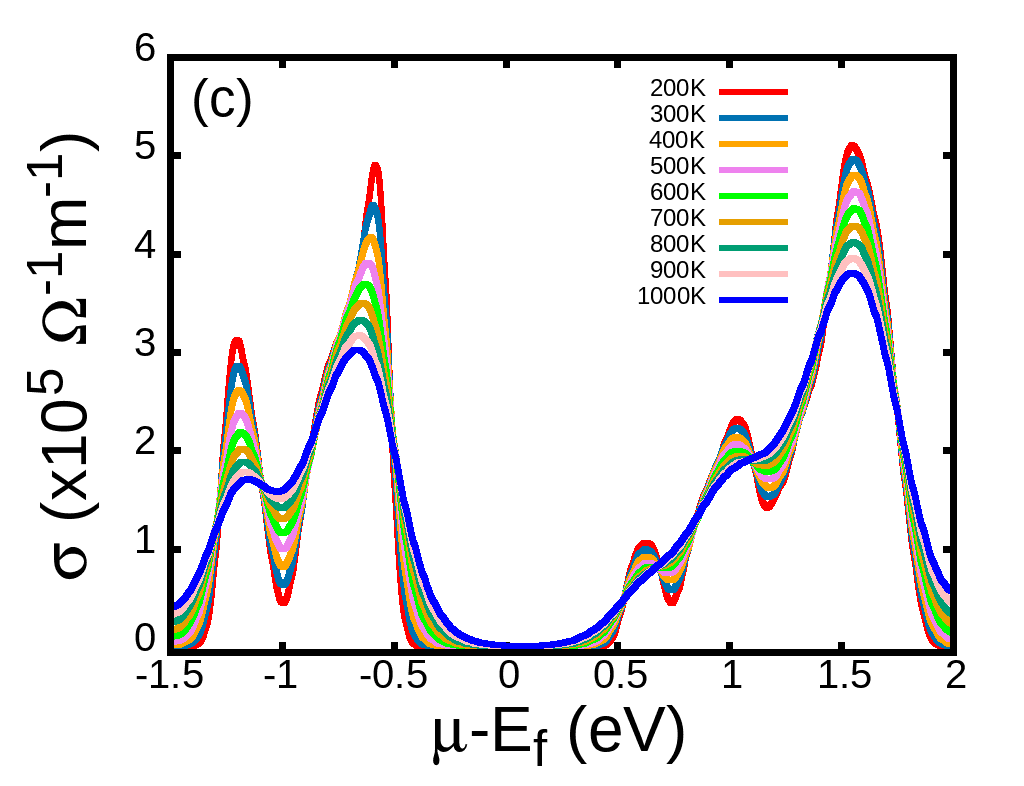}
\includegraphics[scale=0.15]{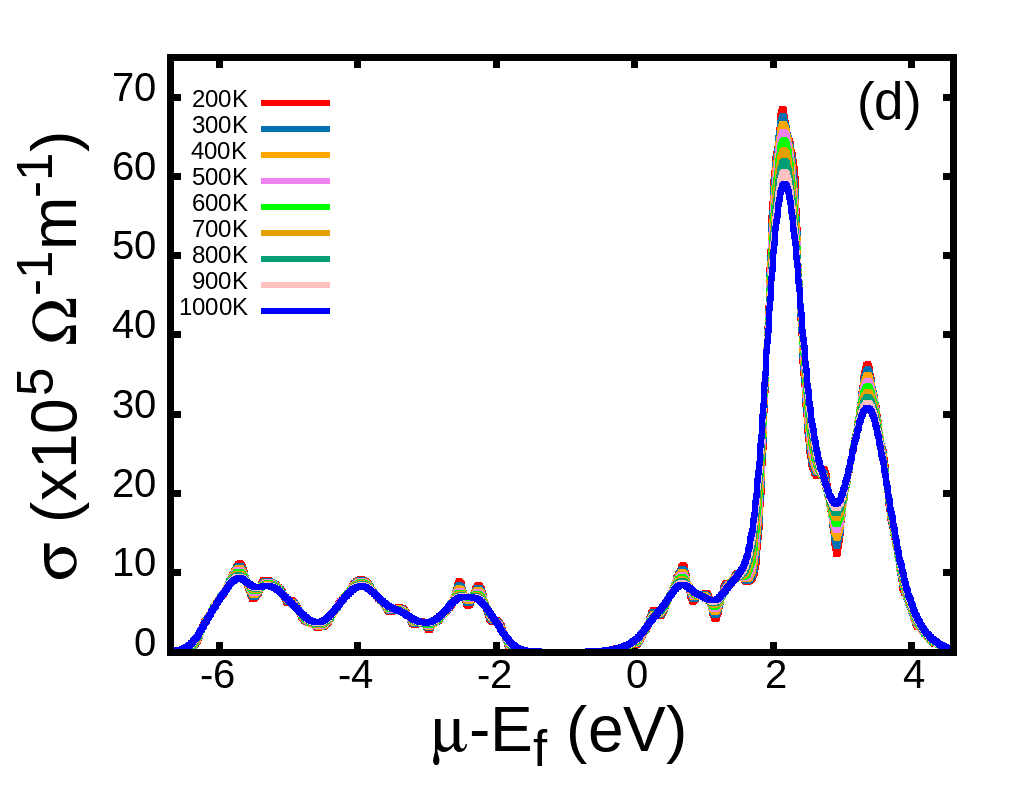}
\includegraphics[scale=0.15]{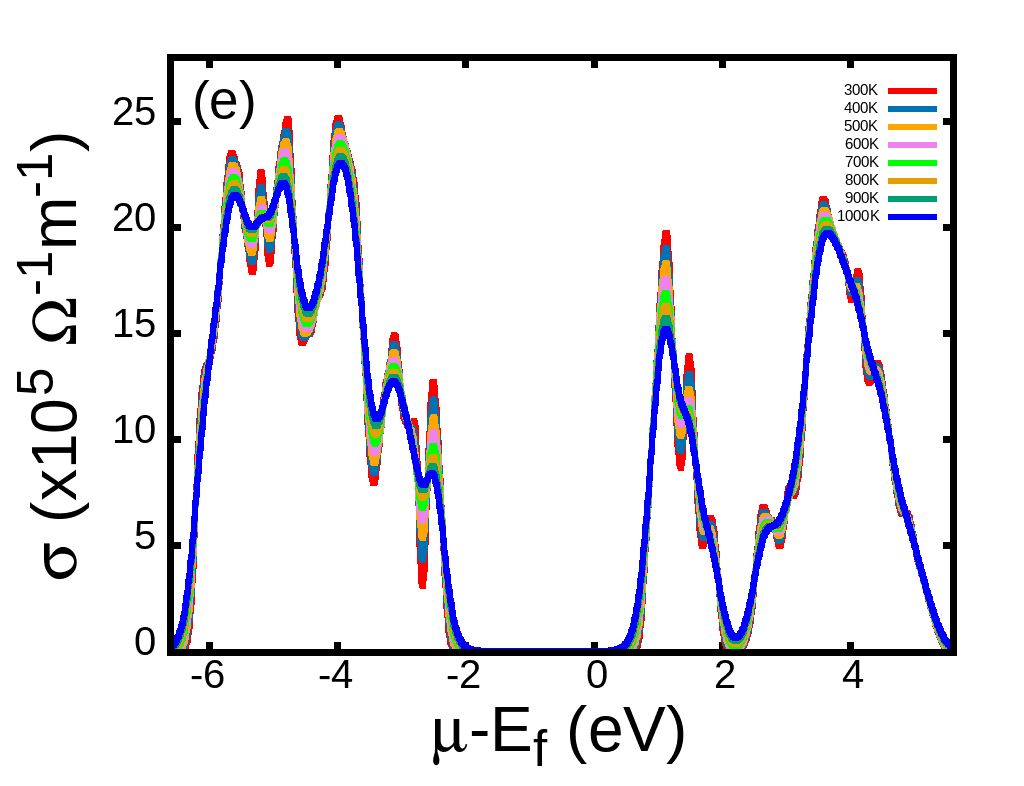}
\includegraphics[scale=0.15]{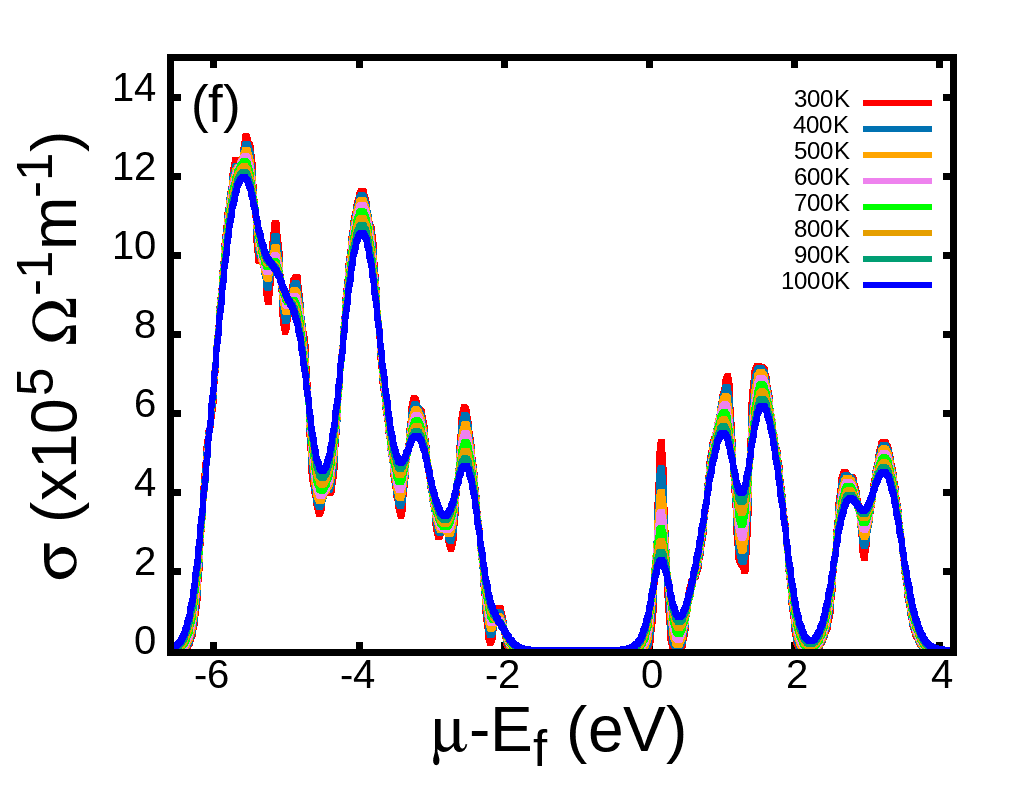}
\caption{Calculated electrical conductivity of BCZT as a function of chemical potential for (a) a6, (b) a7, (c) p6, (d) p7, (e) $r6$ and (f) r7.}
\label{sigma}
\end{figure}
This is one of the important transport coefficients of a material that depends on carrier density and mobility. Eventually, mobility depends directly on relaxation time and it can be increased and decreased with ionized impurities and by neutral impurities respectively. In the $a6$ structure, the electrical conductivity for negative chemical potential is $20.3\times 10^{5} \Omega^{-1}m^{-1}$ and $18.16\times 10^{5} \Omega^{-1}m^{-1}$, whereas for positive chemical potential, it is $17.25\times 10^{5} \Omega^{-1}m^{-1}$ and $15.7\times 10^{5} \Omega^{-1}m^{-1}$ at 300K and 1000K, respectively. This trend signifies that the n-type composition exhibits higher electrical conductivity than the p-type counterpart. A similar behavior is observed in the $a7$ and rhombohedral structures. In 	$a6$, $a7$, and $r6$ composites, the electrical conductivity ($\boldsymbol\sigma$) shows minimal variation between the valence and conduction bands. However, in the $r7$ structure, $\boldsymbol\sigma$ is reduced by half in the conduction band (as illustrated in Fig. \ref{sigma}). In the tetragonal structure, the p-type composition exhibits superior electrical conductivity compared to the n-type. Specifically, $p7$ attains a conductivity of $6.6\times 10^{6} \Omega^{-1}m^{-1}$ at room temperature, which decreases to $5.8\times 10^{6} \Omega^{-1}m^{-1}$ with increasing temperature, making it nearly ten times greater than $p6$. Electrical conductivity for all structures at 300 K, 600 K, and 1000 K is provided in Table S V. In this study, $p7$ demonstrates the highest electrical conductivity, while $p6$ exhibits the lowest \cite{Bano18}. Among $a6$, $a7$, $r6$, and $r7$, $r6$ possesses the highest electrical conductivity. At high temperatures, the significant reduction in the bandgap of the $p6$ structure suggests a transition toward metallic behavior. Additionally, $\boldsymbol\sigma$ exhibits an inverse linear dependence on temperature, which is characteristics of semiconductor behavior in the compound.

\begin{figure}
\includegraphics[scale=0.15]{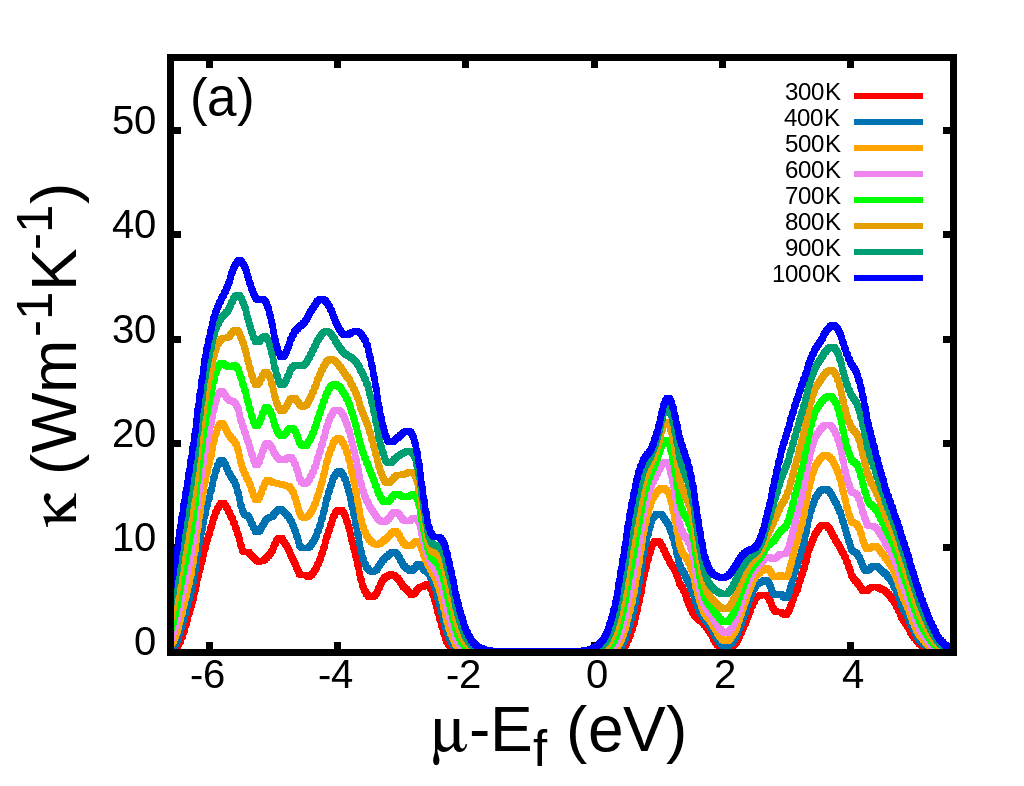}
\includegraphics[scale=0.15]{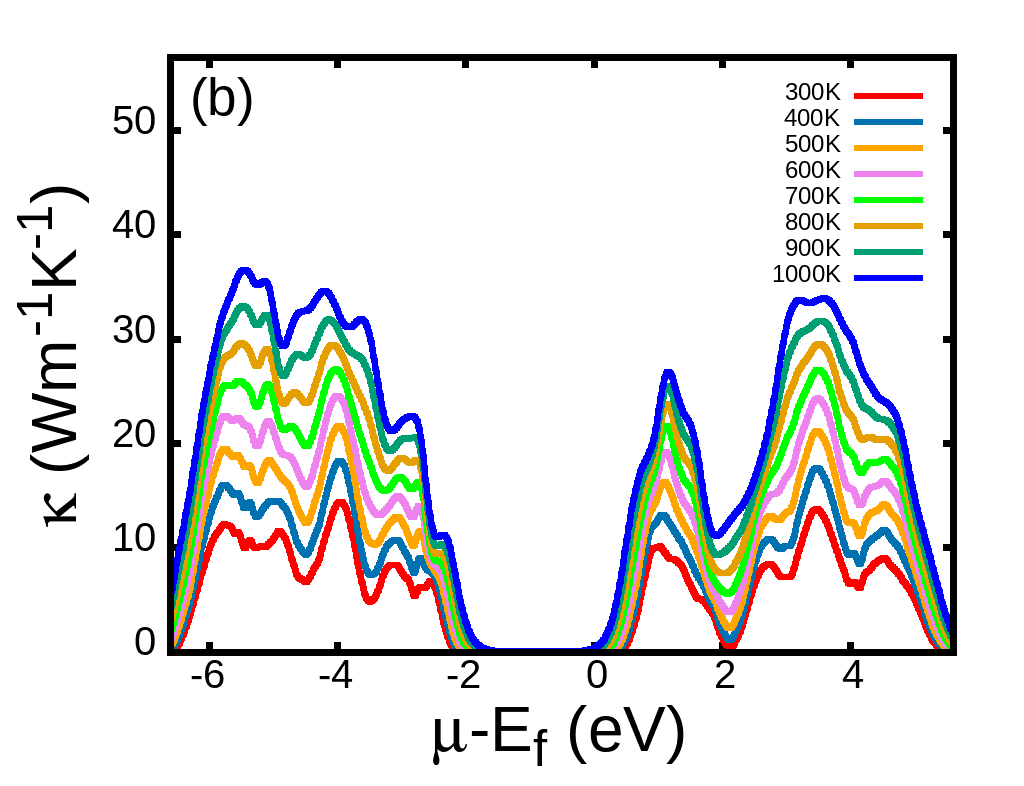}
\includegraphics[scale=0.15]{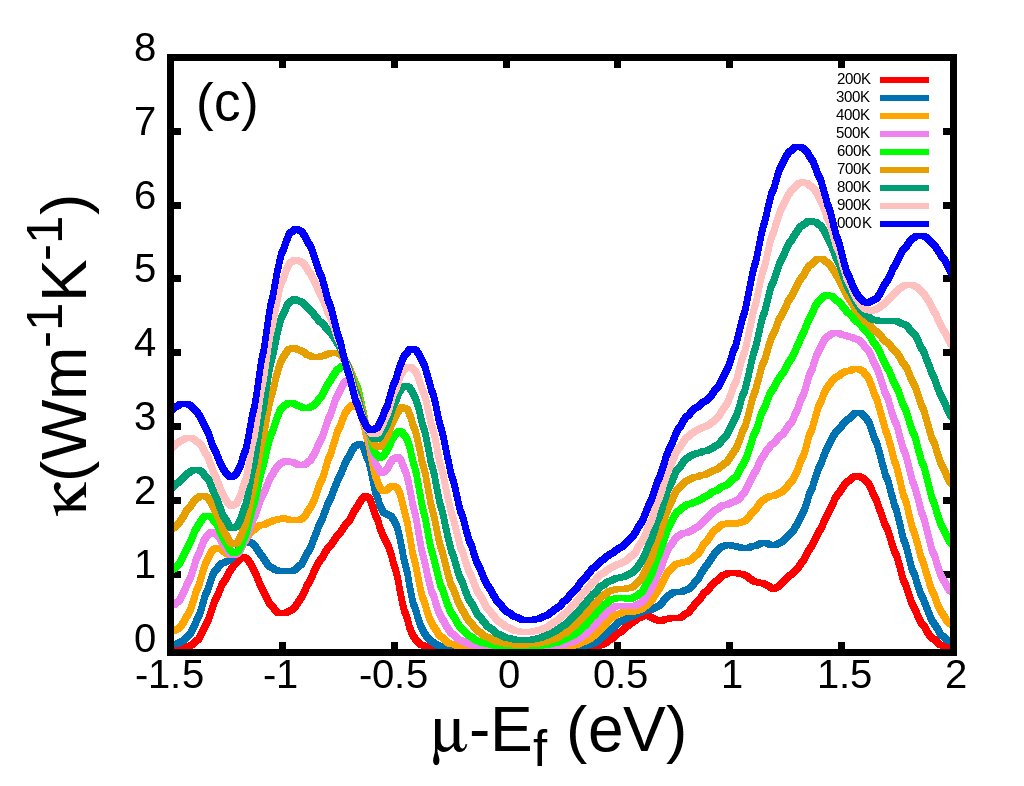}
\includegraphics[scale=0.15]{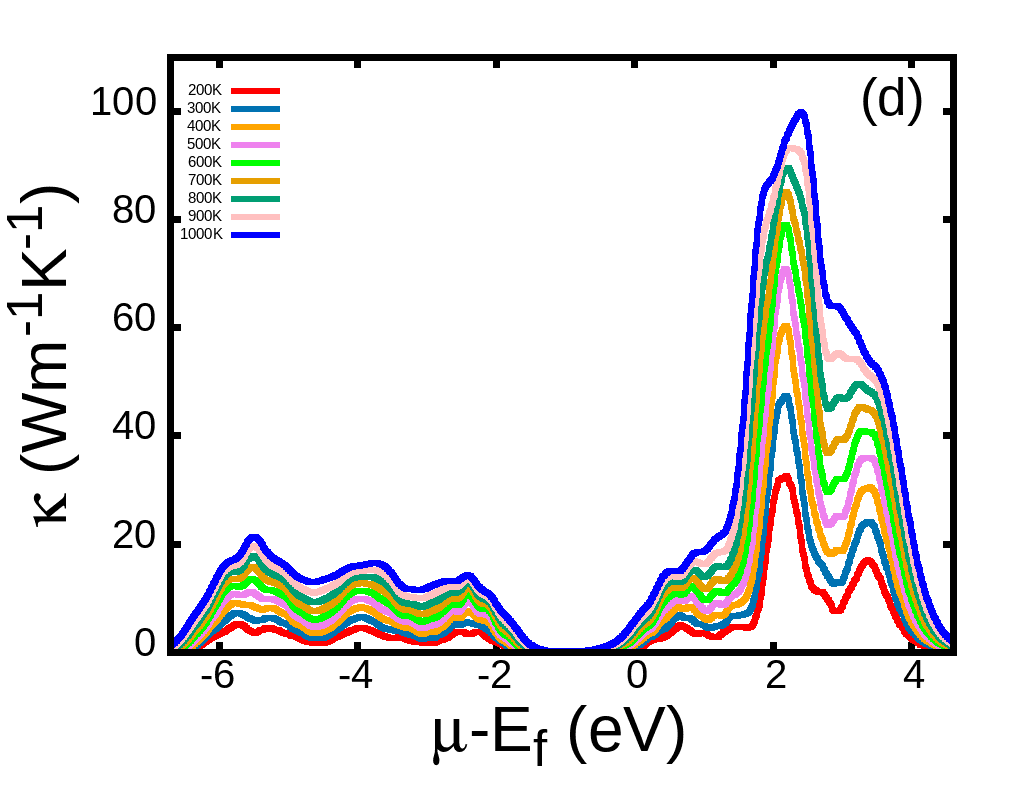}
\includegraphics[scale=0.15]{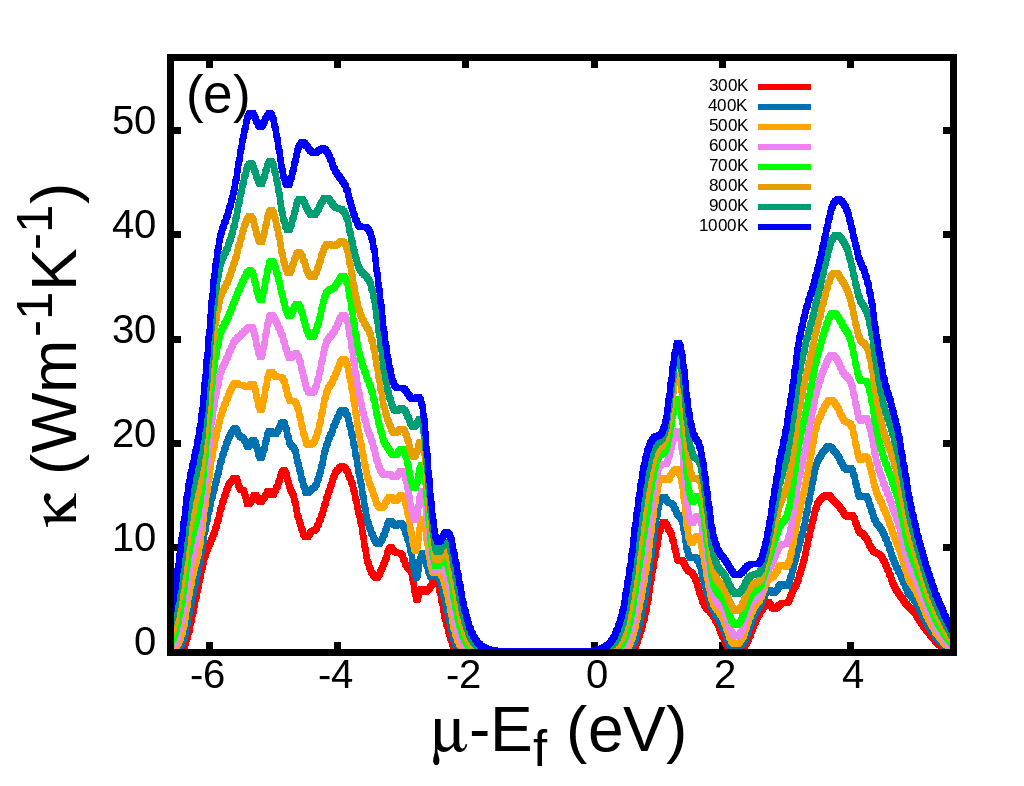}
\includegraphics[scale=0.15]{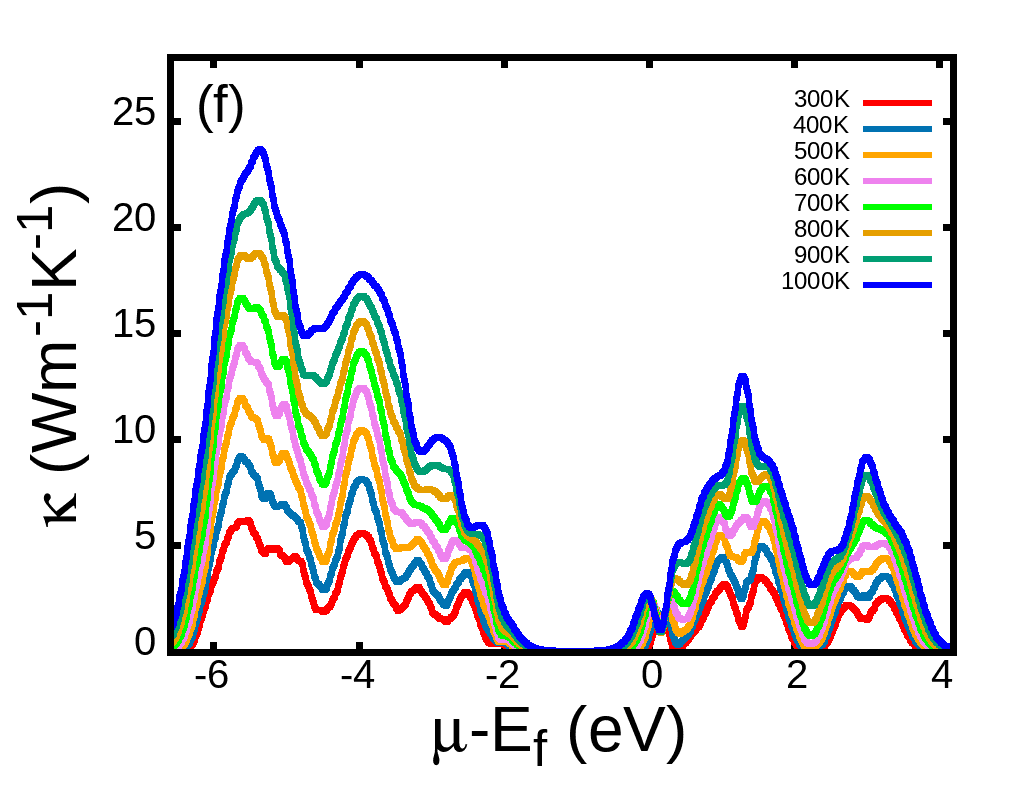}
\caption{Thermal conductivity of BCZT for (a) a6, (b) a7, (c) p6, (d) p7, (e) $r6$ and (f) r7.}
\label{kappa}
\end{figure}
Thermal conductivity profile is similar to electrical conductivity and has a major role in thermoelectric power generation. Thermal conductivity can be obtained from the following relation,
\begin{equation}
[\boldsymbol{\kappa}]_{ij}(\mu, T) = \frac{1}{T} \int_{-\infty}^{+\infty} dE \left( - \frac{\partial f (E, \mu, T)}{\partial E}\right)(E - \mu)^2 \sum_{ij}(E)
\end{equation}
Thermal conductivity, like electrical conductivity, is influenced by carrier concentration. However, in contrast to electrical conductivity, thermal conductivity generally increases with temperature. In most thermoelectric materials, lattice thermal conductivity dominates over the electronic contribution. Lattice thermal conductivity depends on phonon vibration in the lattice and scattering parameters, which is entirely based on structural architecture of the compound. In this work, only the electronic contribution is considered and represented as thermal conductivity. The thermal conductivity directly depends on the distribution function, similar to electrical conductivity, leading to a similar profile curve. $p7$ has the highest thermal conductivity of 32.26 W/m/K and 99.96 W/m/K at 300K and 1000K respectively, among all structures. $p6$ has the lowest value of 2.3 W/m/K and other structures have comparable magnitude and profile, whereas $r6$ and $r7$ are the high and low thermal conductive material respectively. $r6$ is a better thermal conductor than $r7$ because the maxima lies near to the Fermi level. With temperature it increases and the rate of increase depends on the structure and stability of the lattice. Lowest value of $\boldsymbol{\kappa}$ appeared at 300K for all structures, therefore it is the optimal temperature. The thermal conductivity for all structures at 300 K, 600 K, and 1000 K is presented in Table S V. Efficient thermoelectric materials are designed to minimize $\boldsymbol{\kappa}$ while maintaining a reasonable electrical conductivity, leading to an enhanced figure of merit (ZT) for improved energy conversion efficiency. Thermal conductivity can be reduced by increasing bandgap or by interstitial doping with heavy elements. It can also be reduced by maximizing the surface to volume ratio of the compound \cite{Yu10}, by forming a thin layered structure. Defect engineering and increasing disorderness can significantly lower thermal conductivity for a better thermoelectric material \cite{Mukherjee22}. 
\begin{figure}
\includegraphics[scale=0.15]{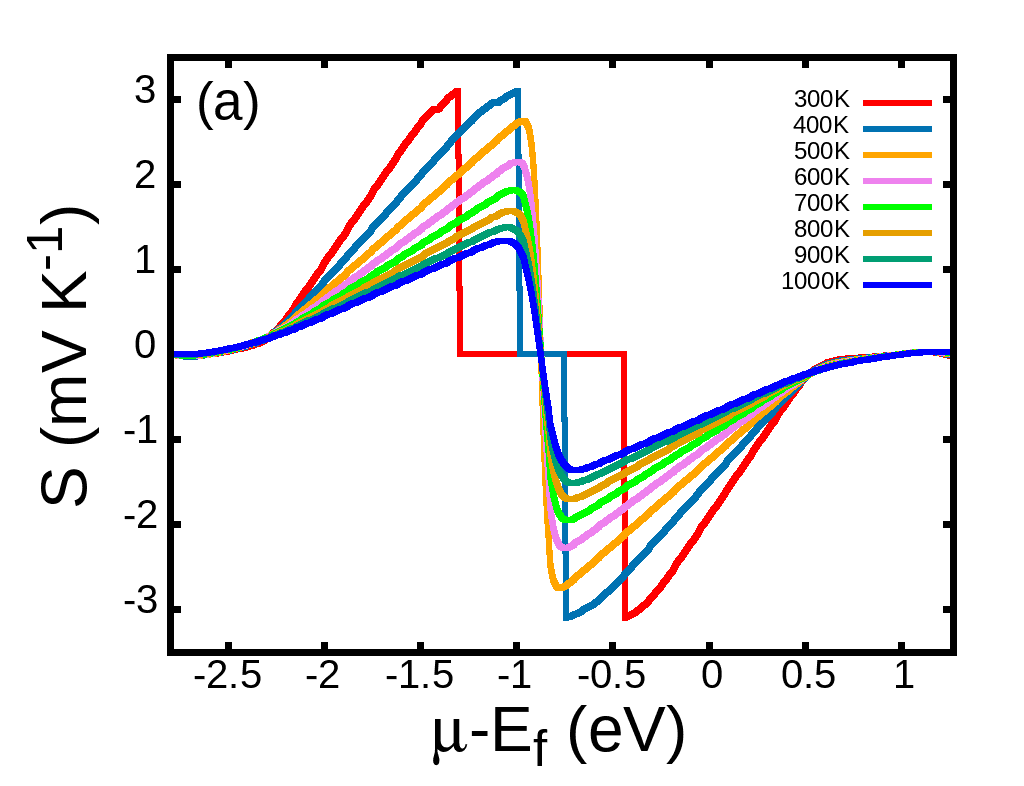}
\includegraphics[scale=0.15]{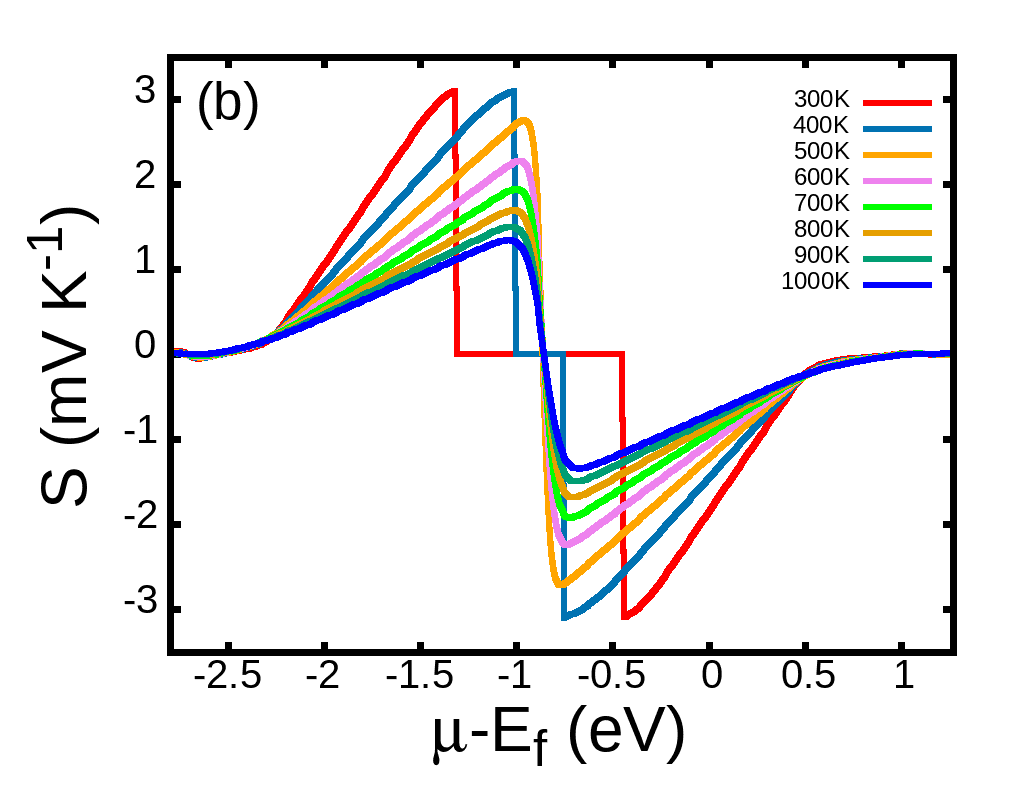}
\includegraphics[scale=0.15]{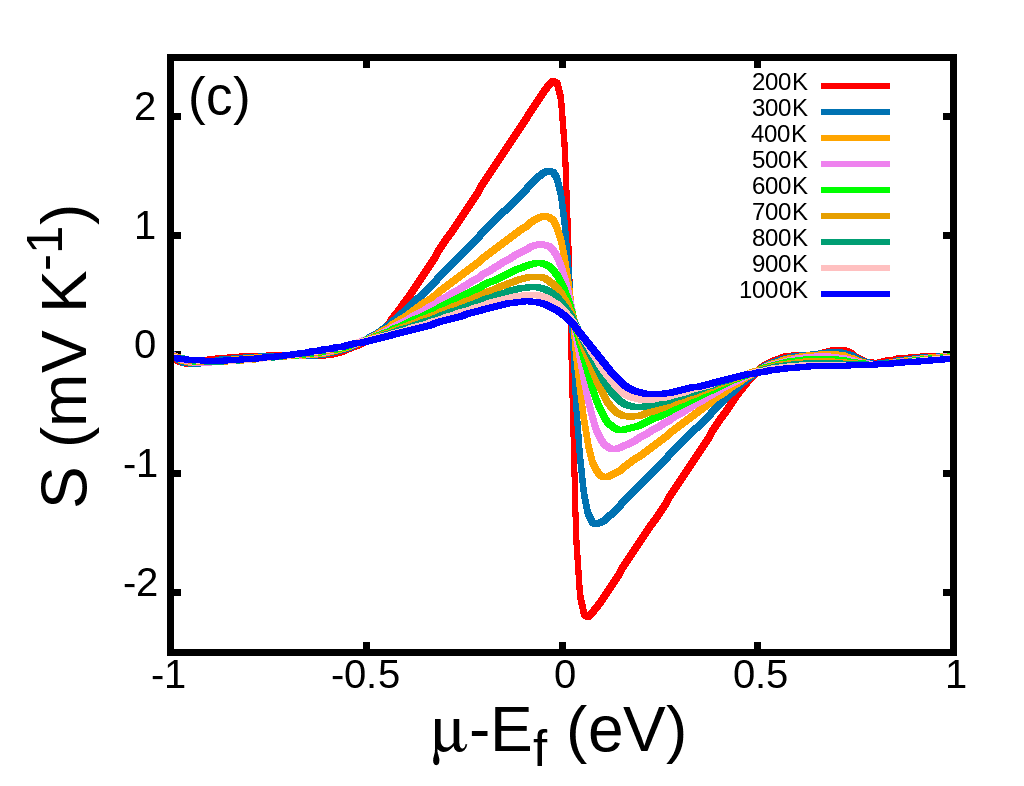}
\includegraphics[scale=0.15]{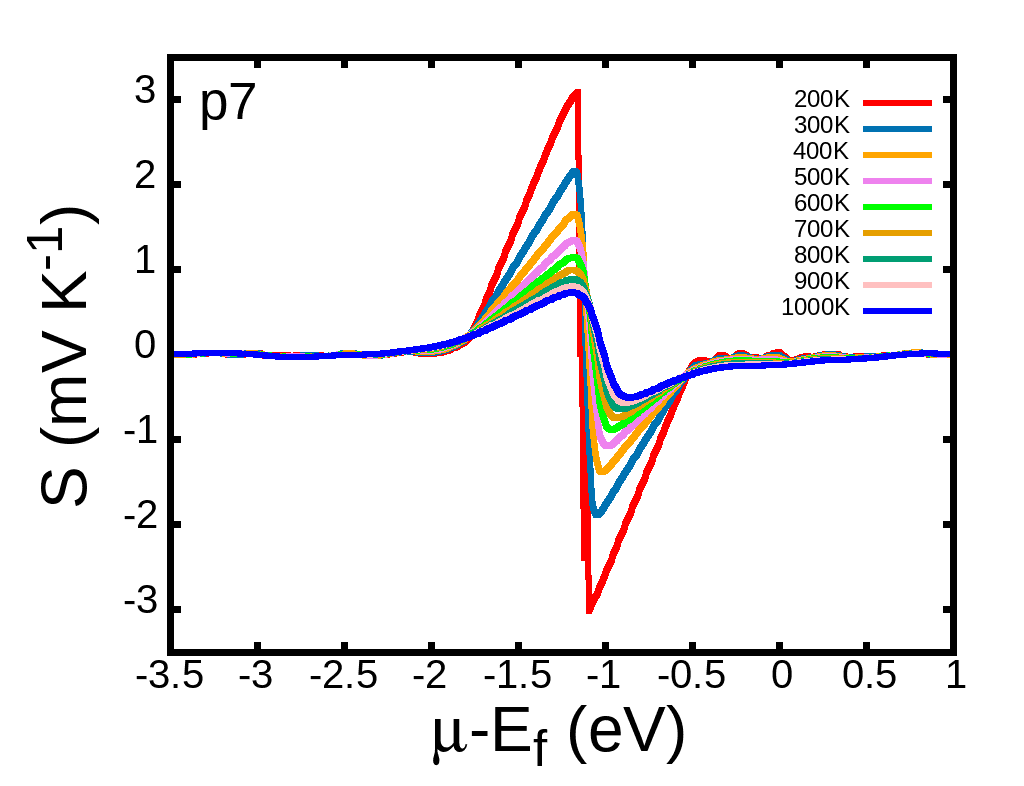}
\includegraphics[scale=0.15]{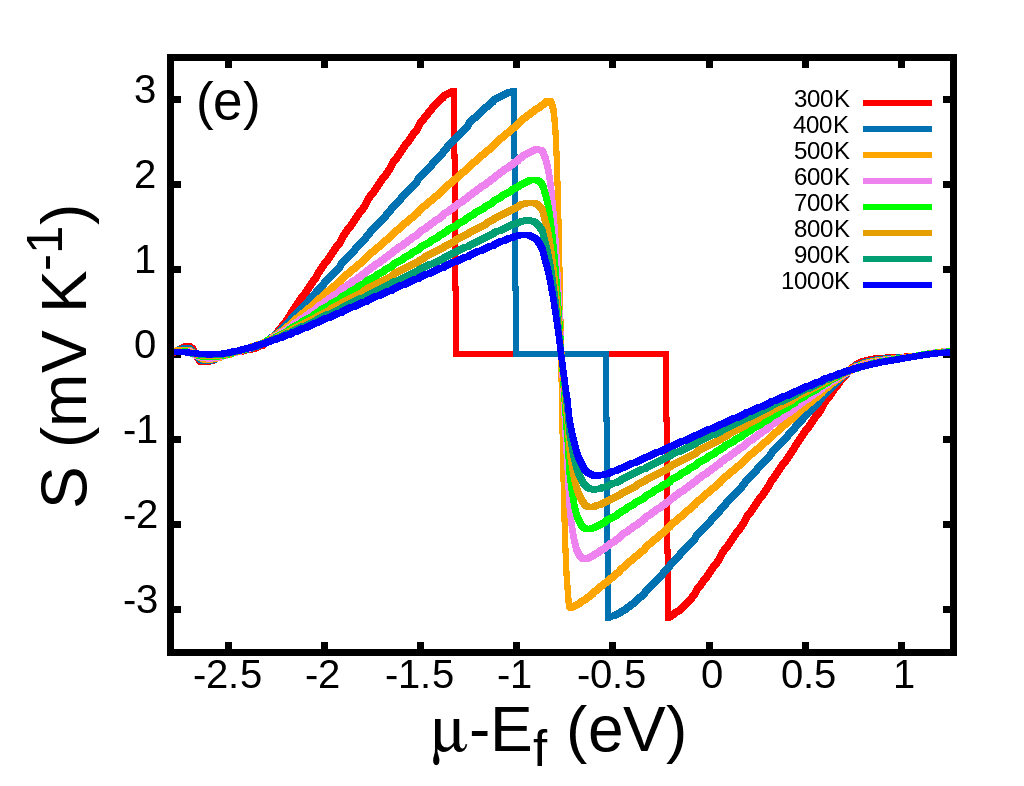}
\includegraphics[scale=0.15]{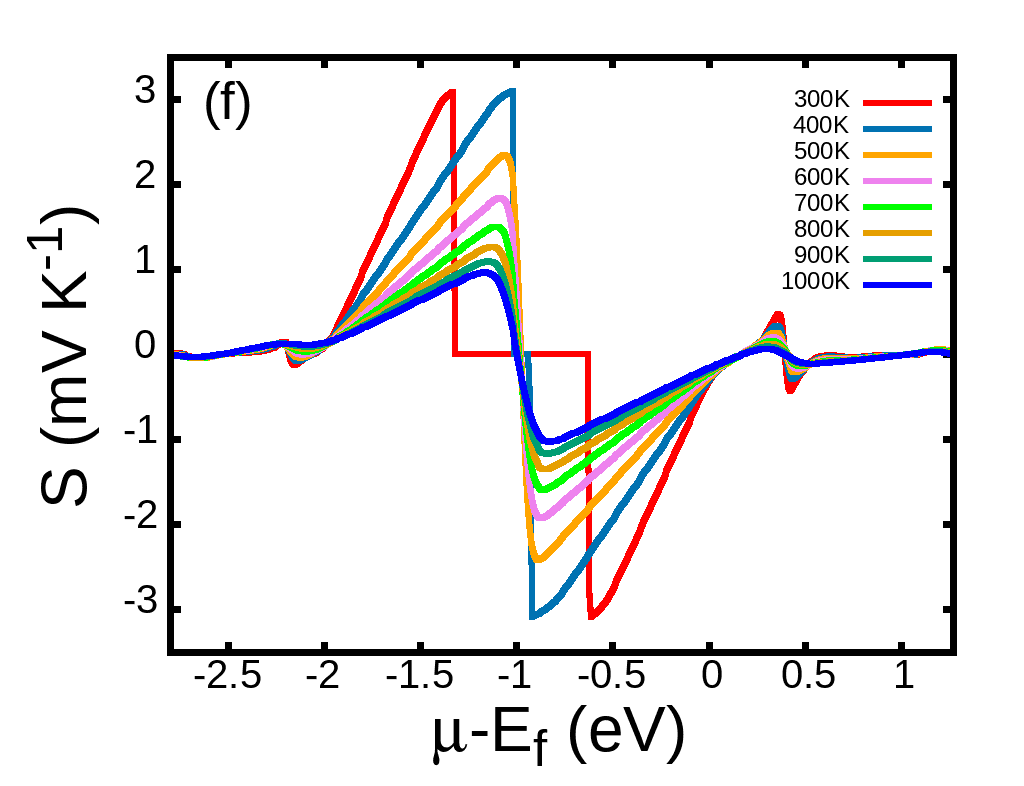}
\caption{Seebeck coefficient of BCZT for (a) a6, (b) a7, (c) p6, (d) p7, (e) $r6$ and (f) r7.}
\label{ Seebeck}
\end{figure}

\begin{figure}
\includegraphics[scale=0.15]{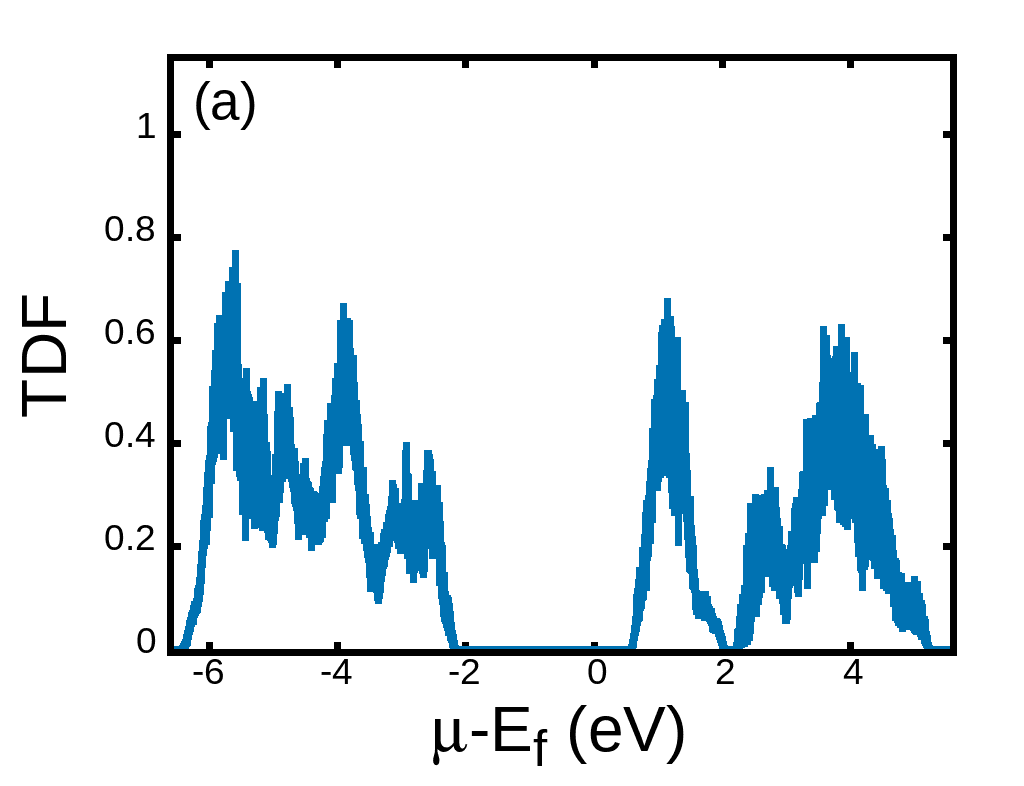}
\includegraphics[scale=0.15]{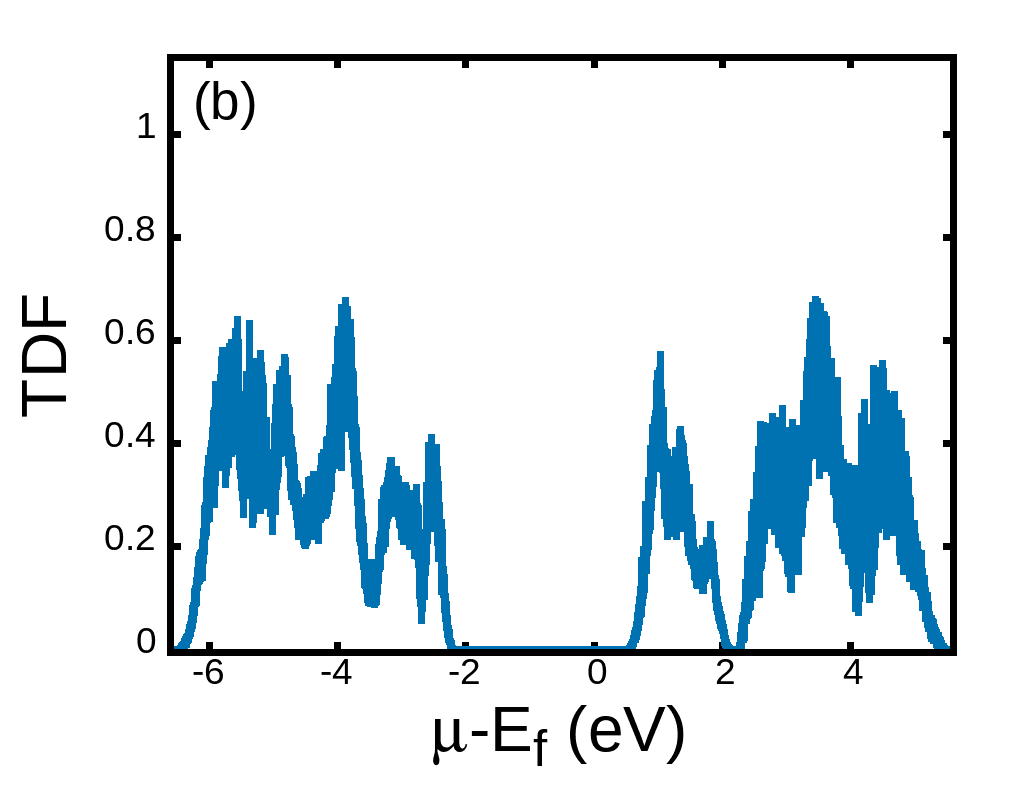}
\includegraphics[scale=0.15]{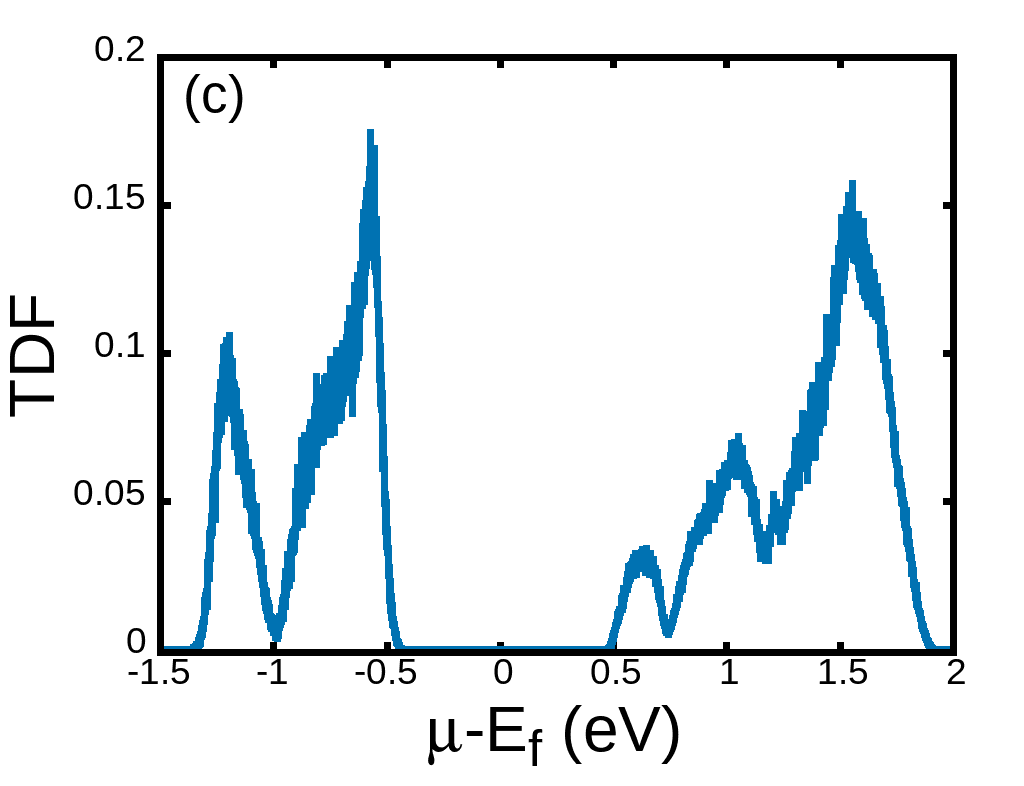}
\includegraphics[scale=0.15]{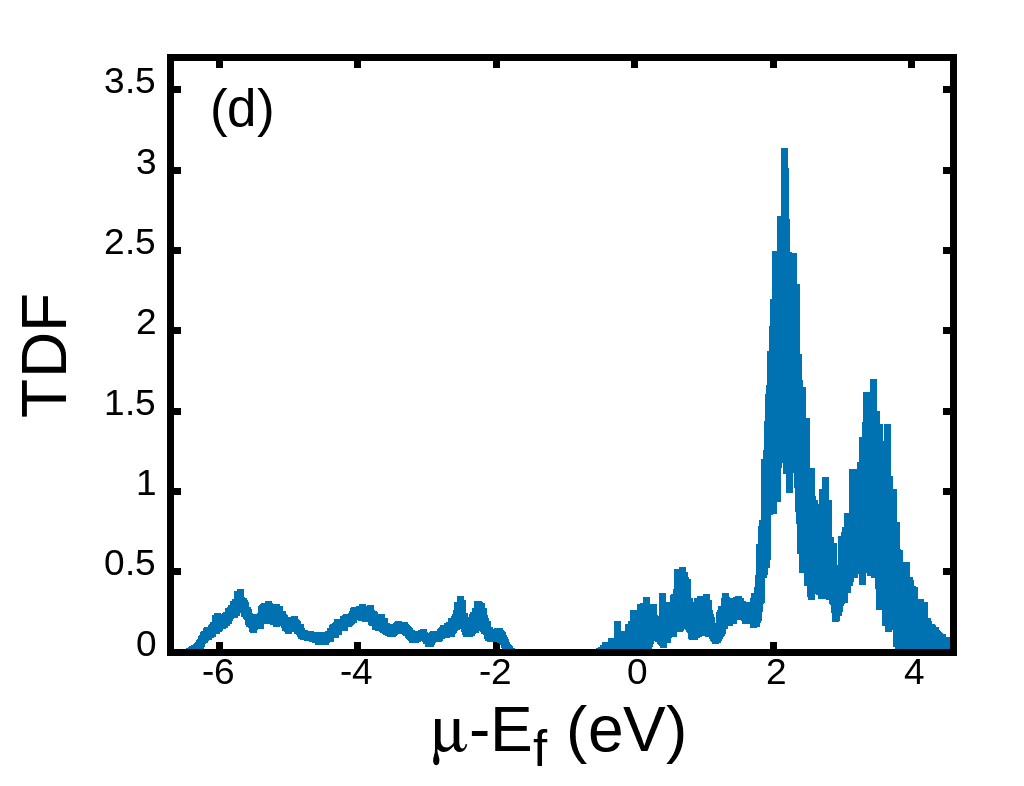}
\includegraphics[scale=0.15]{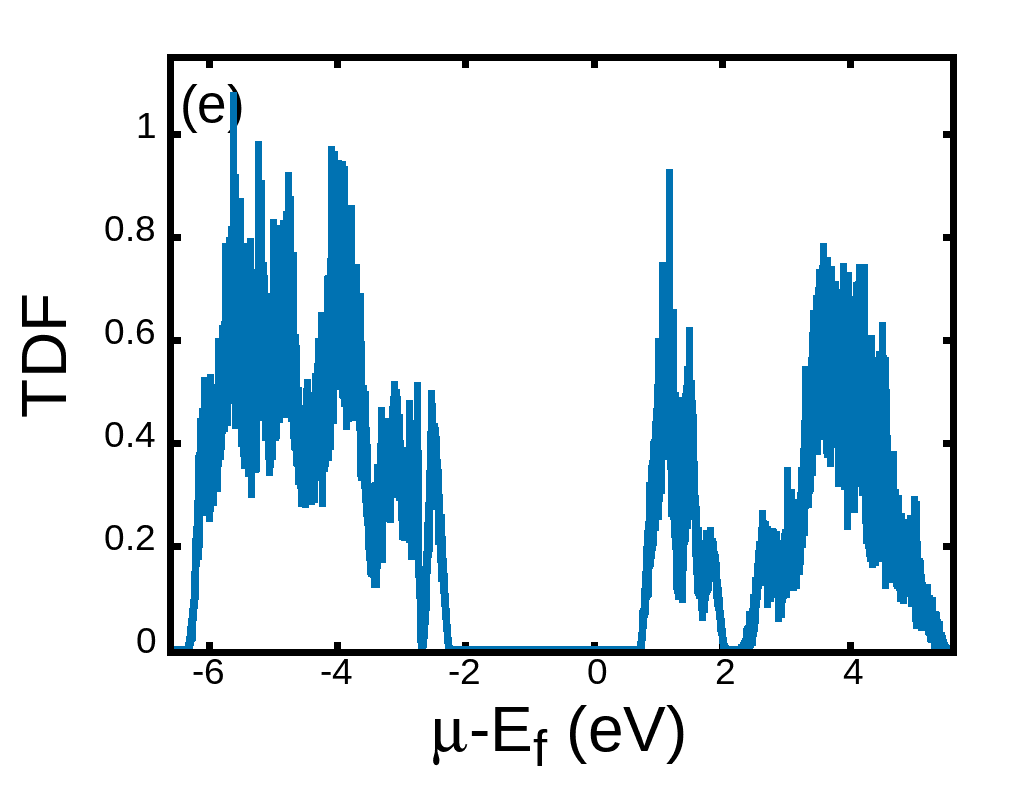}
\includegraphics[scale=0.15]{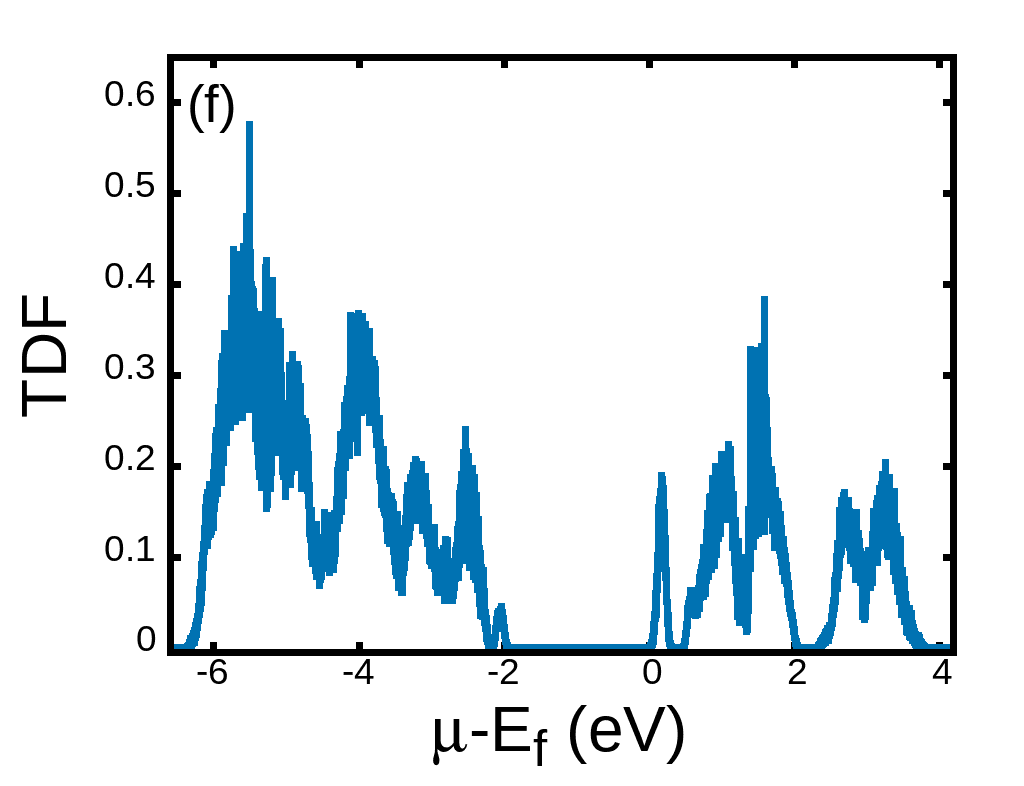}
\caption{Transport distribution function of BCZT for (a) a6, (b) a7, (c) p6, (d) p7, (e) $r6$ and (f) r7.}
\label{bdos}
\end{figure}

Seebeck coefficient is the measure of induced voltage with temperature gradient. When two different metals are connected with the same sink at low temperature at one end, and high temperature at the other end, voltage will be induced across the metals which is proportional to the temperature difference. This voltage is called Seebeck voltage and its ratio with change in temperature is called Seebeck Coefficient, which can be defined as
\begin{eqnarray}
[\boldsymbol{\sigma}\mathbf{S}]_{ij}(\mu, T) = \frac{e}{T} \int_{-\infty}^{+\infty} dE \left(-\frac{\partial f(E, \mu, T)}{\partial E}\right) \nonumber \\ \times (E - \mu) \sum_{ij}(E)
\end{eqnarray}
Commercially available standard thermocouple Seebeck coefficient is $~\pm50\mu VK^{-1}$ \cite{Moffat97}. Seebeck coefficient mainly signifies the performance of the thermocouple and its application is based on how sensitive the material is to small temperature changes. It comprises two contributions: the electrostatic potential term, arising from the diffusion of charge carriers due to a temperature gradient, and the chemical potential term, resulting from the temperature-dependent shift in the Fermi level and carrier distribution.
Here, we have shown the variation of Seebeck coefficient with temperature for orthorhombic, tetragonal and rhombohedral structures of BCZT (see Fig. \ref{ Seebeck}). We have achieved an average colossal Seebeck coefficient of $\pm 2.5 mVK^{-1}$ theoretically, compared to other standard materials \cite{Markov19}. \textbf{S} depends on the grain size and as temperature increases, one can tune it by varying its particle diameter \cite{Byeon19,Kockert19}. Seebeck coefficient reduces with increase in temperature because the carrier concentration increases \cite{Hofmann19} and its variation with chemical potential for different materials are shown in Fig. \ref{ Seebeck} for different temperature ranges $300-1000$K. The estimated value of the Seebeck coefficient is provided in the Table S VI. for different crystals.
Peak shifting and decreasing of Seebeck coefficient is observed in rhombohedral as well as in orthogonal structure with increase in temperature.
Further, the transport distribution function(TDF) $\sum_{ij}(E)$ is defined as,
\begin{equation}
\sum_{ij}(E) = \frac{1}{V} \sum_{n, \mathbf{q}}v_i (n, \mathbf{q}) v_j(n, \mathbf{q})\tau_{n\mathbf{q}}\delta(E - E_{n, \mathbf{q}})
\end{equation}
The calculated TDF are shown in Fig.\ref{bdos} and are well correlated with other thermoelectric properties and transport total density of states [Fig. S 1].

However, the thermoelectric industry is more interested in the power factor (PF = $\boldsymbol{\sigma}\mathbf{S}^{2}$) and figure of merit ($ZT$) for practical purposes.
\begin{figure}
\includegraphics[scale=0.15]{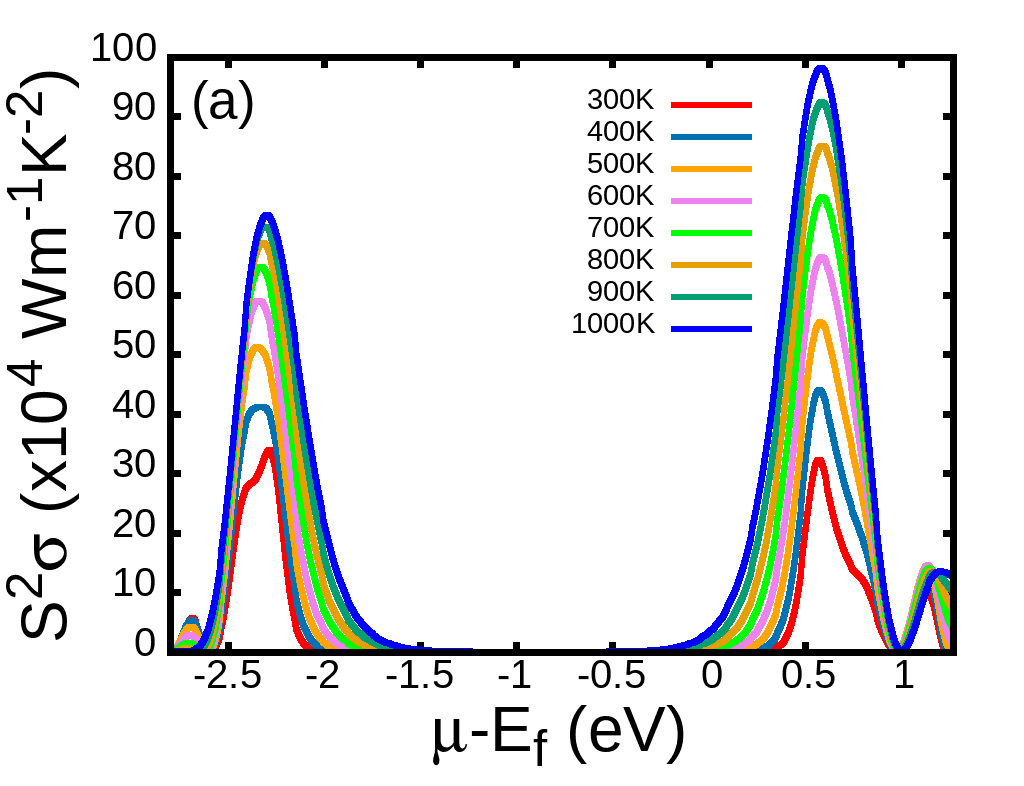}
\includegraphics[scale=0.15]{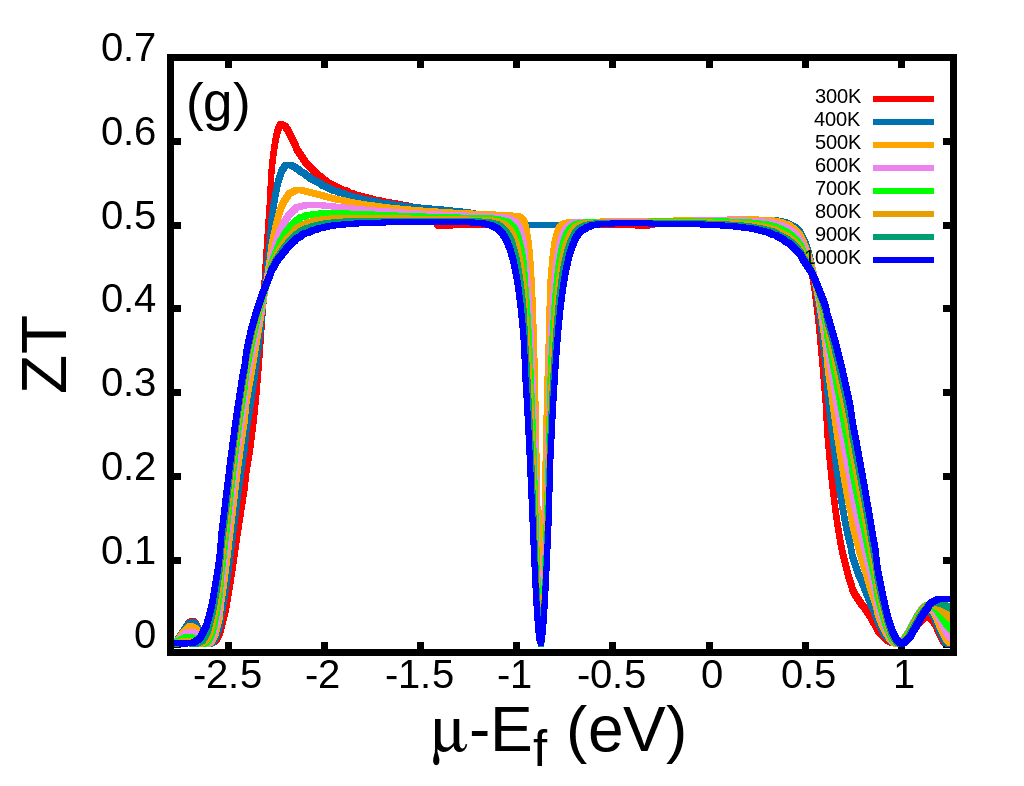}
\includegraphics[scale=0.15]{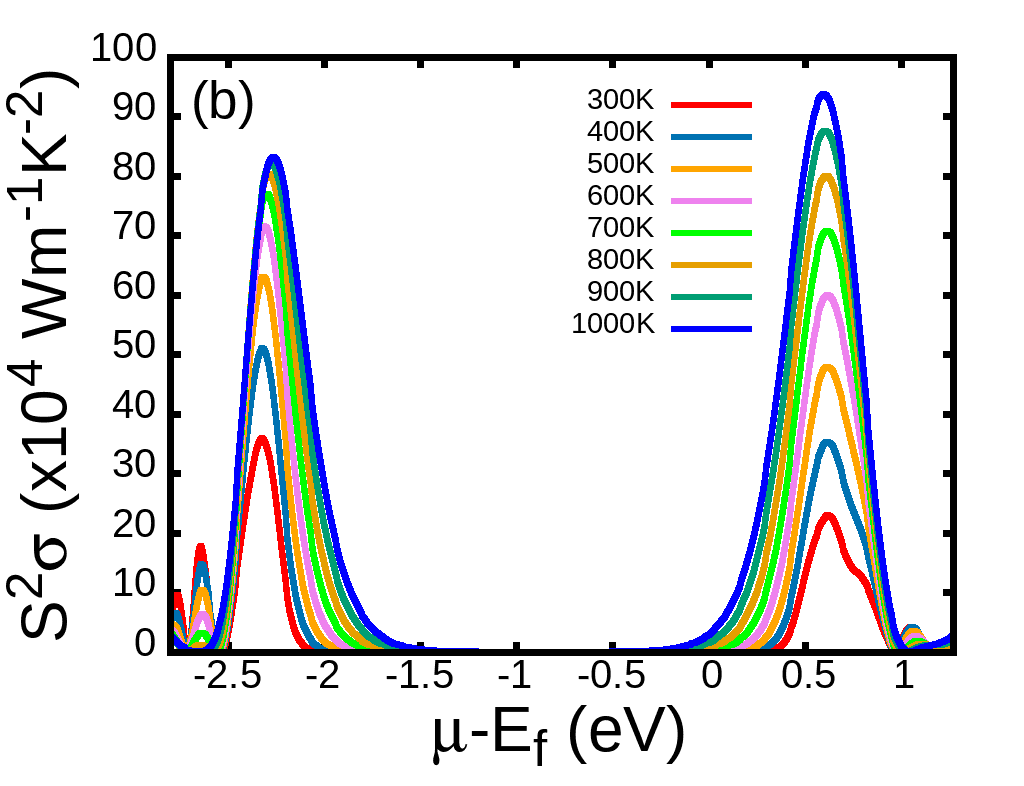}
\includegraphics[scale=0.15]{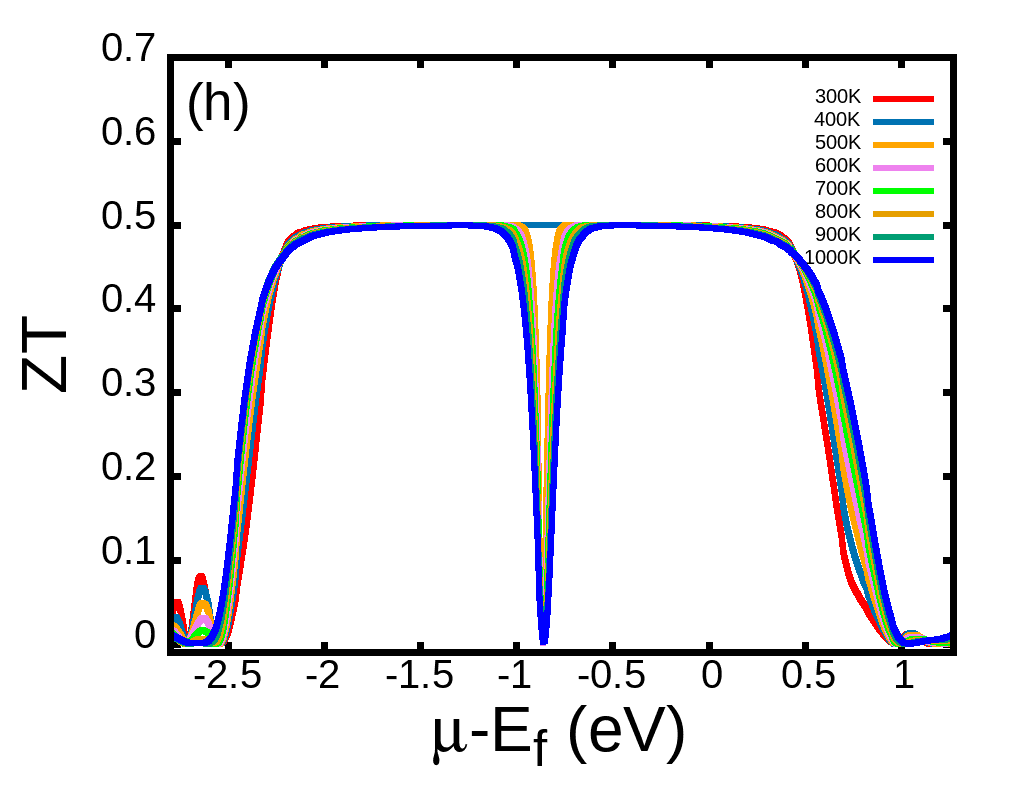}
\includegraphics[scale=0.15]{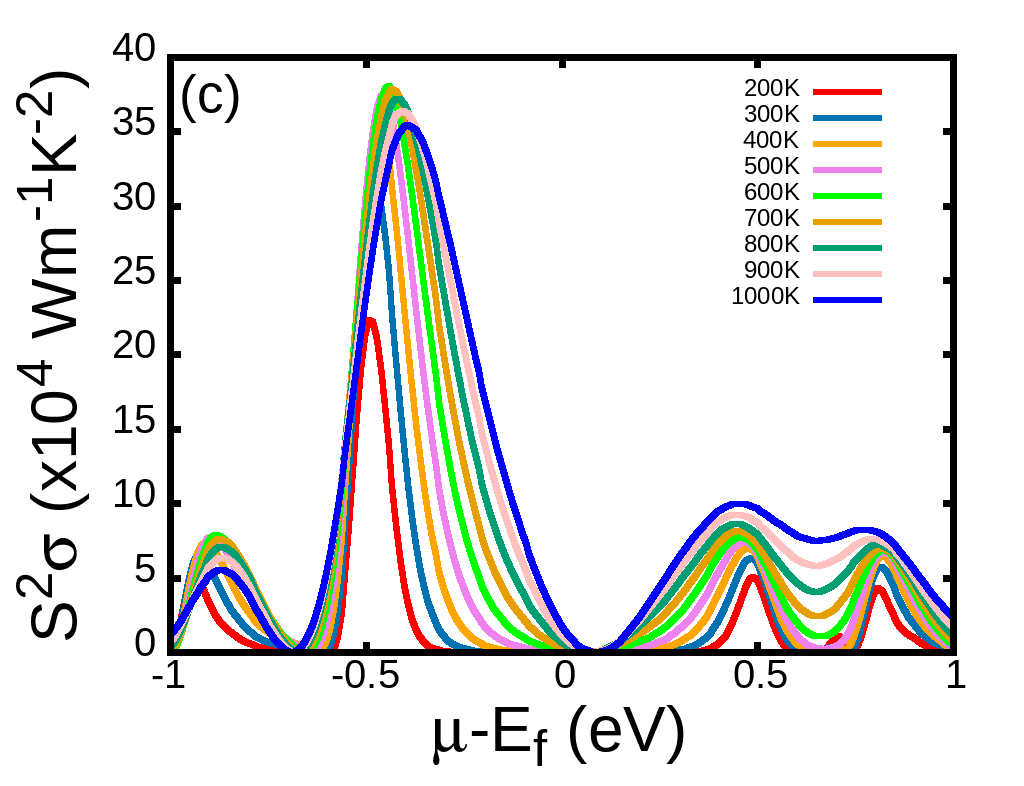}
\includegraphics[scale=0.15]{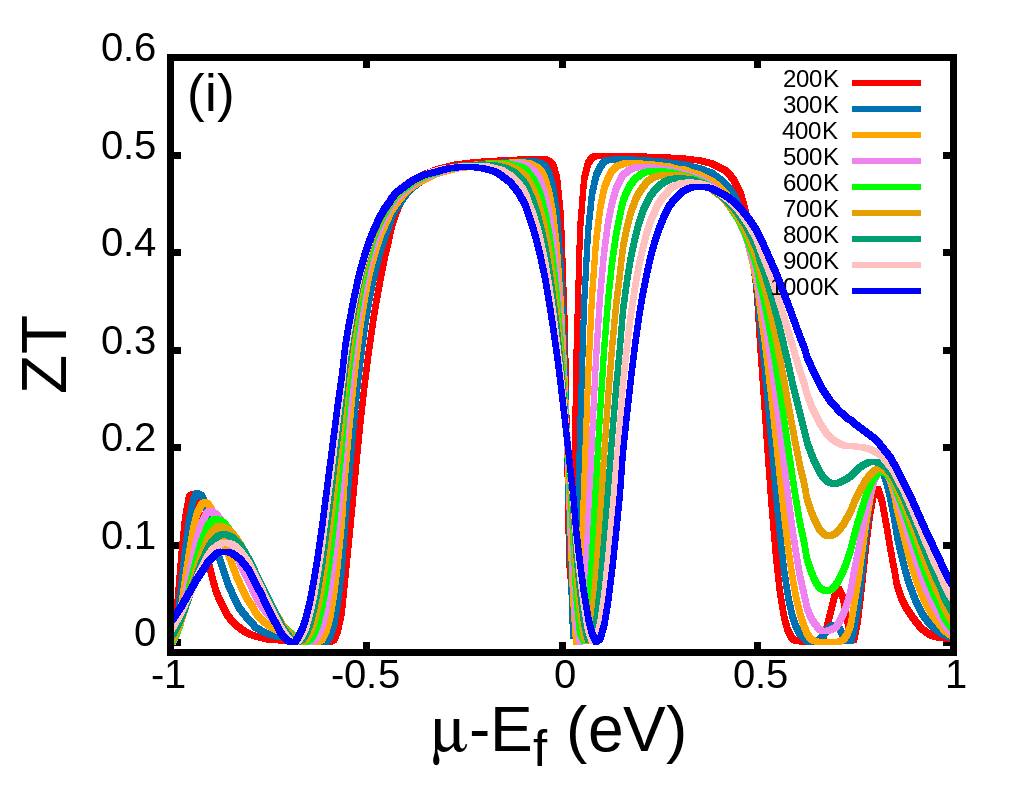}
\includegraphics[scale=0.15]{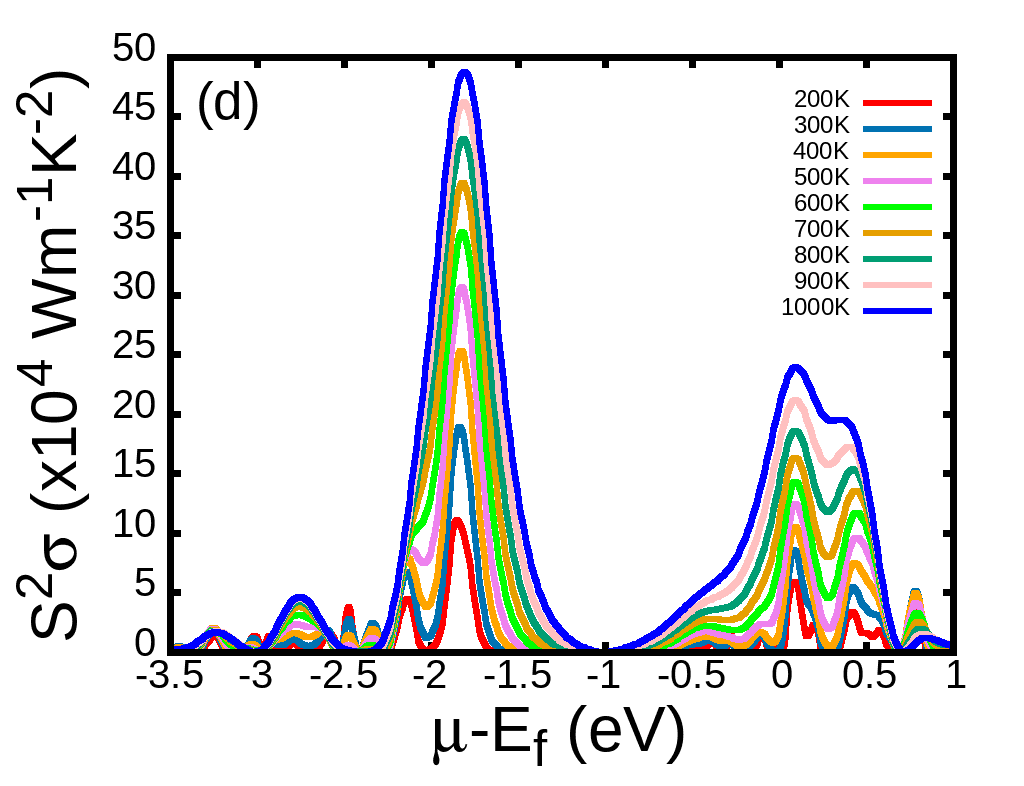}
\includegraphics[scale=0.15]{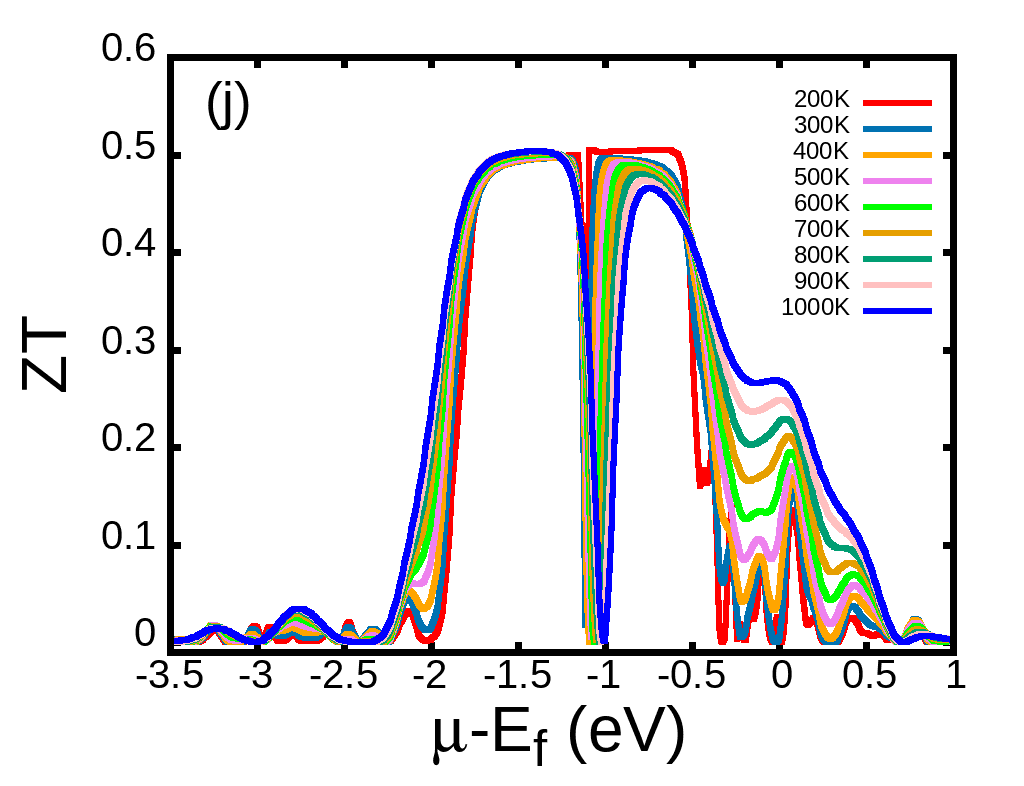}
\includegraphics[scale=0.15]{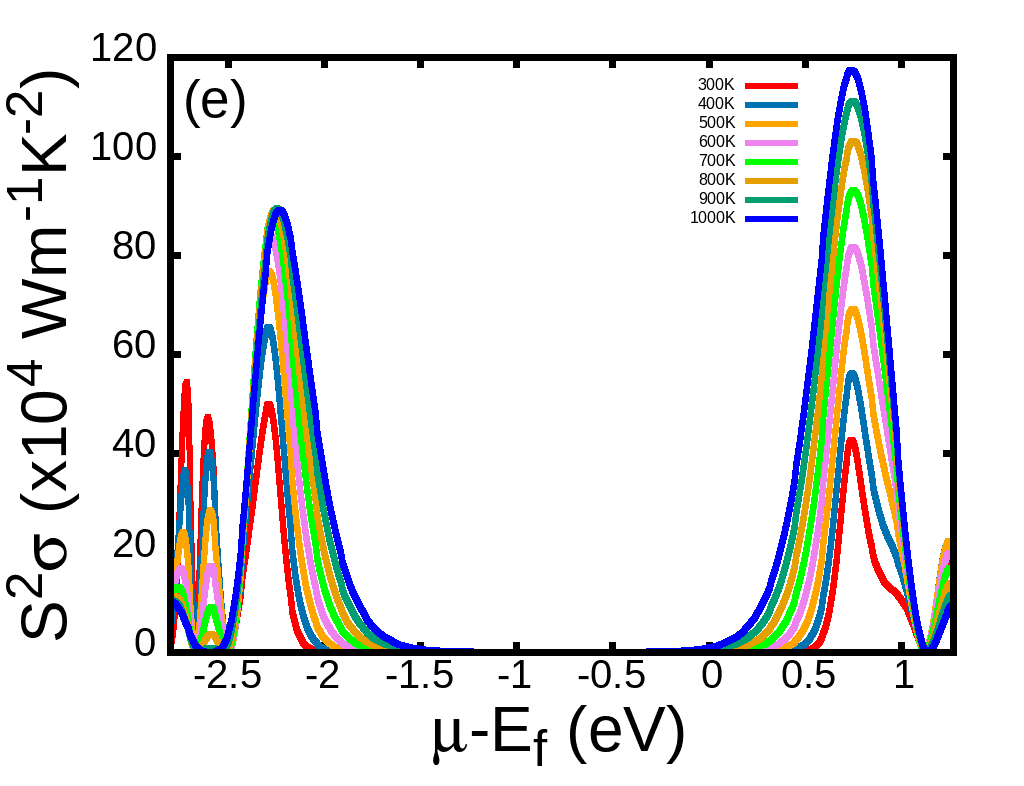}
\includegraphics[scale=0.15]{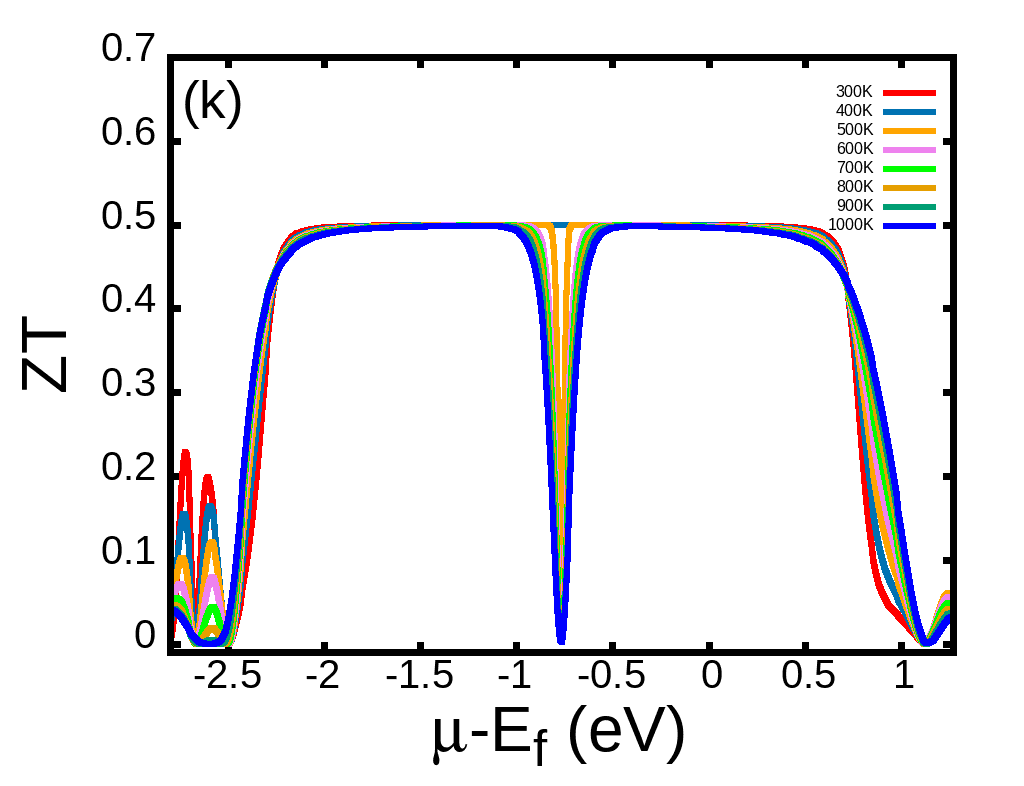}
\includegraphics[scale=0.15]{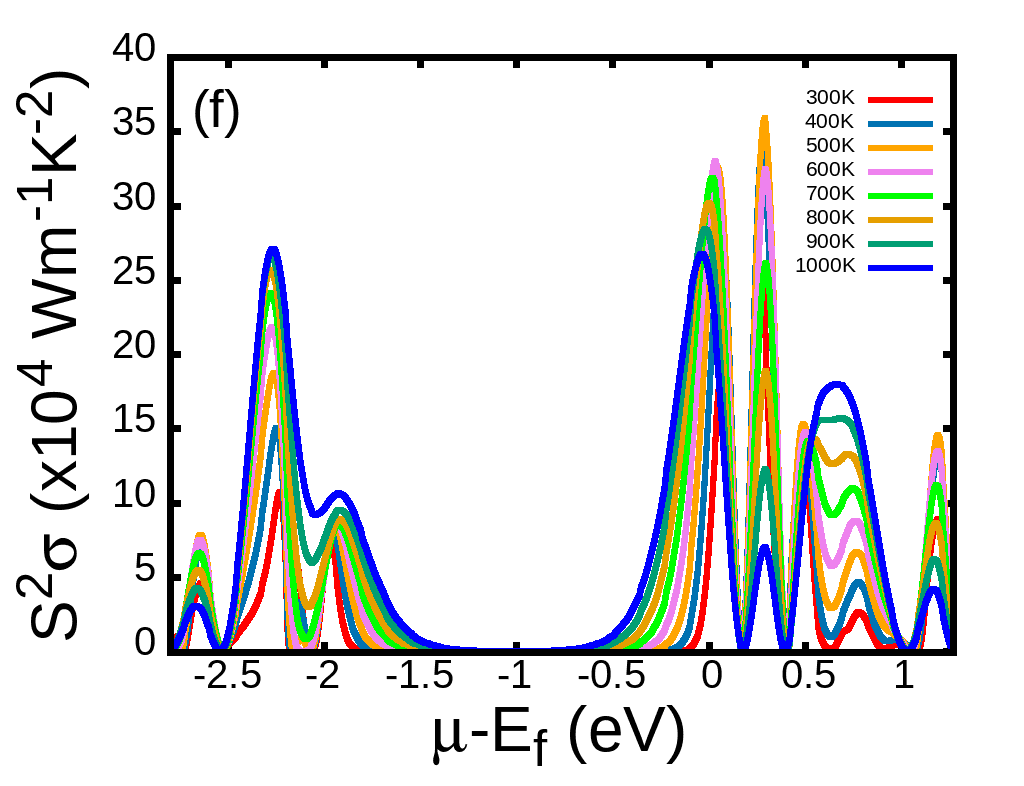}
\includegraphics[scale=0.15]{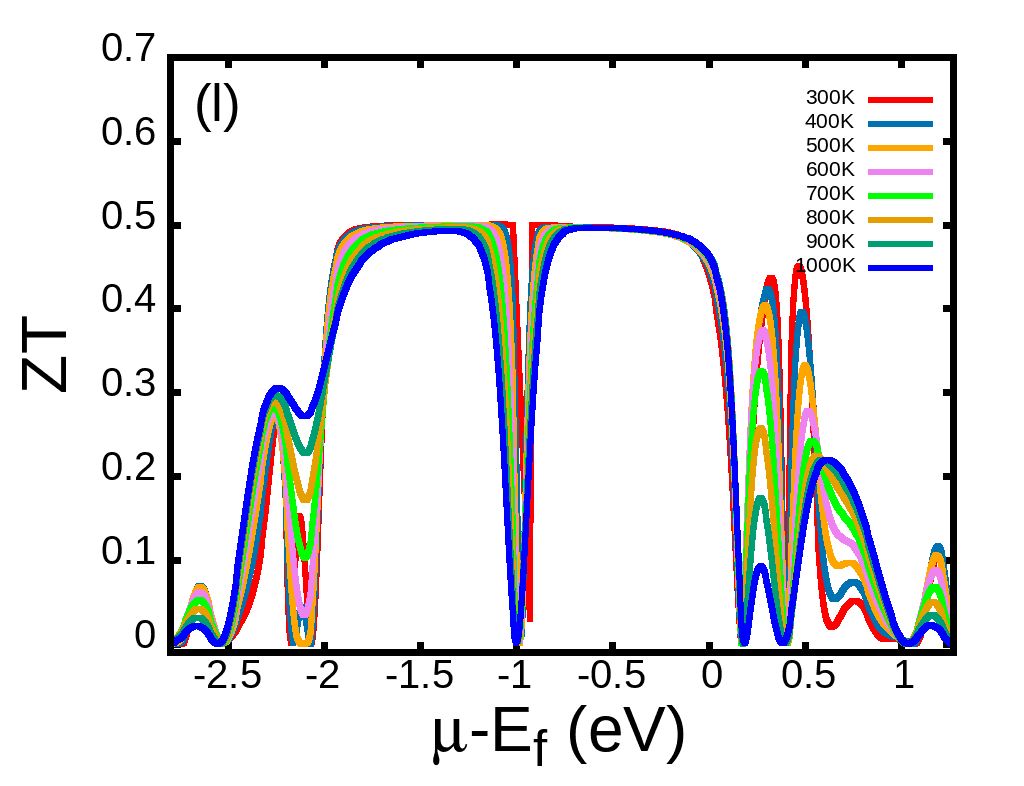}
\caption{Power factor and corresponding figure of merit of BCZT for (a,g) a6, (b,h) a7, (c,i) p6, (d,j) p7, (e,k) $r6$ and (f,l) r7.}
\label{power}
\end{figure}

Power factor $S^{2}\sigma$ is the measure of effective thermopower or thermoelectric efficiency of a material. 
There is a significant reduction in power factor observed in $a7$ as compared to $a6$. 
Despite its high electrical conductivity, $p7$ has low $PF$ value because of the comparatively lower Seebeck coefficient and high thermal conductivity. The power factor is highest in the $r6$ structure among all crystals. Around room temperature, the highest $PF$ is observed in the valence band, but as the temperature increases, the maximum shifts to the conduction band. 
Power factor for $r7$ is robust and highly nonlinear which gets aligned with very high temperature but not feasible in lower or room temperature. For different crystals, power factor plots are shown in Fig. \ref{power} and the values are provided in the Table S VII.

Figure of merit ($ZT$) is a dimensionless quantity and its standard value for a thermoelectric material should be $\geq 1$. This can be improved by increasing the power factor that depends on Seebeck coefficient, electrical conductivity or mobility and low thermal conductivity. Despite multiple characteristics all structures showed nearly the same value of figure of merit. 
$a6, a7$ and $r6$ are more reliable for thermoelectric properties as they have consistent value for a wide range of energy whereas $r7$ comes next in order, followed by $p7$ and then $p6$. For our case the figure of merit is around 0.5 and this is due to the high thermal conductivity. Suitable dopant may be used to tune high \textbf{S}, $\boldsymbol{\sigma}$ and low $\boldsymbol{\kappa}$ for creating the structural distortion, can enhance $ZT$. 
The figure of merit curve becomes smoother with increase in temperature, for example in tetragonal structure, slight distortion in the valence band peaks is observed at low temperature which flattens and reduces at high temperature. The variation of $ZT$ with chemical potential for different crystals is shown in Fig. \ref{power}. 

\section{Conclusion}
A detailed investigation of structural, electronic, vibrational and optical properties of BCZT for three different structures have been carried out using a first principles calculation. The results show that BCZT can be an excellent lead-free ferroelectric/piezoelectric material for device applications. These materials exhibit a wide band gap, ensuring band gap engineering with suitable doping materials, such as Mn, can show strong photo-pyroelectric effect \cite{Lu2024}. These materials are quite stable in different crystallized forms which can be inferred from the tolerance factor, formation and cohesive energies. 
It shows promising absorbing properties with high optical conductivity and low reflectance in the visible range (see Fig. \ref{optic1} \& \ref{optic2}) which is favorable for optoelectronic devices. Within the limit of density functional theory the calculated piezoelectric constant is reasonably high as compared to the commercially available piezoceramics.
The calculation is strongly supported by the Born effective charge, which is particularly high for the Zr atom. The detailed investigation of thermal and thermoelectric properties indicates suitability in thermoelectric devices that can be estimated from power factor and figure of merit. The value of the figure of merit is $\sim 0.5$ which may be due to the significantly high thermal conductivity which can be further tuned by defect and bandgap engineering. Our analysis provides a deep insight to the complex perovskite BCZT in different crystal forms for various potential device applications. 

\section*{Acknowledgements}
The authors acknowledge the High Performance Computing (HPC) facility provided by the Center of Excellence in High Energy and Condensed Matter Physics, Department of Physics, Utkal University, India. DP acknowledges the fellowship from RUSA, Utkal University, India.

\end{document}